\documentclass[12pt,preprint]{aastex}

\newcommand\msun{$M_{\sun}$}

\newcommand\hb{H${\beta}$}
\newcommand\ha{H${\alpha}$}
\newcommand\lya{L${\alpha}$}

\newcommand\si{S~{\sc i}}
\newcommand\sii{S~{\sc ii}}
\newcommand\siii{Si~{\sc ii}}
\newcommand\siiii{Si~{\sc iii}}
\newcommand\siiv{Si~{\sc iv}}
\newcommand\oiii{O~{\sc iii}}
\newcommand\oiv{O~{\sc iv}}
\newcommand\ov{O~{\sc v}}
\newcommand\oii{O~{\sc ii}}
\newcommand\oi{O~{\sc i}}
\newcommand\heii{He~{\sc ii}}

\newcommand\ci{C~{\sc i}}
\newcommand\ciii{C~{\sc iii}}
\newcommand\cii{C~{\sc ii}}
\newcommand\civ{C~{\sc iv}}

\newcommand\neiv{Ne~{\sc iv}}
\newcommand\niv{N~{\sc iv}}
\newcommand\niii{N~{\sc iii}}
\newcommand\nv{N~{\sc v}}
\newcommand\nev{Ne~{\sc v}}
\newcommand\sm{M$_{\odot}$}
\newcommand\feii{Fe~{\sc ii}}
\newcommand\mgii{Mg~{\sc ii}}
\newcommand\mgv{Mg~{\sc v}}

\newcommand\arv{Ar~{\sc v}}

\slugcomment{}
\shortauthors{Stanghellini et al.}
\shorttitle{UV spectroscopy of LMC planetary nebulae} 

\begin{document}

\title{{\it Space telescope
Imaging Spectrograph} ultraviolet spectra of LMC planetary nebulae. A study of
carbon abundances and stellar evolution.
\footnote{Based on observations made with the NASA/ESA Hubble Space Telescope, 
obtained at the Space Telescope Science Institute, which is operated 
by the Association of universities for research in Astronomy, Inc., under
NASA contract NAS 5--26555}}

\author{Letizia Stanghellini\altaffilmark{2}}
\affil{National Optical Astronomy Observatory, 950 N. Cherry Av.,
Tucson, AZ  85719, USA; letizia@noao.edu}

\author{Richard A. Shaw}
\affil{National Optical Astronomy Observatory, 950 N. Cherry Av.,
Tucson, AZ  85719, USA; shaw@noao.edu}

\author{Diane Gilmore}
\affil{Space Telescope Science Institute, 3700 San Martin Drive,
Baltimore, Maryland 21218, USA; dkarakla@stsci.edu}

\altaffiltext{2}{On leave from INAF-Bologna Observatory}

\begin{abstract} 
We acquired spectra of 24 LMC PNe in the 1150--3000 \AA~range in order to 
determine carbon and other ionic
abundances. The sample more than doubles the number of LMC PNe with good
quality UV spectra in this wavelength range, and whose optical images are 
available in the {\it HST}
archive. The {\it Space Telescope Imaging 
Spectrograph} was used with a very large aperture to obtain virtually slit-less
spectra, thus the monochromatic images in the major nebulae emission lines
are also available. The analysis of the data shows extremely good quality spectra.
This paper presents the emission lines identified and measured, and the 
calculation of the ionic abundances of the emitting carbon and other ions, and total carbon abundance. P-Cygni profiles have been
found in a fraction of the nebulae, and the limiting velocities of the
stellar winds estimated. 
The total carbon abundance can be inferred reliably in most nebulae. We found that
the average carbon abundance in round and elliptical PNe is one order of magnitude larger than
that of the bipolar PNe, while elliptical and round PNe with a bipolar core have a 
bimodal behavior. This results confirm that bipolarity in LMC PNe is tightly correlated with high mass progenitors.
When compared to predicted yields, we found that the observed abundance ratio show a shift 
toward higher carbon abundances, that
may be due to initial conditions assumed in the models not appropriate for LMC PNe. 

\end{abstract}


\section{Introduction}

Planetary Nebulae (PNe) have been studied for decades, and through those 
efforts has come an understanding of 
the final phases of stellar evolution of stars with masses in 
the $~1$ -- 8 \msun~range.  
Since PNe are gas shells ejected by evolved stars, they
refuel the ISM with elements processed by stellar evolution. Thus PNe are 
ideal probes to test the theory of stellar evolution itself, and to study 
the ISM enrichment from their parent stars in a quantitative way. 

The scientific importance of studying 
Planetary Nebulae in the Large Magellanic Cloud (LMC) can be
very simply summarized: LMC PNe are absolute probes of 
stellar and nebular brightness, luminosity, and size, because their distance
is known (unlike the distances to Galactic PNe which are typically 
uncertain by $\gtrsim50$\%) 
and because their interstellar extinction is comparatively low 
(unlike Galactic PNe, where the disk and bulge PN populations are 
underrepresented in most samples).

Planetary nebulae are ejected toward the end of the evolution of low-
and intermediate-mass stars, after the thermally-pulsing asymptotic giant 
branch (TP-AGB) phase.  The gas ejected at this phase contains elements 
that have been produced during the previous evolutionary stages, and then 
carried to the stellar surface by the convective dredge-up processes 
(Iben \& Renzini 1983; van den Hoek \& Groenewegen 1997, hereafter HG97).
The single-star evolution models predict a chemical enrichment of the 
outer region of evolved stars in the 1 -- 8 \msun~range, which can be 
summarized in the following four points:
(a) During the first Red Giant phase, the convective envelope penetrates
regions that are partially CNO-processed. This dredge-up results in a $^{13}$C 
and $^{14}$N enhancement, and depletion of $^{12}$C.
Later, He-burning starts in the stellar core, marking the beginning 
of the Horizontal Branch phase.  
Helium burning then continues in the helium shell. 
Finally, the hydrogen and helium burning occur alternately 
in two nuclear burning shells surrounding the CO core, and the star
ramps up to the TP-AGB;
(b) The second convective dredge-up (for stellar masses larger than
$\approx$3 \msun ) occurs at the onset of the AGB,
when the H-burning is quiescent. This process carries $^4$He, $^{13}$C,
and $^{14}$N-rich material to the stellar surface;
(c) During the TP-AGB phase, the envelope is able to dredge-up material 
after each thermal pulse, carrying $^4$He, $^{12}$C, and the s-process elements 
to the surface;
and (d) For stars with masses larger than $\approx$3 \msun, one final process 
is thought to occur that alters the chemical composition: the so-called 
"hot-bottom burning" (HBB) that processes most of the carbon into nitrogen. 

Stellar evolution models predict the sequence of processes given 
above for Galactic and LMC PN progenitors alike (HG97). 
The key to assessing the above predictions is to measure the abundances 
of the processed elements, particularly C, N, and O, which is straightforward 
to do with PNe. 
Carbon depletion and nitrogen enrichment 
also depend on the progenitor mass, which forms a direct connection between
observed progenitor mass (i.e. Population) and chemical content.
Evolutionary models show that carbon depletion and nitrogen enrichment in higher
mass progenitors occur 
in models with a variety of initial chemistry,
thus, by measuring the C and N abundances in PNe one can 
at once validate key elements of stellar evolution theory, and 
make a more robust measure of the contribution of low- and intermediate-mass 
stars to the enrichment of the local ISM.  

It is important to note that most central stars are found at low luminosity 
on the HR diagram, where even a very small error in the luminosity 
translates into a large error in the inferred mass \citep{2000ApJ...542..308S}. 
The great advantage of performing such an abundance study on LMC PNe 
is that we know their distance, thus a direct 
comparison of models and data is possible, in principle, by determining the mass of the
central stars from their location on the HR diagram and comparing 
predicted and observed elemental abundances. 

Abundance calculations require optical and ultraviolet nebular lines 
to support a reliable analysis of all the ionized stages of the relevant
ions.  Optical spectra of both LMC and Galactic PNe can be acquired 
from the ground. 
By studying extra-Galactic PNe, we have shown that morphology 
is a surprisingly useful indicator of 
the progenitor stellar population (Stanghellini, Shaw, Balick, \& Blades 2000, hereafter SSBB).
From a combination of morphology (determined from the images)
and abundances, SSBB demonstrated that bipolarity 
in LMC PNe is indicative of a younger stellar 
progenitor. This was shown by the high abundance of alpha-elements
(Ne, S, Ar), and 
other elements not altered by stellar evolution of these stars. 
SSBB found that nitrogen 
enhancement and carbon depletion are also mildly correlated with bipolarity
(see Figures 1 and 2 in SSBB). 
These two main results were broadly consistent with the predictions of stellar evolution 
if the progenitors of bipolar PNe have on average larger masses than the progenitors of round 
and elliptical PNe. 
The SSBB results are based on a rather small sample of objects: 
The carbon abundance was known at that time for only 18 LMC PNe with known morphology. 

While the UV diagnostics emission lines are 
available in the IUE archive for several bright LMC PNe, and 
in the HST archives for a few others acquired with the FOS, most
of these LMC PNe do not have spatially-resolved optical
counterparts. More importantly, {\it none of their central stars have been
observed directly}. The first set of central stars of LMC PNe available in the
literature has been acquired with HST program 8271 \citep{sha01,sta02}, and the stellar 
physics has been studied by \citet{cspaper}. 
Ultraviolet spectra
of most of these targets are not available in the literature, nor
in the HST and IUE Data Archives.  To study how the chemical evolution
of PNe relates to the evolution of their stars and morphology, we have acquired low dispersion UV 
STIS spectra of those LMC PNe for 
which narrow-band images have been acquired with {\it HST}, but for 
which no UV spectroscopy existed. 
This total of 24 targets greatly improves the statistics of LMC PNe whose 
carbon abundance has been determined, and it constitutes the first data set 
ever available 
to study central stars of PNe and their nucleosynthesis without the distance 
bias well known in Galactic PNe. 
This paper deals with the observed nebular properties in the UV, and presents limited 
comparison of the data to theoretical yields.
In $\S$ 2 we present the observations and the data analysis of the
24 target PNe. Section 3 includes the study of the emission lines and the calculation
the ionic and carbon abundances, including the comparison between the derived abundances
and the theoretical yields, and the comparison of our abundances to those of Galactic PNe. 
In $\S$4 we summarize and discuss the conclusions
of our findings.
The analysis of the stellar
continua, the detailed modeling of the central stars winds, and the detailed
modeling for each nebula will be published in future papers.

\section {Observations and data analysis} 

We observed 24 LMC PNe 
using the STIS MAMA
detectors with the first-order G140L and G230L gratings for long-slit
spectroscopy, covering the NUV-FUV wavelength range from 1150 to 3100
\AA. The spatial scale of the spectrograms is 2.44$\times$10$^{-2}$
arcsec~pix$^{-1}$, an excellent resolution to observe
the morphology of the emission lines in LMC PNe that are typically
half an acrsec wide. 
The G140L grating has central wavelength 
$\lambda_{\rm c}$=1425 \AA~~and the spectral range is 590 \AA, with nominal 
point source
spectral resolution of 1425 at $\lambda_{\rm c}$ \citep{STIS_HB}. The G230L grating 
has $\lambda_{\rm c}$=2376 \AA~~and $\Delta \lambda$=1616 \AA, with 
spectral resolution of 740 at $\lambda_{\rm c}$. The spectral 
resolution of our near slit-less
spectra depends on the extension of the feature, and it is typically 
much lower than the nominal point source resolution.
The observing log is presented in Table~\ref{ObsLog}.

We observed the nebulae using
a large 6\arcsec~x 6\arcsec~aperture to produce monochromatic
images of the nebulae in each
emission line. In these images, spatial information is convoluted with
spectral information in the dispersion direction, practically as in
a slit-less spectrum.  However, most
spectral emission lines are well-separated, providing a
clear picture of the spatial morphology in the UV emission lines. 
Also, because most LMC PNe are angularly small, total
uncontaminated fluxes are easily determined. 
While no comparison arcs were acquired related to these spectra, their pattern of
separations, together with the known wavelength limits of the gratings,
provide for unique identification of the lines in all but a few cases. 
For the rest, identification was based on the most probable features
expected. 

To measure the fluxes of emission features, images were collapsed in
the spatial direction by extracting a spectrum using the {\bf x1d} task in
the IRAF\footnote{IRAF is the Image Reduction and Analysis Facility, a 
general purpose software system for the reduction and analysis of astronomical data. 
It is distributed by the National Optical Astronomy Observatories, which are operated 
by the Association of Universities for Research in Astronomy, Inc., under cooperative 
agreement with the National Science Foundation.}  {\bf stis} package \citep{CALSTIS1, CALSTIS2}.  
The flux-calibration of extracted spectra
relies, in part, on the quality of the spectral wavelength calibration.
For each observation the position of the source in the dispersion
direction is unknown. 
  
Since the reference wavelengths are suitable only for a source
centered in the aperture and with  velocity, it was necessary to
determine the offset between the reference wavelength positions and the
source wavelengths prior to flux calibration. The process we used is
outlined below in section 2.2.

\subsection{Data Reduction}
 We obtained the bias-subtracted and dark-corrected spectral images,
as processed with {\it calstis} \citep{CALSTIS1,CALSTIS2} from the HST archive. Spectra obtained
with large apertures were not automatically flat-corrected during OTFR
since it was believed at one time that the flat-field might be
wavelength-dependent. The flat-field correction was applied separately to the data 
using the most appropriate reference files available (those generated from slit data).  
This is a two-step process. This correction is small for typical PN fluxes.  
High-resolution PFLATS, or pixel-to-pixel flats, have 5-6$\%$
peak-to-peak variations in alternating columns, but these are rebinned
by a factor of two in each dimension when applied to the science data.
Rebinning to a lower resolution smoothes out the effect, so that only
the smallest, brightest emission line fluxes are affected 
at the 1-2$\%$ level. This is comparable to the flux errors of only the
highest S/N PN spectra. Typical PN emission features are more extended
and less bright, and flat-field correction has a much smaller effect.

   In addition to the PFLATS, LFLATS, which contain the large-scale
flat-field variations were also applied to the science data during
flat-field correction. These remove the spatially-dependent 
sensitivity variations, and are normalized to a value of one in the 
standard extraction region. The spatial dependance is negligible in the
G230L data, and LFLAT calibration for these is routinely not performed,
and was not for our G230L PN data. The flat-field effects are more
pronounced in the FUV (the G140L data). All our spectra are extracted
from a region where there are at most a few percent variations in the
LFLAT. The LFLATs contain variations on scales larger than about 60
pixels, and are by definition 1.0 in the standard extraction region.
They contain only the spatial-dependance of the flat-field correction.
The task
{\bf x1d} removes the spectral-dependance.

Furthermore, {\bf x1d} corrects for sensitivity variations across the field
(interpolated between certain values of A2CENTER in the 
SPTRCTAB), but
it assumes the spectrum is at the aperture center in the dispersion
direction. We used an iterative procedure to correct for the 
wavelength offset.  If the source is not centered well, the effects on
the flux calibration are worse at the edges of the spectrum.  Applying
the wavelength offset correction to SHIFTA1 to rest wavelength as we do
when we extract the spectrum with {\bf x1d} should remove most of the effect.
The other potential effect on fluxes is due to the wavelength offset
caused by the radial velocity of the source, which was not accounted 
for.  That could potentially affect the goodness of the sensitivity
correction that is applied with {\bf x1d}. However, we calculated a
difference of only a few pixels for the radial velocity offset, so we
believe that it will not affect the fluxes at a significant level.

\subsection{Spectral extraction}

The height and width of the extraction region were determined at the
image center in the dispersion direction. The y-height was taken as the
visual centroid of continuum emission, where present.  Continuum
emission from either the central star, or in some cases, the nebula
itself, were used. In some cases it was possible to further refine the
spatial center during the extraction by cross-correlation with the
appropriate line profile. 

 The extraction width was determined by visual examination of the
brightest nebular emission line feature, and was later enlarged if it was
found to be smaller than 1.5 times the FWHM of the measured feature in the
extracted spectrum.  For a gaussian intensity distribution, about 96$\%$
of the light falls within this limit.  For flatter distributions, more
of the light falls outside the limit, and the measured flux represents
a conservative estimate of the total flux in the line. For those
sources with flatter wings, we used a Voigt function to more accurately
measure the flux. 

 Since the two-dimensional spectra are not perfectly aligned with the
image rows, spectra extracted from different y-heights of the
two-dimensional  image have non-parallel spectral traces along the
dispersion direction. The traces for a sampling of heights are provided
in the reference file SPTRCTAB found in the image header
and are used by the extraction routine,
{\bf x1d}, to appropriately extract spectra. The traces are interpolated for
heights between listed table values.

Background-subtraction was performed during spectral extraction. The
background contribution was measured in large areas away from the
nebular spectrum and any continuum spectral sources in the field. This 
process helped remove the geo-coronal lines that are prominent in the G140L
data.

For some PNe, like SMP~19 and SMP~71, 
different emission lines of the same nebula have different spatial extensions.  
In most cases where this is true, the
centroid of the intensity distribution is the same, but the gradient is
different. For a few of the nebulae we chose to extract spectra
using two different widths in order to more appropriately measure the
fluxes in the various  features.  Where larger extraction regions are
used, the measured flux is subject to larger noise errors.

In several cases the stellar continuum is clearly distinguished in the 2D spectra.

\subsection{Aperture correction}

The aperture correction is appropriate for point
sources, since by extracting 1D spectra with very narrow widths
one may undercount the total flux by not including the flux scattered
outside the aperture itself. 
The
correction is applied in  {\bf x1d} for specific extraction widths.
The task {\bf x1d} selects the most closely-matching extraction 
width and applies the corresponding aperture correction found in PCTAB.  
 However, since the nebulae are all of different spatial extent (few
of which are point-sources) and all have been measured with various
aperture sizes, it was decided to measure the fluxes without aperture
corrections.   We instead try to estimate the size of the aperture
correction for a few different nebulae, and incorporate it into a flux
error. 

\subsection{Feature identification and flux measurement}

The extracted spectra were plotted and the
locations of the gaussian centers of detectable features were measured
with the IRAF task {\bf splot}.
Wavelength offsets were determined by comparing 
feature centers
measured with {\bf splot}
 to the wavelengths of lines identified with our own IDL
routine. The IDL routine plots an array of marks representing the locations of
possible emission features onto the two-dimensional spectral image. The
array is shifted until the best visual alignment is achieved. More
weight is given to matching prominent UV lines predicted  by current
theory of PN evolution (Cox 2000; Feibelman, Oliversen, Nicholsbohlin,
\& Garhart 1988). Once the line identifications are
established, comparison of the feature centers measured with {\bf splot} and
the rest wavelengths of the identified lines yields the wavelength
shift of the extracted spectrum. Conversion to pixel offsets was done
using the average dispersion of the spectrum for the given grating, and
the image header keyword SHIFTA1 was updated with the computed
offset.  For PNe with few emission lines, stellar emission or
absorption features were used to help wavelength-calibrate the spectra
whenever possible. Additionally, the sky emission at \lya~ and \oi~ 1304
aided wavelength calibration of the G140L spectra.  Sky emission from
these lines fills the aperture and creates large boxes of emission
superimposed on the nebular spectrum.  These geocoronal lines were used
primarily as a guide to exclude alternate wavelength solutions when the
solution was otherwise uncertain.

 Spectra were then re-extracted with {\bf x1d} which used the header
SHIFTA1 values to more accurately flux-calibrate the extracted spectrum.
This method allows for a first-order correction of the wavelengths. 
Two other factors add to the wavelength-solution uncertainty: the
spatial extent of the source in the dispersion direction (especially
diffuse or complex sources like SMP~30 and SMP~93), 
and the source's radial  velocity.  The
line-of-sight heliocentric velocity to the LMC is about 270 km sec$^{-1}$. 
For the gratings
used, this velocity amounts to no more than a few pixels offset in the
dispersion direction.  Sources with emission features having a
non-symmetric spatial distribution will further complicate the
determination of unique wavelengths across the spectrum. 
The resultant wavelength correction then will depend on several factors: the
number of features present, their spatial extent, intensity, and
symmetry.  For some nebulae, we used the interstellar absorption lines in
the spectra of nearby stars
to help determine the wavelength solution whenever
possible.

To measure the flux in the line, 
Gaussian fits were performed. During
measurement, the cursor is placed at the level of the continuum on
either side of the feature before the feature is fit with a Gaussian.
This procedure removes the contribution of the underlying stellar
continuum, if present, from the flux measurements.  When close-lying 
spectral features appeared in the extracted spectra, de-blending
was done using the {\bf d} feature in {\bf splot}, which solves for several
superimposed spectral features with the same baseline.

We have identified several nebular emission lines and a few stellar
P-Cygni features in our sample of PNe. 
We observed the helium recombination lines 
at 1640, 2511, and 2733 \AA, corresponding respectively to the \heii~
Balmer-$\alpha$, Paschen-$\gamma$, and Paschen-$\beta$ lines.
We saw, as expected, many of the carbon features that were the 
essential motivation for this science program, namely
\cii~ $\lambda$1335/36 and 2837/38 (both in emission and as a P-Cygni profile); 
\cii]~ $\lambda$ 2325-29; 
\ciii~ $\lambda$1175/76 (possibly collisionally populated),
a feature at $\lambda$1247, possibly a \ciii~ emission line; 
\ciii]~ $\lambda$1907/09; 
and \civ~ $\lambda$1548/50.
We observed the following emission lines from the oxygen ions:
[\oii]~ $\lambda$2470; 
\oiii]~ $\lambda$1658-1666; 
[\oiii]~ $\lambda$2321/31; 
the Bowen fluorescence emission of \oiii~ $\lambda$3043/47 and 3133;
the possible \oiv~ $\lambda$1342 in SMP~102 only; 
\oiv]~ $\lambda$1407; and 
\ov~ $\lambda$1371, with a P-Cygni profile in SMP~79.
We have also identified the following features :
\niii]~ $\lambda$1747-1754; 
\niv]~ $\lambda$1483/87; 
\nv~ $\lambda$1239/43 (with P-Cygni profile, typical of O stars);
[\neiv]~$\lambda$1602 and 2423/25; 
[\nev]~ $\lambda$1575; 
[\siii]~ $\lambda$2334-50 (in SMP~28); \siiii]~ $\lambda$1883 and 1892; 
\oiv]~ $\lambda$1397/1407. 
Finally, some lines whose identification remains uncertain are 
\sii~ $\lambda$1915 (in SMP~28); 
a possible \feii~ feature at 2629 \AA~~(in SMP~28); 
\mgii~ $\lambda$2796/2803, and \mgv~ $\lambda$2784/2929.

In Table~\ref{Fluxes} we give the complete set of observed emission lines. Column (1)
gives the PN name, column (2) and (3) the laboratory wavelength and identification of
the emission line(s) respectively, column (4) gives the flux ratio of the
emission line relative to \hb, columns (5) and (6) the extinction correction
of the targets for the Galactic and LMC contributions, and column (7)
gives the corrected line ratio (see $\S$2.5). Finally, column (8) gives comments on the measured 
fluxes. The \hb~ fluxes in these ratios are from the STIS/{\it HST} data from \citet{sta02}. These \hb~ 
fluxes were neasured by extracting 1D spectra from the 2D (slitless) optical 
spectral images with the same apertures for given PNe than used in the 1D extraction of the UV spectra.

The P-Cygni profile analysis is given
in Table 3. The terminal velocities were measured at the
violet edge of the P-Cygni absorption feature. This gives a good indication
of the terminal velocity for well-saturated lines. In the cases that we present here,
most of the P-Cygni lines are not totally saturated, but they are very strong, and their
measurement
gives a good estimate of the velocity \citep{lamers}. In most cases we have two measurements
of the terminal velocity for each of the nebulae in this Table. A good
estimate of the terminal velocity could then be the average of these two measurements. 
In some cases, one of the
features had very low S/N, as noted in the table notes, and the other velocity measurement
should be taken for further use.

The 1D spectra of all targets 
are plotted in Figures 1 through 24. In order to make the plots more legible, we have 
smoothed the spectra with a 10 pixel box-car (naturally all measurements were 
obtained from the original data). We use the excitation constant (EC) 
from Morgan (1984), when available (see $\S$3.2 and Table 4).
Information on the central star temperature is taken from the Zanstra 
analysis by \citet{cspaper}.

While stellar and detailed nebular models are not the intent of this paper,
we could get an estimate of the effective temperature of the central stars in the cases
where the stellar continua were clearly present in the 2D spectra and, short-ward of 2200 \AA,
could be fitted with a simple black body
function. We use these estimates only to double check the consistency between Zanstra 
temperatures and emission spectra. Detailed nebular and stellar models, 
including  the UV and optical data, will be presented in a future paper.

In the following text we describe in some detail the spectra of our targets.

{{\bf J~41}:
This is an intermediate excitation nebula with central star temperature T$_{\rm Z (He II)}$=
60,000 K. The G140L spectrum is very noisy and does not show other features than a
possible trace of the \civ $\lambda$1548/50 line. Black-body fitting of the continuum confirms the Zanstra temperature.
There is a trace of emission at 1220 \AA.
The G230L spectrum clearly shows the \ciii] $\lambda$1907/09
feature, but no other obvious emission lines (see Fig.~1). Absorption lines
are identified with $\lambda$2380 \heii~ and $\lambda$2799 \mgii.}

{{\bf SMP~4}:
A moderately high excitation nebula, whose Zanstra temperature, T$_{\rm Z (He II)}$=90,000 K. 
the stellar continuum is prominent in the
2D image, and the Zanstra temperature
agrees with the black-body fitting of the stellar continuum. 
The nebular
morphology of the 2D line images is similar to that of the optical images \citep{sha01}.
We detected and measured the \cii~ $\lambda$1335/36 resonance line, with 
superimposed interstellar absorption. 
Possible absorption features are detected at \lya, \siii~ 1309 \AA, \cii~ 1335/36 \AA,
and \sii~ 1253 \AA. Evidence of dust absorption is evident at about 2200 \AA. The flux in the [\nev] 
emission at 1575 \AA~ is marginal, while its presence
agrees with the nebular high excitation (see fig.~2). } 

{{\bf SMP~9}:
This is a moderately high excitation PN, with EC=6-7. (Fig.~3).
The 2D image shows nebular continuum and almost no stellar continuum.
The major UV lines are all as spatially extended as the
optical emission lines. We marginally detect a feature around 
$\lambda$1400 \AA, probably \siiv, or \oiv], or a blend of the two components.
We also marginally detect the \mgii~ emission around $\lambda$2800 \AA, corresponding
to the blend of the $\lambda$2796/2803, 2800 features. }

{{\bf SMP~10}:
An intermediate excitation nebula (Fig.~4), with EC=6-7 and Zanstra
temperature
T$_{\rm Z (He II)}$=75,000 K. The marginal detection
of the [\nev] $\lambda$1575 line agrees with the relatively
high stellar temperature. The 2D G140L spectrum clearly shows the continuum
of the central star.
We detect several emission features. The \nv~$\lambda$1239 emission
line is measured, while an additional component of the \nv~doublet
is barely detected at 1243 \AA.
Absorption features are evident at 
\lya, $\lambda$1334 (possibly \cii), and others.
The signal-to-noise ratio (S/N) is low for this extended, point-symmetric PN.}

{{\bf SMP~16}:
This is a high excitation nebula (T$_{\rm Z (He II)}$=142,000 K,
EC=8) with plenty of emission lines in its
spectra (see Fig.~5). The stellar continuum is almost completely
absorbed in the G140L spectrum, while there is evident nebular continuum.
The \oiii~ $\lambda$1658-1666 line is measured above the wings of the \heii~$\lambda$1640 feature. A feature around $\lambda$2150 \AA~ 
is unidentified.
We marginally detected a group of emission lines around 2740 \AA, possibly \oii.}

{{\bf SMP~18}:
The G140L spectrum of this 
nebula is dominated by the absorption lines (Fig.~6). The spectra are
characteristic of a low-excitation nebula (EC=2-6), where the \heii~ lines are barely seen below
the 5 percent level.
The black-body temperature, measured by UV continuum fitting, is around 40,000 K.
We detect two P-Cygni features, corresponding to \nv~$\lambda$1239/43 and
 \civ~ $\lambda$1548/50. We also detect sky absorption lines 
at \lya, $\lambda$ 1304 \oi, 1335/36 \cii, and 1371 \ov, and possibly others. 
A possible emission at $\lambda$ 1600 \AA~ may correspond to the \neiv~ line.
The prominent \ciii]~ $\lambda$1907/09
line is the only extended emission in this PN. Other absorption lines, notably
\mgii~ at $\lambda$2799, and some Fe features, are present in the G230L spectrum.
A possible emission feature blended with \ciii] is detected at $\lambda$1993 \AA.}

{{\bf SMP~19}:
A high excitation nebula (T$_{\rm Z (He II)}$=144,000 K, EC=7),
it shows faint nebular continuum (Fig.~7). 
The marginal presence of [\nev] $\lambda$ 1575 confirms the high-excitation
temperature.
Both spectrograms show several emission lines, 
including the \oiii~ lines. 
Most of the nebular lines are extended in the 2D spectra, similarly to the
optical emission lines. As in SMP~4, there is evidence of dust absorption
around 2200 \AA. }

{{\bf SMP~25}:
Similarly to SMP~18, the G140L spectrum (Fig.~8) presents P-Cygni profiles corresponding
to the photospheric \nv~ $\lambda$1239/43 and \civ~$\lambda$1548/50 winds. 
In the G230L spectrum, the nebular emission
features corresponding to \ciii]~ $\lambda$1908 and \cii]~ $\lambda$2325-29 are 
as spatially extended as the optical emission lines. The stellar continuum 
is visible in the 2D spectra. }

{{\bf SMP~27}:
Another photosphere-dominated G140L spectrum for this low excitation (EC=3-6)
quadrupolar PN (Fig.~9). The continuum fitting gives a stellar
temperature around 30,000 K. the G140L spectrum 
presents P-Cygni profiles corresponding
to the photospheric \nv~$\lambda$1239/43 and \civ~$\lambda$1550 winds. 
Several absorption lines are observed in the G140L spectrum, including
\lya.
Very interesting is the G230L spectrum, where we can see the nebular \ciii]~
emission at $\lambda$1908, and several absorption features including an
absorption feature around $\lambda$2598, that could be identified as
\feii.}

{{\bf SMP~28}:
Very rich spectra (Fig.~10) of a moderately high excitation PN (EC=6-7).
The group of features around 1400 \AA~~are hard to identify unambiguously. 
It could be the \siiv~plus two or more of the \oiv~emission lines around 1400
\AA. The blend around 1900 \AA~~is probably a combination of the \siiii~ lines 
($\lambda$1883, 1892), \ciii] $\lambda$ 1907/09, and \si~$\lambda$1915. 
The G230L spectrum presents an unidentified
two-component feature at 2145 \AA,
possibly \feii~ $\lambda$2630, and  \oiii~$\lambda$3133.
It looks like there are two components to most emission lines. The optical slit-less
images also look double, thus both the optical and UV emission originates from two nebular blobs.}

{{\bf SMP~30}:
The spectra of this very extended PN have low S/N (Fig.~11). This is a high excitation PN, with EC=8.
The [\nev] feature at $\lambda$ 1575 that would confirm the high excitation status could
not be correctly measured, but its presence
can't be excluded given the low signal to noise ratio.
The continuum is typically nebular, thus no black body temperature could be estimated. 
The feature at 1403 \AA~ could be interpreted either as a blend
of the \oiv~multiplet, or as \siiv. The G230L spectrum is 
characterized by nebular continuum, and by the \ciii] $\lambda$1909 line. The
wavelength calibration of the G230L spectrum is somewhat
uncertain, since the only three features available are very broad and 
have low S/N. An unidentified features has been observed at $\lambda$2135 \AA.}

{{\bf SMP~34}:
A low excitation nebula (Fig.~12), with EC=2. While Zanstra analysis gives
T$_{\rm Z (He II)}$=68,000, the UV continuum fitting suggests a somewhat lower temperature
($\approx$40,000 K).
As with SMP~18, SMP~25, and SMP~27, the G140L spectrum shows P-Cygni profiles
corresponding to the photospheric $\lambda$1239/43 \nv~and $\lambda$1548/50 \civ~ 
wind features. Several sky lines are seen in absorption, overlying a strong stellar
continuum.
}

{{\bf SMP~45}:
Very large nebula, and low S/N spectra with few clear emission features (Fig.~13). 
The excitation is moderately high (EC=6). The wavelength calibration of the G230L
spectrum is uncertain, thus the features listed in Table 2 could
be misidentified. As in SMP~30, a diffuse nebular continuum is present in both spectrograms.}

{{\bf SMP~46}:
A high excitation nebula (Fig.~14) with T$_{\rm Z (He II)}$=119,000 K
and EC=6, its high S/N spectra
shows all the major carbon emission lines. The spectrogram also shows a diffuse nebular continuum 
like SMP~45 (Fig.~14).}

{{\bf SMP~48}:
Absorption is present at \lya. The P-Cygni profiles correspond
to \nv~ $\lambda$1239/43 and \civ~ $\lambda$1548/50 (Fig.~15).
The G230L spectrum shows the \ciii]~ and \cii]~ lines, an absorption feature at 2381
\AA~~(possibly \heii), a possible absorption feature around 2596 \AA~~(\feii),
and
\mgii~ at $\lambda$2800. This is a low excitation nebula (EC=3-4) with central star
temperature T=30,000 K inferred from the continuum fitting. The continuum
shows a dominant
stellar component.}

{{\bf SMP~59}:
Extended quadrupolar PN with high central star temperature 
(T$_{\rm Z (He II)}$=98,000 K) and high excitation (EC=7-9) (Fig.~16). 
There is marginal
detections of [\nev] at $\lambda$ 1575. The 2D spectra are adjacent to the edge of the aperture. 
The extracted 1D spectrum may not include the whole nebular flux
in the emission lines. We estimate, from the 2D image, that less or 
about 20 $\%$ of the flux may be not accounted for in the 1D extracted spectra. }

{{\bf SMP~71}:
Moderately high excitation PN (T$_{\rm Z (He II)}$=83,400 K and EC=6-7), with a 
very rich spectrum (Fig.~17). Like SMP~45 and SMP~46, the G140L spectrum
shows a diffuse
nebular continuum, and the stellar continuum. An unidentified emission feature is evident at 1520 \AA,
in the wing of the \civ~ feature. }

{{\bf SMP~72}:
An extended PN whose optical image was taken with the early PC1 \citep{dop96}, 
and whose UV
spectra show very low S/N (Fig.~18). The 2D spectra indicate the presence of stellar continuum. 
Both the \civ~ $\lambda$1550 and the \heii~ $\lambda$1640 
emission lines are broad, due to the large spatial extent of the nebula, and marginal splitting. This is a relatively 
high excitation nebula (EC=7), though [\nev] $\lambda$1575 is not detected.}

{{\bf SMP~79}:
A moderate excitation PN (EC=5, Fig.~19), with a rich emission spectrum. 
The P-Cygni feature at 1371 \AA~~is associated with stellar \ov.
We do not know
the Zanstra temperature of the central star, but the presence of \ov~
indicates it to be an early O star, and the presence of [\nev] in the spectra 
indicates high excitation. A fit to the prominent stellar continuum gives T$\approx$60,000 K.
Another possible P-Cygni feature is shown 
in the G230L spectrum, corresponding to the \cii~ $\lambda$2837/38 emission.
Like SMP~71, and possibly SMP~19, the spectrum shows emission at $\lambda$2783 \AA~
(\mgv~ or [Ar V]?)}

{{\bf SMP~80}:
The G140L spectrum shows hydrogen in absorption, and possible
P-Cygni lines corresponding to \nv~ $\lambda$1240, and \civ~ $\lambda$1550. 
The almost featureless G230L spectrum
shows only a possible low intensity \ciii] 1909 \AA~~feature.
The G230L spectrum has sloping continuum. The excitation is low in this
PN (EC=2-4, T=30,000 K from continuum fitting, spectra in Fig.~20).}

{{\bf SMP~81}:
A very compact PN, showing \lya~absorption. The wing of
this absorption line merges with the P-Cygni profile corresponding to 
photospheric \nv~
$\lambda$1239/43 emission (Fig.~21). The determination of the strength of these
lines is thus uncertain, given the particular configuration. All the major nebular 
emission lines are present.
Absorption features are detected at 1304, 1335/36, 2380, 2591, and 2799,
 \AA, possibly corresponding to  \siiii, \cii, \heii, \feii,
and  \mgii. The nebula has intermediate optical excitation (EC=5) and T=45,000 K
from the stellar continuum fitting.}
 
 {{\bf SMP 93}:
 An unusually extended PN, its Zanstra temperature, T$_{\rm Z (He II)}$ =372,000 K, is
 likely to be an overestimate \citep{cspaper}. The nebular excitation from the optical
 lines is deemed to be high (EC=8). The 
 spectra have very low S/N so that the extracted fluxes and wavelengths
 are uncertain (Fig.~22).
 The line profiles are similar to those of SMP~72. Like SMP~45 and others, the spectrum shows
 diffuse nebular continuum. As for SMP~59, the nebula is at the edge of the aperture, and its line 
 intensities may be underestimated.}

{{\bf SMP~95}:
A moderately high excitation elliptical nebula (EC=5-7, 
T$_{\rm Z (He II)}$=146,000 K), where
the major UV lines have the same two-dimensional morphology as the optical lines. 
Many nebular emission lines appear in the spectra (Fig.~23). The continuum
is weak and appears to be of nebular origin from the 2D spectra analysis.}

{{\bf SMP~102}:
A high excitation nebula (T$_{\rm Z (He II)}$=132,000 K, EC=7),
shows a host of emission lines (Fig.~24). Absorption features are evident at \lya,
\oi~ or \sii~ at $\lambda$ 1301, and elsewhere.
SMP~102 shows a diffuse ellipsoidal light distribution in
most lines with a small circular enhancement on the redward side of a
few emission features.  This enhanced region is more prominent in
the \ciii] 1908 and \cii~ 2325-29 lines than in any other feature seen.
Neutral helium is seen in emission at $\lambda$2734. The stellar continuum is prominent in both its spectra.}

\subsection{Extinction correction}

The flux measurements have been corrected both for Galactic foreground
and for the LMC extinction. It was assumed that the internal extinction within the nebulae is negligible. Since the extinction curves
for the LMC and the Galaxy are different in the UV, we have proceeded
as follows. The relation between observed and de-reddened fluxes, 
scaled to \hb, can be written as:

$$ I_{\lambda} / I_{\beta} = (F_{\lambda} / F_{\beta})  10^{(c  f_{\lambda})} \eqno(1)$$

where {\it c} is the target-dependent logarithmic extinction at \hb~ and f$_{\lambda}$ is the 
reddening function at wavelength $\lambda$. The reddening term includes a foreground Galactic 
and a LMC component. We can express the double origin of extinction as:

$$ c  f_{\lambda} = c_{\rm G}  f_{\lambda, \rm G} + c_{\rm LMC}  f_{\lambda, \rm LMC} \eqno(2)$$

where  c$_{\rm G}$ is the reddening constant for the Galactic foreground in the direction of the LMC,
f$_{\lambda, \rm G}$ is the Galactic extinction law,  c$_{\rm LMC}$ is the reddening constant
for the LMC, and  f$_{\lambda, \rm LMC}$ is the LMC extinction law. Since the Galactic and LMC extinction
laws are different in the UV, we need to correct for the two terms independently.

We use the values of the Galactic color excess, E$_{\rm B-V}$, for 
the line-of-sight to each of our targets from 
Gochermann (private communication), and we calculate c$_{\rm G}$ =1.47  E$_{\rm B-V}$
for each target, as listed in Table 2, column (5). 
In a couple of cases the foreground reddening was not available from Gochermann 
and we have estimated E$_{\rm B-V}$ by eye, from the LMC foreground 
reddening map of \citet{1991A&A...246..231S}.

By using f$_{\lambda, \rm G}$ from
\citet{1989ApJ...345..245C}, we correct the UV fluxes by foreground extinction, and obtain:

$$(I_{\lambda} / I_{\beta})_{0} = F_{\lambda} / F_{\beta}  10^{(c_{\rm G}  f_{\lambda, \rm G})}. \eqno(3)$$

If we apply Equation (3) to the (\ha/\hb) ratio, we obtain (\ha/\hb)$_0$, an essential ingredient for
evaluating the LMC reddening:

$$ c_{\rm LMC}=2.875 \times {\rm log} [(H\alpha/H\beta)_0/2.85] \eqno(4)$$

where 2.875 is the inverse of the f$_{\lambda}$ value at \ha, and 2.85 is the assumed intrinsic ratio of \ha/\hb~ 
in the case of T$_{\rm e}$=10,000 and 
N$_{\rm e}$=10,000 K \citep{ost}.

Values of c$_{\rm LMC}$ are listed in Table 2, Column
6. If the computed foreground optical extinction constant is larger than the total optical extinction constant
from \citet{sta02}, implying that the foreground correction overshoots the
actual reddening, we set c$_{\rm LMC}$=0 instead of using Eq.~(4).

The final step of the de-reddening is the correction for the LMC extinction, by using:

$$I_{\lambda} / I_{\beta} = (I_{\lambda} / I_{\beta})_0  10^{(c_{\rm LMC}  f_{\lambda, \rm LMC})}, \eqno (5)$$

where f$_{\lambda, \rm LMC}$ is parametrized by \citet{1983MNRAS.203..301H}. 
The line intensity ratios in Table 2 have been corrected with the above procedure.

In Figure 25 we show the histograms of the foreground and LMC extinction 
constant, c$_{\rm G}$ and c$_{\rm LMC}$. We omitted one datum from the
LMC plot, corresponding to c$_{\rm LMC}$=0.96 of SMP~45, for compactness of the Figure.
The averages of the
 extinction constants are respectively $<$c$_{\rm G}>$=0.11$\pm$0.06,
 and $<$c$_{\rm LMC}>$=0.09$\pm$0.20. While the foreground correction tends
 to be more evenly distributed than the LMC factor, it is clear that adopting
 the same foreground constant for all PNe in our target list would be
 inappropriate. In studies of LMC PNe, a common
 c$_{\rm G}$ for the foreground correction is usually assumed. For example,
 \citet{1998ApJS..114..237V}
 use c$_{\rm G}$=0.074 for the Galactic foreground to the LMC PNe, which is
a moderately appropriate median value, but not at all representative
 of extreme PNe. 
 We estimate that by using the point-by-point foreground extinction we
 have lowered errors in the measured de-reddened intensities as much as
 50$\%$ in cases where the LMC reddening is very low. For example, in the
 case of SMP~79, using the foreground correction
 of c$_{\rm G}$=0.074 would underestimate the \civ~ flux by 52$\%$.
 
 A sanity check for the extinction correction is the comparison between the
 \heii~ $\lambda$1640 and the optical $\lambda$4686 lines. We will not be able
 to derive an extinction constant from this comparison,
 given the two-step procedure that we use to correct our fluxes,
  but we can compare the two de-reddened intensities 
 for each nebula where these are available.
 We correlate the UV and optical \heii~ line intensities.
 The UV and optical line intensities correlate tightly (R$_{\rm xy}$=0.97),
 and their least square fit gives I$_{\rm 1640} = 5.011 \times {\rm I}_{4686}$, in 
 broad agreement with the theoretical predictions
 \citep{ost}. This result demonstrates that the UV extinction corrections
 we have applied are appropriate. There are two
 straggler points in this relation corresponding to 
 SMP~59 and SMP~93, the two very extended PNe whose intensities are uncertain due to their positions within the aperture.

\section{Results}

\subsection{Flux ranges, errors, and comparisons}

In Figure 26 we plot the cumulative distributions of the
line intensities (relative to \hb) of the four most commonly
observed emission lines 
in our spectra: \civ~$\lambda$1540/50, \heii~$\lambda$1640,
\ciii]~$\lambda$1907/09, and \cii]~$\lambda$2325-29. 
These plots show the normal distribution of the line ratios in our PN sample.
The plots also show the 
the median intensity ratios in the prominent lines.
The carbon lines
are very prominent in LMC PNe, and they are typically several times the 
intensity of \hb. In Galactic PNe, the \civ~ line is typically half
of the \hb~ flux, while the \ciii] line is only slightly higher than \hb~ 
\citep{2000ApJ...531..928H}. The carbon emission can be an important
coolant in LMC PNe, as shown by \citet{sta03}.

In order to determine the errors associated with the line intensities,
we should consider the many sources of possible
errors in the data analysis, namely, (1) the sky subtraction, (2) the spectral
extraction, 
(3) the measure of the actual spectral line by line fit, 
(4) the extinction correction, and (5) the 
uncertainty due to the S/N within the nebular spectra.
We are less concerned about the errors due to the line identification. 
Typically, the emission lines observed in the spectra are well-recognized, 
with the possible exceptions of SMP~45 and SMP~30. 

One of the tools that we can use to grossly constrain the error sources
is the direct comparison of the emission
line \heii~1640 \AA, which is often detectable both in the G140L and the 
G230L spectra. Since this line appears at the blue edge of the G230L spectra,
its measure is uncertain, thus our
comparison gives a super-conservative upper limit to the flux errors.
We find that the \heii~ fluxes from the two grisms
are correlated with R$_{\rm xy}>$0.96, and that the
average difference between the \heii~ flux measured on the 
G140L and the G230L grisms is
28$\pm$0.09$\%$. We also find a mild relation between the
photometric radii of the PNe (from optical data) and the
G140L to G230L \heii~ flux difference, with correlation
coefficient R$_{\rm xy}$=0.73; PNe with radii smaller than about 0.4\arcsec~
have \heii~ flux errors smaller than about 25$\%$.
In general, for the central
parts of the grisms, we expect that the flux errors are much smaller
than this.

We should expect a marginal source of error in the flux determination in
the cases in which the spectral lines did not have gaussian profile.
This source of errors, though, is marginal with respect to other
sources such as (1), (2) and (5) above. As an upper limit, we have the case of SMP~28
where the Balmer-$\alpha$ line was measured with a gaussian in the G230L 
spectrum and with a Voigt function in the G140L spectrum. The relative error
between the two measurement is around 35$\%$, which is on the upper end
of the errors mentioned above.

We assume that the typical flux error from fitting a gaussian profile via the
IRAF routines is $<$5$\%$ \citep{sta02}. This is the major error source 
for the UV fluxes of compact nebulae, with high S/N spectra, and for emission
lines near the center of the grisms. 
Another check on the reliability of our calibrations is by comparing the UV and optical 
\heii~ line emission, when the latter are available from the optical measurements in the literature.
For this comparison we used the G140L fluxes in this paper (\heii~ $\lambda$1640) and the \heii~
$\lambda$4686 optical data either from Leisy (private communication) or from our
own optical observations with the ESO telescopes (Shaw et al., in preparation).
The intensities of the UV and optical \heii~ fluxes for the 15 PNe used in the comparison
correlate to better than 95$\%$. If we exclude the large bipolar PNe (SMP~93 and SMP~59), whose
fluxes carry larger errors due to the fact that
the aperture of extraction do not include 100 $\%$ of their fluxes, 
the correlation coefficient goes up to 98$\%$.  This result 
indicates that the calibration is sound for what the emission lines are concerned. 

The continuum level of the G140L and G230L spectra are the same in the common part of the
spectra if the extraction box used is the same. In general, the extraction boxes are 
optimized for the emission lines, and this may effect the continuum level (but not the 
reliability of the line intensities). We take into account the different continuum level 
when we fit the UV continua with a Plankian function to obtain the (color) temperature of the
central star.

\subsection{Abundances}

The main goal of these {\it HST} observations was that of deriving the 
carbon abundance for the target nebulae. All of the major emission lines 
that are commonly used to determine the C abundance in PNe fall in the 
satellite UV range between 1200 and 2000 \AA. In this paper we calculate 
the abundances of the observable ions of C, N, O, and Ne, and from the 
C ions we derive approximate total C abundance for our targets. 

Two fairly common approaches to computing chemical abundances in ionized 
nebulae are the ``ICF'' method and photo-ionization modeling 
\citep[see, e.g., ][]{aller}. In the ICF approach it is 
necessary to compute the electron temperature (T$_e$) and density (N$_e$) 
for the ionized gas from standard diagnostics, determine the abundances 
of the observed ions based upon the inferred T$_e$ and N$_e$, and then 
use empirical relations to correct for the unseen ionization stages in 
order to derive total elemental abundances. The other common approach is 
to build a photo-ionization model of the nebula by iteratively positing 
physical characteristics (T$_e$, N$_e$, chemical abundance, physical 
size of the nebula, the spectral energy distribution of the ionizing 
source, etc.) and refining those assumptions in order to match the 
computed spectrum with that observed. Often these approaches can be 
combined to good effect \citep[see, e.g., ][]{Hyung94}. In this paper we 
focus on the carbon abundance in these planetary nebulae, where the 
relevant ionization stages can generally be observed and where the 
temperature and density information is taken from other sources. For the 
modest accuracy required for the present purpose, the ICF method will 
suffice; more detailed nebular modeling will be addressed in a future 
paper. 

The bulk of the ionic abundances for this study were computed using the 
{\bf nebular} package in STSDAS, developed by \citet{1995PASP..107..896S},
as updated by \cite{1998adass...7..192S}. We used as input the relative 
intensities from collisionally excited emission lines presented in 
Table~\ref{Fluxes}, corrected for extinction. We adopted the simple model 
used in the {\bf nebular} package for the variation of T$_e$ and N$_e$ 
within the nebula, namely, that the various plasma diagnostics are most 
applicable within one of three zones of ionization potential. 
For example, the densities and temperatures derived from [\ion{S}{2}] and 
[\ion{O}{2}] apply to a zone of low ionization, whereas the 
temperature derived from [\ion{O}{3}] and the density derived from 
[\ion{Ar}{4}] apply to a zone of moderate ionization. The T$_e$ and 
N$_e$ applied within each target nebula is given in Table~\ref{Diags}. 
These diagnostics were derived using {\bf nebular} from standard T$_e$ 
and N$_e$ diagnostics and based upon fluxes from our own ground-based 
optical observations at ESO and MSSSO (Shaw et al. in preparation), and 
from the observations by Leisy (private communication). For a few nebulae 
where these diagnostic  line fluxes were not available we have assumed 
typical values of electron temperature and density (also given in the table).

The ionic abundances presented in Table~\ref{Abund} were derived from the 
fluxes in the emission lines given in Table~\ref{LinesUsed}. The abundance 
of some ions must be derived from recombination lines, which were estimated 
based on approximate relations from Aller (1984, and Table 5-3 therein), as 
they are not computed within the {\bf nebular} package. In columns (2) 
through (4) of Table 6 we list the carbon ionic abundances C$^+$, C$^{2+}$, and 
C$^{3+}$, and in column (5) we give the sum of the carbon ionic abundances. 
Columns (6) through (10) give the abundances of the nitrogen ions, of 
O$^{3+}$, and of Ne$^{3+}$ (though they are not used further in this paper). 

We labeled the sum of the observed carbon ions in column (5) of 
Table~\ref{Abund} as the total C/H abundance, though in principle this sum 
must be corrected for the presence of C$^{4+}$, which is not observed. The 
magnitude of this correction can be estimated using the prescriptions given 
by \citet{1994MNRAS.271..257K}. 
In the case of the very high excitation nebulae, \ion{C}{5} emission should 
be accompanied by \ion{N}{5} $\lambda 1240$ emission since this latter ion 
has a much lower ionization potential. (Note that, normally, the presence of 
a \ion{N}{5} $\lambda 1240$ feature with a P-Cygni profile is not an 
indication of a high excitation PN, since the origin of this feature is a 
stellar wind.) The absence of this nebular feature in many of our target 
nebulae is a strong indication that the C$^{4+}$ abundance is very low or 
absent. In addition, the absence of \ion{He}{2} $\lambda 1640$ strengthens 
the low-excitation classification for most PNe in our sample that do not 
show \ion{N}{5} emission. 

In our PN sample, SMP~4, SMP~10, SMP~16, SMP~28, and SMP~79 may be of high 
enough excitation to require a finite correction for C$^{4+}$. Here we 
examine the carbon content of these PNe in more detail. If we assume that 
C$^{2+}$/O$^{2+}$$\approx$ C/O \citep{1996A&AS..116...95L}, and multiply 
that ratio by oxygen abundances obtained from the literature 
\citep{1998A&A...336..667S}, we obtain C/H values that are comparable to 
the C/H ratios in Table~\ref{Abund}. With this further assurance, and the 
extremely high potential of ionization of \ion{C}{5}, we conclude that 
the carbon abundances in column (5) of Table~\ref{Abund} are good 
approximations for all of the high excitation PNe in this sample. 
In the low and intermediate excitation PNe, sometimes it happens that the 
C$^+$ $\lambda$2326 emission line intensity is weak and below the 
noise limit of the spectrum (SMP~30, SMP~45, SMP~72, and SMP~102).
In these cases, the total carbon abundances in Table~\ref{Abund} may be
underestimated by $\sim10\%$. We have also checked whether there are correlations between the derived
abundances and the excitation constant, EC, and found none. This is a 
good sign that the abundance analysis is sound and the results do not depend
on the excitation level of the PN.

In Figures 27 and 28 we used our C/H abundances from Table 6 and the O/H, N/H, and He/H abundances from \citet{1998A&A...336..667S}. 
In addition, we also plot all data that were available a few years ago, and that have been presented in SSBB.
In all we have a sizable sample of LMC PNe whose morphology are known from direct {\it HST} observations
and whose carbon abundances have been derived by means of either {\it HST} or IUE spectroscopy. Prior to this paper, there were 16 PNe in the LMC
whose morphology and carbon abundances were known, though the
morphologies only approximate since the {\it HST} images were from the pre-COSTAR era.
The sample of 24 PNe presented in this paper more than doubles the original sample and improves enormously the reliability of the analysis.

In Figure 27
we show the C/H versus N/H relation in a plot where data points are coded by morphological type (open circles: round,
asterisks: elliptical, filled triangles: round or elliptical with bipolar core, squares: bipolar, and filled circles: point-symmetric PNe).
This is a quantitative improvement to the plot in Fig.~4 by SSBB. The additional nebulae shown here are
the 11 nebulae studied in this paper whose nitrogen abundances are available in the literature \citep{1998A&A...336..667S}. This plot largely
confirms 
what was anticipated in SSBB: Bipolar and Point-symmetric PNe show low carbon and high nitrogen abundances with
respect to hydrogen, while round and elliptical PNe show the opposite behavior. Interestingly, we also find that round and elliptical PNe
with a bipolar core are split into two separate groups, one that is
chemically more like bipolar, the other closer to the domain of elliptical PNe. 
From this plot, it looks like all bipolar (and point-symmetric) PNe of this sample have gone through the process of carbon depletion during the 
HBB, which also partially enriches nitrogen; round and elliptical PNe have not gone through that process, 
while the bipolar core PNe are a mixed group.

In Figure 27 the morphology of the bipolar PN with lowest nitrogen abundance (SMP~45), and the bipolar core PN with second to
lowest nitrogen abundance (SMP~100), are ambiguous; erasing these two
targets from the plot does not change any of the above results.
A similar situation is shown when the C/O and N/O ratios are plotted (Fig. 28). Note that the arrow indicates the 
range of C/O that defines the domain of carbon stars, from which it is clear that bipolar PNe are
not the progeny of carbon stars in the LMC (HG97).

In Table~\ref{AvgAbund} we give the average carbon abundances for each morphological type for the whole sample of PNe whose carbon abundance
is known. The sample consists now of 40 LMC PNe, enough to populate each morphological class, except the point-symmetric, reliably.  From
the table we infer that the median
carbon abundance of bipolar PNe is about one order of magnitude lower than those of round and elliptical PNe, while the content of 
carbon in bipolar core PNe is about half than that of
round and elliptical PNe. We will explore the nature of BC PNe with detailed individual
photo-ionization models in a future paper.

\subsection{Comparison to theoretical yields}

The carbon abundances that we derived can be used successfully to constrain the 
stellar evolution models for stars that undergo the AGB phase. 
Only with the comparison between the observed 
abundances and the theoretical yields one can improve the current knowledge 
of stellar evolution and its variation with initial metallicity.

Several authors have modeled the evolution of stars that go through the
AGB phase, and then calculated the yields of the principal elements such as
H, He, C, N, and O \citep{1981A&A....94..175R,1997A&AS..123..305V,1997A&AS..123..241F,2001A&A...370..194M,karakas}. 
For the comparison with LMC PN data we look for models
that have initial LMC composition. Furthermore, we looked for authors that
evaluated the mass return into the
ISM both from the total evolution, and in the final thermal pulses, the latter mass return
being directly
comparable to the PN ejecta. The synthetic AGB 
models by \citet{2001A&A...370..194M} (hereafter M01) cover a wide initial mass range, and initial 
Galactic, LMC, and SMC composition.
\citet{2001A&A...370..194M} also evaluated the expected PN composition by averaging the
abundances in the mass ejected during the final 30,000 yr of the AGB. Such
yields can be compared to our data directly. Van den Hoek \& Groenewegen (1997) also produced 
yields for Magellanic and
Galactic initial composition for a wide initial mass range of AGB evolutionary 
models, giving 
the total and final (25,000 yr) yields. By knowing the initial composition 
we can derive the expected PN composition to compare with our data, following
their formulation. 

It is worth noting that the models by M01 and HG97 
are not directly comparable. Van den Hoek \& Groenewegen (1997) uses the empirical mass loss
prescription by \citet{1994ApJS...92..125V}, different mixing-length 
parameters for different initial mass, and Y=0.25 for the initial
set of LMC models. Van den Hoek \& Groenewegen (1997) use \citet{1975psae.book..229R}'s
mass loss
rate with efficiency $\eta$=4 for the LMC models, independent of initial mass, 
Y=0.264 for the initial LMC model. Finally, M01's use of extra-dredge up makes
the carbon yields higher that those of HG97. 

There are other evolutionary models in the literature that estimate
atomic yields from AGB stars. \citet{1981A&A....94..175R} gave surface and PN 
abundances derived from evolutionary models, but they did not perform their analysis for 
an LMC composition. \citet{karakas} calculated complete AGB models for stars
in the appropriate mass range, and with adequate initial composition, but did not
evaluate the mass fraction of the major atoms ejected in the final
thermal pulses. Finally, \citet{1997A&AS..123..241F} provided the yields
in the final 4 thermal pulses, thus comparable with PN abundances, with
extrapolated nucleosynthesis. The initial composition of their models 
is limited to Z=0.02 and Z=0.005, thus we will make only marginal use of these
models in this paper. 

In Figure 29 we show the log C/O vs. log N/O trends similarly to those of Fig. 28,
including also those PNe whose morphology is uncertain or 
not known (small open circles).
In this Figure we show the loci of the yields from the stellar evolution models
models. M01 and HG97 PN yields for initial LMC composition have been plotted 
respectively with solid lines and broken lines.
The lower curves connect log C/O and log N/O  for initial 
masses in the 0.85 -- 3.5 \sm~ (M01) and 0.8 -- 3.5 \sm~ (HG97) ranges. 
The upper curves are for M$>$3.5 \sm. The segmented high mass solid
curve corresponds to M01 models with M$>$3.5 \sm~ and $\alpha$=1.68, 2, and 2.5.

The models seem to generally encompass the data. The group of PNe 
with high C/O and low N/O are in the same general area of the low-mass
models. We confirm that no bipolar PNe are in the general area of the low-mass models.
On the other hand, it seems that the high mass (M$>$ 3.5 \sm) models predict carbon and nitrogen
yields that
are higher than observed in LMC PNe. While the different assumptions in the M01 and HG97 models 
make the relative yields
slightly different, they both over-predict the elemental abundance of C/O and N/O in most high-mass
cases. The effect of over-prediction of carbon and nitrogen in the case of high-mass progenitors
is even more evident in Figure 30, where we plot the elemental abundances of C and N relative to 
hydrogen. None of the examined models yields to log C/H $<$ 4, observed in roughly half of the PNe.

We believe that there are three possible explanations to account for the inconsistency of the prediction 
with respect to the observed abundances. First, the initial conditions used in the M01 and HG97 models
may not be appropriate for the correct description of our data.
HG97 obtains all the initial abundance ratios (C/H, N/H, etc) from the observed solar abundance ratios
\citep{1996coab.proc..117G}, scaled according to Y. 
When comparing the initial conditions 
chosen by GH97 for the LMC to actual LMC abundance measurements, for example in H II regions or supernova
remnants \citep{1990ApJS...74...93R,1996A&AS..116...95L}, we found $\Delta ({\rm log} C/H) \le 0.5$ and 
$\Delta ({\rm log} N/H) \le 0.63$, where $\Delta$ represent the differences between the HG97 model assumptions and
the observed initial abundance in the LMC. 
By looking at Figure 30 one can see that 
this initial abundance bias may account for the observed inconsistency. Obviously, one must get the
correct yields from the models with the appropriate initial conditions to obtain the right comparison.
\citet{karakas} used in her LMC models appropriate initial conditions, as derived from the observations.
We can not compare her models to our data here, since she does not provide with the final yields. Such a
comparison would be very important for future developments.

Second, while we have considered only models of single star evolution, some of the PNe might originate from 
close binary evolution. This may be true especially in the case of aspheric PNe, whose morphology may derive from
common envelope evolution. Although complete binary models that involve an AGB star, with initial composition 
similar to that of LMC, does not exist, \citet{2003PASA...20..345I,2004MNRAS.350L...1I} have calculated yields 
from binary evolution with solar composition.
They found that, if we exclude the evolution through the nova and supernova stages, the final yields of C, N, and O
from binary evolution are respectively 0.86, 0.69, and 1 times the yields from single star evolution. If such
results could be extrapolated in the case of the LMC models, we see that would not be enough to account for the
observed inconsistencies. Nonetheless, this is an interesting explanation that is related to PN morphology, 
and we know that the discrepancy is certainly higher for aspherical PNe.

Third, the theoretical yields in PNe  represent the ejecta of the final
stages of the evolution of the AGB stars, while the observations determine the elements ratios 
in the ionized gas. Within the ionized region of a PNe there may be additional carbon, confined
in carbonaceous dust particles \citep{2004Natur.430..985K}. While the dust contribution to the
C/H ratio may be very low, it may be a factor to consider in comparing models and data.
Finally, the presence of CO in PNe may alter the comparison of theoretical yields and observed abundances, 
depending on the
geometry of the nebulae. In a convex, regular (such as round or elliptical) PN, CO is dissociated 
by the ionization front, and it will not coexist with the ionized gas. As a consequence, 
the abundances measured within the ionized gas well represent the actual elemental concentration.
On the other hand, in the case of concave (such as bipolar) PNe, CO may lie in 
a ring orthogonal to the lobal axis \citep{2000A&A...353..363J}. The inclusion of the ring CO in bipolar
PNe in the abundance analysis is not possible to date, lacking an accurate model for bipolar mass loss
that follows stellar evolution from the AGB onward. 

\subsection{Comparison with the Galactic sample}

In Figure 31 we plot [C/H] against [N/H] for the LMC (filled circles, this paper) and the Galactic 
\citep{2000ApJ...531..928H} (crosses) PN samples, and compare them with 
the M01 yields.  
Both yields and data have been scaled to the corresponding solar values given by \citet{1996coab.proc..117G}, and
we use the standard notation [X/H]=(X/X$_{\odot}$)/(H/H$_{\odot}$).
The thick line connects the yields of [C/H] and [N/H] for 0.85 $<$ M/M$_{\odot}$ $<$ 5 and $\alpha$=1.68 (LMC models), 
while
the thin lines connects the yields of [C/H] and [N/H] for 0.85 $<$ M/M$_{\odot}$ $<$ 5  and $\alpha$=1.68 (Galactic models).
From the Figure one can infer that neither the Galactic nor
the LMC model yields reach down to the observed [C/H] and [N/H] concentration of, respectively, the Galactic 
and the LMC PN samples. Similar results
are obtained when comparing the data of both the LMC and Galactic samples
with HG97 models, as shown in Figure 32. Here we reproduce the same data as of Fig. 31, except
we superimpose to them the HG97 models, for LMC (thick broken line) and Galactic (thin broken line) AGB evolution.
Both Figs. 31 and 32 give the clear impression of a systematic
offset between observations and models. 

In Figure 33  we plot [C/H] versus [O/H] for the LMC
and the Galactic PN samples. Here we superimpose the the theoretical yields by 
M01 (LMC: thick solid line; Galaxy: thin solid line) and HG97
(LMC: thick broken line; Galaxy: thin broken line). While the yields by HG97 seem to better encompass both LMC and 
Galactic data, none of the models predict the observations satisfactorily. 

We use Figs. 31 through 33 to compare the group characteristics of the LMC 
and Galactic PNe, especially for what carbon abundances are concerned. The carbon abundance seem to
peak at lower oxygen (and nitrogen) values in the LMC than in the Galactic PNe. Also, carbon, nitrogen,
and oxygen abundances of LMC and Galactic PNe seem to follow the same trends 
but LMC abundances are typically lower for all three elements. For the given samples,
the median C/H abundance is 2.63$\times$10$^{-4}$ in the LMC and 4.11$\times$10$^{-4}$ in the Galactic PNe.
It is worth noting that the models are
unable to predict the actual abundance ratios correctly both for LMC and Galactic PNe, but they do
predict quite accurately the trends of both C/H vs. O/H and C/H vs. N/H for both PN samples, including their
offsets.
While it is not straightforward to compare directly the
LMC and Galactic PN samples, due to the selection biases that strongly affect Galactic observations,
it seems clear that the samples analyzed here show similar characteristics, with a shift of 
(initial) conditions, where the LMC PN progenitors originate from a 
lower metallicity environment.

\section{Summary and Discussion}

We acquired the UV spectra of 24 LMC PNe with {\it HST}/STIS, by using the G140L and G230L grisms to obtain spectral information
from 1150 to 3000 \AA. The interstellar
extinction was measured both for the Galactic and the LMC contributions case-by-case for all targets.
Since none of the present PNe was observed before in the
UV range, we could not compare our results with those available in the literature. Instead,  we checked for internal consistency of the 
data set and for UV to optical consistency, finding that our line intensities are sound and errors reasonable, with the exception of 
residual errors in the extinction correction, that in all cases are much lower than the internal instrumental errors related to the observations.

For each nebula we determine the consistency between the excitation class from the optical images, the Zanstra temperature 
(determined by Villaver, Stanghellini, \& Shaw 2003) and, in some cases, the color temperature of the central stars determined by a Planck fit
of the UV continuum. The excitation of the nebula was related to the emission lines present in the UV spectra. Eight PNe present P-Cygni 
profile typically in the \civ~  and \nv~ features. We have measured the terminal velocities relative to these
P-Cygni features. 

We used the measured carbon lines to determine the ionic abundances, and we present the ionic abundances for carbon and other
ions. The ionic abundances of carbon have been used to infer elemental carbon abundances. The data presented here more than double 
the sample of LMC PNe whose carbon abundance have became available.
We found that the median (and mean) carbon abundance varies considerably
across morphological types, being
almost one order of magnitude larger in bipolar than in round PNe.
These relations, while hinted previously (SSBB), are now supported by a sizable and reliable
data set. We also confirm the previously determined carbon and nitrogen abundance trends that disclose the occurrence of the 
third dredge-up and the HBB events in the more massive progenitor stars. This is the first time that carbon and nitrogen abundances can
be directly correlated for a large sample of PNe whose morphological type is known. We confirm the link between bipolarity and carbon 
depletion, and between carbon enrichment and spherical or elliptical symmetry, in the LMC.

By comparing the abundances studied in this paper to stellar yields calculated from evolutionary models 
we found that the predicted carbon abundances are generally higher that those observed. Our data have also been compared to those of
Galactic PNe, and a similar mismatch between data and predictions has been found in both samples. We suspect that the discrepant 
predicted yields (especially of carbon) are at least partially due to the choice of initial composition in the leading models for whose the final yields
are available to date (M01, HG97). The comparison of the data to the yields calculated with adequate initial conditions (e.g., those
of Karakas 2003) will allow us to determine other possible causes of discrepancy, including binary star evolution. 
Models with a set of different HBB efficiencies will also be useful to determine whether an increased HBB efficiency 
may reconcile theoretical yields and our data.
A possible source of discrepancy
is also the limitation of the carbon abundance studies to the ionized gas. Future quantitative analysis of
solid (dust) carbon will help to constraint the evolutionary models. More accurate modeling of the aspheric mass loss at the
tip of the TP-AGB is also needed to completely account for the different components of the PN that may contain carbon in
ionized and neutral gaseous, or solid form. 

One of the main reasons to study the LMC PNe is that their distance is known, and consequently the determination of the central star masses
can be performed more reliably than for Galactic PNe. 
The determination of the central star masses, and their relation to chemistry (especially carbon and nitrogen) and morphology, allows 
to account for the evolutionary path between the low- and intermediate-mass star in the AGB and the PNe. 
The sample of LMC PNe whose central star masses has been measured \citep{cspaper} and whose carbon abundance are known (this paper)
consists of only seven objects, too small a sample
for meaningful statistical analysis. Nonetheless
there are a few individual comparison that are worth mentioning.

The central star mass of SMP~4 and SMP~10 is identical within the errors (M$_{\rm CS}$=0.58 \sm);
the carbon abundance of the former, an elliptical PN, is a factor of 20 higher than the carbon abundance of the latter, a point-symmetric PN. From the carbon depletion 
detected in SMP~10 it is reasonable to assume that its progenitor mass was much higher than the progenitor mass of SMP~4. The similarity of the 
central star masses may indicate that a much more powerful mass-loss process had occurred to strip the envelope of SMP~10 than that of SMP~4.
Other such examples are available. SMP~19 (an elliptical PN with a bipolar core) and 
SMP~30 (a bipolar PN, likely) have very similar CS masses, yet the former has much higher carbon content than the latter, suggesting that SMP~30
is the progeny of a higher mass progenitor, that underwent heavy mass-loss. 

If the carbon depletion really occurs mainly during the HBB process, we would conclude that SMP~10 and SMP~30 have much higher progenitor masses than, respectively, SMP~4 and SMP~19. Yet their central stars end up with the same masses. The evolutionary path that produces the bipolar and
point-symmetric PNe, SMP~10 and SMP~30, seems to be related to a much higher total mass-loss than the processes that produces the elliptical PNe SMP~4 and SMP~19. 

A much larger data set of central stars of LMC PNe, and
a similar data set for SMC PNe, are being collected to further explore these ideas. In addition, efforts are in order to develop hydrodynamic 
and photo-ionization models for the individual Magellanic planetary nebulae whose central star mass, morphology, and carbon abundance have become available.

 \section{Acknowledgements}

Many thanks to Chris Blades, Bruce Balick, Eva Villaver, and Anabel Arrieta for scientific
discussion and their help in this project, and to
Tom Brown, Charles Proffitt, Phil Hodge, and Max Mutchler for their 
help in the data analysis.
It is a pleasure to thank Josef Gochermann for providing us with the color excess for 
our targets, using his data prior of publication, and 
Pierre Leisy for allowing us to use his $\lambda$4686 \AA~ emission line intensities 
prior of publication. Thanks are due to an anonymous Referee for 
important suggestions.
This project was supported by NASA to grant GO-09120.01-A from the Space Telescope 
Science Institute, which is operated by the Association of Universities for Research 
in Astronomy (AURA), Inc., under NASA contract NAS 5-26555.

\clearpage

\figcaption{G140L (a) and G230L (b) spectrograms of J~41}
\figcaption{As in Fig.~1, for SMP~04}
\figcaption{As in Fig.~1, for SMP~09}
\figcaption{As in Fig.~1, for SMP~10}
\figcaption{As in Fig.~1, for SMP~16}
\figcaption{As in Fig.~1, for SMP~18}
\figcaption{As in Fig.~1, for SMP~19}
\figcaption{As in Fig.~1, for SMP~25}

\figcaption{As in Fig.~1, for SMP~27}
\figcaption{As in Fig.~1, for SMP~28}
\figcaption{As in Fig.~1, for SMP~30}
\figcaption{As in Fig.~1, for SMP~34}
\figcaption{As in Fig.~1, for SMP~45}
\figcaption{As in Fig.~1, for SMP~46}
\figcaption{As in Fig.~1, for SMP~48}
\figcaption{As in Fig.~1, for SMP~59}

\figcaption{As in Fig.~1, for SMP~71}
\figcaption{As in Fig.~1, for SMP~72}
\figcaption{As in Fig.~1, for SMP~79}
\figcaption{As in Fig.~1, for SMP~80}
\figcaption{As in Fig.~1, for SMP~81}
\figcaption{As in Fig.~1, for SMP~93}
\figcaption{As in Fig.~1, for SMP~95}
\figcaption{As in Fig.~1, for SMP~102}

\figcaption{Distributions of the Galactic foreground extinction constant (top)
and the LMC extintion constant for the nebulae in our sample.}
\figcaption{Cumulative gaussian distributions of the de-reddened intensities
of the major emission lines in the nebulae of our sample. The emitting ion and 
its wavelength are indicated in the figure legend. The cross correspond to 
the mean flux in the given ion. The solid lines correspond to the domain
$<F_{\lambda}> \pm \sigma$.}
\figcaption{N/H vs. C/H abundances for the sample od this paper (nitrogen abundances from \citet{1998A&A...336..667S}) 
plotted with the sample in \citet{2000ApJ...534L.167S}. The symbols code the morphological types: open circles=
round; asterisks=elliptical, filled triangles: round or elliptical with bipolar core; filled squares: bipolar,
filled circles: point-symmetric PNe.}

\figcaption{N/O vs. C/O abundances for the sample od this paper (nitrogen and oxygen abundances from \citet{1998A&A...336..667S}) 
plotted with the sample in \citet{2000ApJ...534L.167S}. The symbols are codes as in Figure 27. Note that the
arrow indicates the range of abundances in carbon stars.}

\figcaption{Same as in Fig. 28, with superimposed the final yields from the LMC M01's models (solid lines; upper lines are for M$\>$3.5 \sm
and $\alpha$=1.68, 2, and 2.5; lower line is for M$\le$3.5 and $\alpha$=1.68) and from the LMC HG97 models (broken lines; upper line is
for M$\>$3.5 \sm, lower line for M$\le$3.5).}

\figcaption{Same as in Fig. 29, but for the N/H vs. C/H ratios.}

\figcaption{[C/H] vs. [N/H] for LMC (filled circles) and Galactic (crosses) PNe. Galactic PN data are from 
\citet{2000ApJ...531..928H}.
Thick solid line: LMC M01 yields for $\alpha$=1.68; thin solid line: Galactic M01 yields for $\alpha$=1.68.}

\figcaption{[C/H] vs. [N/H] for LMC (filled circles) and Galactic (crosses) PNe, as in Fig.~30 . Theoretical yields from HG97. 
Thick broken line: LMC HG97 yields; thin broken line: Galactic HG97 yields.}

\figcaption{[C/H] vs. [O/H] for LMC (filled circles) and Galactic (crosses) PNe. Galactic PN data are from 
\citet{2000ApJ...531..928H}.
Thick solid line: LMC M01 yields for $\alpha$=1.68; thin solid line: Galactic M01 yields for $\alpha$=1.68.
Thick broken line: LMC HG97 yields; thin broken line: Galactic HG97 yields.}

\clearpage

\begin{deluxetable}{lccr}

\tablewidth{10truecm}
\tablecaption {Observing log\tablenotemark{a} \label{ObsLog}}

\tablehead {
\colhead {Name} & \colhead{data set} & \colhead{obs. date} & 
\colhead{t$_{\rm exp}$}\\
\colhead{}& \colhead{}& \colhead{}& \colhead{[s]}\\}

\startdata

   J~41  &   o6cn01ngq  & 2002-01-31  & 1320  \\
 &   o6cn01niq  & 2002-01-31  &  133  \\
 SMP~04  &   o6cn02ncq  & 2001-10-11  & 1440  \\
 &   o6cn02ndq  & 2001-10-11  & 1445  \\
 SMP~09  &   o6cn03ceq  & 2002-08-03  & 2880  \\
 &   o6cn03ciq  & 2002-08-04  & 1930  \\
 SMP~10  &   o6cn04gdq  & 2002-04-15  & 1320  \\
 &   o6cn04gfq  & 2002-04-15  & 1260  \\
 SMP~16  &   o6cn05o0q  & 2002-01-31  & 1320  \\
 &   o6cn05o2q  & 2002-01-31  & 1181  \\
 SMP~18  &   o6cn06o6q  & 2002-02-01  & 1320  \\
 &   o6cn06o8q  & 2002-02-01  & 1195  \\
 SMP~19  &   o6cn07ocq  & 2002-02-01  & 1320  \\
 &   o6cn07oeq  & 2002-02-01  & 1320  \\
 SMP~25  &   o6cn08pqq  & 2002-07-16  & 1320  \\
 &   o6cn08psq  & 2002-07-16  & 1260  \\
 SMP~27  &   o6cn09u5q  & 2002-02-02  & 1320  \\
 &   o6cn09u7q  & 2002-02-02  &  953  \\
 SMP~28  &   o6cn10l5q  & 2001-12-09  & 2964  \\
 &   o6cn10l7q  & 2001-12-09  & 3013  \\
 SMP~30  &   o6cn11vlq  & 2002-02-02  & 1320  \\
&   o6cn11vnq  & 2002-02-02  & 1260  \\
 SMP~34  &   o6cn12y0q  & 2002-07-31  & 1320  \\
  &   o6cn12y5q  & 2002-07-31  & 1235  \\
 SMP~45  &   o6cn13r9q  & 2002-01-19  & 1320  \\
 &   o6cn13rbq  & 2002-01-19  & 1260  \\
 SMP~46  &   o6cn14wfq  & 2002-02-02  & 1320  \\
  &   o6cn14whq  & 2002-02-02  & 1260  \\
 SMP~48  &   o6cn15tvq  & 2001-12-11  & 1080  \\
 &   o6cn15txq  & 2001-12-12  &  753  \\
 SMP~59  &   o6cn16l2q  & 2002-01-17  & 2956  \\
&   o6cn16l4q  & 2002-01-17  & 1891  \\
 SMP~71  &   o6cn17wmq  & 2002-02-02  & 1320  \\
 &   o6cn17woq  & 2002-02-02  & 1320  \\
 SMP~72  &   o6cn18adq  & 2002-02-03  & 1320  \\
  &   o6cn18afq  & 2002-02-03  & 1320  \\
 SMP~79  &   o6cn19fzq  & 2001-12-08  & 1440  \\
 &   o6cn19g0q  & 2001-12-08  & 1445  \\
 SMP~80  &   o6cn20ags  &   2001-10-15  &  1320 \\
  &    o6cn20ais &   2001-10-15         &   1320   \\
 SMP~81  &   o6cn21n7q  & 2001-10-11  & 1440  \\
 &   o6cn21n8q  & 2001-10-11  & 1445  \\
 SMP~93  &   o6cn22q2q  & 2002-01-24  & 2880  \\
&   o6cn22q4q  & 2002-01-24  & 1055  \\
 SMP~95  &   o6cn23lbq  & 2002-07-19  & 1320  \\
  &   o6cn23leq  & 2002-07-19  & 1260  \\
SMP~102  &   o6cn24l0q  & 2002-07-19  & 1320  \\
&   o6cn24l3q  & 2002-07-19  & 1260  \\
\hline
\enddata
\tablenotetext{a}{Format of table is G230L grating spectrum
followed by G140L spectrum for each target}
\end{deluxetable}

\clearpage

\begin{deluxetable}{lllccccr}

\tabletypesize{\scriptsize}
\tablewidth{18truecm}
\tablecaption {Observed and Corrected Emission Line Flux Ratios \label{Fluxes}}

\tablehead {
\colhead {Name} & \colhead{$\lambda$} & \colhead{ID} & 
\colhead{F$_{\lambda}$/F$_{\beta}$} & \colhead{c$_{\rm G}$} & 
\colhead{c$_{\rm LMC}$} & \colhead{I$_{\lambda}$/I$_{\beta}$}& \colhead{note}\\
}

\startdata
   J~41  &  1907/09  &  \ciii]  &  1.49  &  0.096  &  0.000  &  1.99 &  \\
  SMP~4  &  1175/76  &  \ciii &  0.74  &  0.131  &  0.000  &  1.43 & \\
         &  1239/43  &    \nv &  0.25  &  0.131  &  0.000    &  0.44 & two components \\
         &  1335/36  &   \cii &  0.29  &   0.131 &   0.000   &  0.47 & emission overlying absorption \\
         &  1548/50  &   \civ &  2.49  &  0.131  &  0.000    &  3.62 & deblended from [\nev] \\
	 &  1575     &   [\nev] &  0.53  & 0.131   &   0.000   &  0.76 & deblended from \civ, id uncertain \\
         &  1640     &  \heii &  1.51  &  0.131  &   0.000   &  2.16 & Balmer-${\alpha}$, Voigt profile, G140L \\
         &  1907/09  &  \ciii] &  3.09  & 0.131   &   0.000   &  4.59 & \\
  SMP~9  &  1335/36   &  \cii   &  0.14  &  0.079  &  0.154  &  0.43  & \\
         &  1397/1407 & \oiv]   &  0.06  &  0.079  &  0.154  &  0.18  & \\
         &  1548/50   &  \civ   &  2.33  &  0.079  &  0.154  &  5.68  & Voigt profile \\
         &  1640      & \heii   &  0.91  &  0.079  &  0.154  &  2.08  & Balmer-${\alpha}$, Voigt profile, G140L\\
         &  1640      & \heii   &  1.37  &  0.079  &  0.154  &  3.11  & Balmer-${\alpha}$, Voigt profile, G230L\\
         &  1907/09   & \ciii]  &  3.75  &  0.079  &  0.154  &  7.94  & Voigt profile \\
         &  2325-29   &  \cii]  &  1.23  &  0.079  &  0.154  &  2.43  & \\
         &  2423/25   & [\neiv] &  0.12  &  0.079  &  0.154  &  0.22  & \\
         &  2470      & [\oii]  &  0.07  &  0.079  &  0.154  &  0.13  & \\
         &  2733      &  \heii  &  0.06  &  0.079  &  0.154  &  0.09  & Paschen-${\beta}$\\
         &  2796/2803 &  \mgii  &  0.23  &  0.079  &  0.154  &  0.34  & could be 2784 [\mgv]\\
 SMP~10  &  1239/43  &    \nv  &  0.09  &  0.103  &  0.072  &  0.22 & possible separate 1239 feature, P-Cygni, Voigt profile\\
         &  1548/50  &   \civ  &  0.32  &  0.103  &  0.072  &  0.58 &  \\
         &  1640     &  \heii  &  0.19  &  0.103  &  0.072  &  0.33 & Balmer-${\alpha}$, G140L\\
         &  1907/09  &  \ciii] &  1.59  &  0.103  &  0.072  &  2.76 & single feature, uncertain id\\

 SMP~16  &  1239/43   &    \nv  &  0.36  &  0.088  &  0.063  &  0.80 & \\
         &  1397/1407 &  \oiv]  &  0.20  &  0.088  &  0.063  &  0.37 & \\
         &  1483/87  &   \niv]  &  0.58  &  0.088  &  0.063  &  1.01 & \\
         &  1548/50  &   \civ   &  0.41  &  0.088  &  0.063  &  0.70 & \\
         &  1640     &  \heii   &  1.40  &  0.088  &  0.063  &  2.29 & Balmer-${\alpha}$, deblended from \oiii], G140L\\
         &  1658-1666 &  \oiii] &  0.14  &  0.088  &  0.063  &  0.23 & deblended from Balmer-${\alpha}$ \\
	 &  1750     &  \niii]   &  0.53  &  0.088  &  0.063  &  0.84 & \\
         &  1907/09  &  \ciii]  &  0.92  &  0.088  &  0.063  &  1.49 & \\
         &  2325-29  &   \cii]  &  0.17  &  0.088  &  0.063  &  0.26 & deblended from [\neiv] and [\oii], Voigt profile \\
         &  2423/25  &  [\neiv] &  1.01  &  0.088  &  0.063  &  1.47 & deblended from \cii] and [\oii], Voigt profile \\
         &  2470     &  [\oii]  &  0.05  &  0.088  &  0.063  &  0.07 & deblended from \cii] and [\neiv], Voigt profile \\
 SMP~18  &  1239/43     &  \nv   &  0.69  &  0.076  &  0.000 &  0.96 & P-Cygni \\
	 &  1548/50  &  \civ    &  0.49  &  0.076  &  0.000 &  0.61 & P-Cygni \\
         &  1907/09  &  \ciii]  &  2.89  &  0.076  &  0.000 &  3.64 & Voigt profile, G230L \\
         &  1993     &  \ci     &  0.21  &  0.076  &  0.000 &  0.27 & marginal, deblended from \ciii], Voigt profile, G140L\\
         &   2325-29  &   \cii] &  0.14  &  0.076  &  0.000 &  0.17 & marginal, high errors \\
 SMP~19  &  1247     &  \ciii   &  0.11  &  0.076  &  0.117  &  0.31 & could be 1239/43 \nv \\
         &  1335/36  &   \cii   &  0.07  &  0.076  &  0.117  &  0.17 & \\
  	 &  1397/1407 &  \oiv]  &  0.10  &  0.076  &  0.117  &  0.23 &  \\
   	 &  1483/87  &   \niv]  &  0.07  &  0.076  &  0.117  &  0.15 & \\
   	 &  1548/50  &   \civ   &  2.44  &  0.076  &  0.117  &  5.03 & Voigt profile \\
   	 &  1640     &  \heii   &  1.22  &  0.076  &  0.117  &  2.38 & Balmer-${\alpha}$, G140L, Voigt, deblended from \oiii] \\
   	 &  1640     &  \heii   &  1.68  &  0.076  &  0.117  &  3.28 & Balmer-${\alpha}$, G230L, extreme edge of spectrum \\
   	 &  1658-66  &  \oiii] &  0.21  &  0.076  &  0.117  &  0.40 & deblended from \heii, G140L \\
   	 &  1907/09  &  \ciii]  &  6.11  &  0.076  &  0.117  & 11.31 & \\
   	 &  2325-29  &   \cii]  &  1.15  &  0.076  &  0.117  &  2.01 & \\
   	 &  2423/25  &  [\neiv] &  0.34  &  0.076  &  0.117  &  0.56 & deblended from [\oii] and \heii \\
   	 &  2470     &   [\oii] &  0.11  &  0.076  &  0.117  &  0.17 & deblended from \cii]  and \heii \\
   	 &  2511     &  \heii   &  0.05  &  0.076  &  0.117  &  0.07 & Paschen-${\gamma}$, deblended from [\neiv] and [\oii] \\
   	 &  2733     &   \heii  &  0.07  &  0.076  &  0.117  &  0.09 & Paschen-${\beta}$, deblended from \arv~ and \cii \\
   	 &  2786     &   \arv   &  0.06  &  0.076  &  0.117  &  0.08 & deblended from Paschen-${\beta}$ and \cii \\
   	 &  2837/38  &   \cii   &  0.08  &  0.076  &  0.117  &  0.10 & Bowen fluorescence, deblended from Paschen-${\beta}$ and \arv \\
   	 &  3043/47  &  \oiii   &  0.14  &  0.076  &  0.117  &  0.18 & Bowen fluorescence, two components deblended \\
   	 &  3133     &  \oiii   &  0.55  &  0.076  &  0.117  &  0.70 & Bowen fluorescence, two components deblended \\
 SMP~25  &  1239/43  &   \nv    &  0.82  &  0.050  &  0.080  &  1.70 & P-Cygni \\
	 &  1335/36  &   \cii   &  0.22  &  0.050  &  0.080  &  0.41 &  \\
	 &  1548/50  &   \civ   &  0.97  &  0.050  &  0.080  &  1.58 & P-Cygni \\ 
	 &  1658-66  &  \oiii]  &  0.63  &  0.050  &  0.080  &  0.99 & marginally P-Cygni \\ 
	 &  1907/09  &  \ciii]  &  1.56  &  0.050  &  0.080  &  2.37 & \\
         &  2325-29  &   \cii]  &  0.29  &  0.050  &  0.080  &  0.43 & \\
         &  2470     &   [\oii] &  0.05  &  0.050  &  0.080  &  0.07 & \\
 SMP~27  &  1239/43  &   \nv    &  0.98  &  0.184  &  0.000  &  2.20 & P-Cygni\\
	 &  1548/50  &   \civ   &  0.59  &  0.184  &  0.000  &  1.01 & P-Cygni \\ 
	 &  1907/09  &  \ciii]  &  1.85  &  0.184  &  0.000  &  3.23 & \\
        &  2325-29  &   \cii]  &  0.17 &    0.184  &  0.000  &  0.30 & \\
 SMP~28  &  1239/43  &    \nv   &  0.86  &  0.046  &  0.281  &  6.24  & \\
         &  1397/14037&  \oiv]  &  0.05  &  0.046  &  0.281  &  0.26  & uncertain id, 3-line blend \\
         &  1397-1407 &  \oiv] &  0.08  &  0.046  &  0.281  &  0.37  & uncertain id, 3-line blend \\
  	 &  1397-1407 &  \oiv] &  0.08  &  0.046  &  0.281  &  0.36  & uncertain id, 3-line blend \\ 
   	 &  1483/87  &   \niv]  &  0.27  &  0.046  &  0.281  &  1.14  & \\
   	 &  1548/50  &   \civ   &  0.08  &  0.046  &  0.281  &  0.30  & \\
   	 &  1640     &  \heii   &  0.34  &  0.046  &  0.281  &  1.19  & Balmer-$\alpha$, G140L, Voigt profile \\
   	 &  1640     &  \heii   &  0.63  &  0.046  &  0.281  &  2.19  & Balmer-$\alpha$, G230L \\
   	 &  1658-1666 &  \oiii] &  0.15  &  0.046  &  0.281  &  0.52  & G140L \\
   	 &  1658-1666 &  \oiii] &  0.22  &  0.046  &  0.281  &  0.75  & G230L \\
   	 &  1747-1754 &  \niii] &  0.45  &  0.046  &  0.281  &  1.40  & marginally double feature \\
   	 &  1892     & \siiii]  &  0.16  &  0.046  &  0.281  &  0.47  & deblended from \ciii] \\
   	 &  1907/09  &  \ciii]  &  0.13  &  0.046  &  0.281  &  0.37  & deblended from \siiii \\
   	 &  2145     &   $\dots$    &  0.08  &  0.046  &  0.281  &  0.23  & not identified\\
   	 &  2334-50  &  [\sii]  &  0.12  &  0.046  &  0.281  &  0.29  & \\
   	 &  2423/25  &  [\neiv] &  0.12  &  0.046  &  0.281  &  0.28  & Voigt profile \\
   	 &  2470     &   [\oii] &  0.09  &  0.046  &  0.281  &  0.19  & Voigt profile \\
   	 &  2511     &  \heii   &  0.03  &  0.046  &  0.281  &  0.05  & Paschen-${\gamma}$ \\
   	 &  2629     & \feii?   &  0.06  &  0.046  &  0.281  &  0.11  & \\
   	 &  2837/38  &   \cii   &  0.12  &  0.046  &  0.281  &  0.20  & Bowen fluorescence \\
   	 &  3133     &  \oiii   &  0.16  &  0.046  &  0.281  &  0.24  & Bowen fluorescence  \\
 SMP~30  &  1397/1407 &  \oiv]  &  0.19  &  0.176  &  0.000  &  0.35 & \\
   	 &  1483/87  &   \niv]  &  0.80  &  0.176  &  0.000  &  1.36 & \\
   	 &  1548/50  &   \civ   &  0.35  &  0.176  &  0.000  &  0.58 & \\
   	 &  1640     &  \heii   &  2.26  &  0.176  &  0.000  &  3.66 & Balmer-$\alpha$, Voigt profile, G140L \\
   	 &  1747-1754 &  \niii] &  1.56  &  0.176  &  0.000  &  2.51 & noisy, smoothed with box=3 \\
   	 &  1907/09  & \ciii]   &  0.86  &  0.176  &  0.000  &  1.46 & noisy, smoothed with box=3 \\
   	 &  2423/25  &  [\neiv] &  0.67  &  0.176  &  0.000  &  1.03 & \\
 SMP~34  &  1239/43  &    \nv   &  0.07  &  0.047  &  0.024  &  0.10 & P-Cygni \\
   	 &  1548/50  &   \civ   &  0.12  &  0.047  &  0.024  &  0.15 & P-Cygni \\
	 &  1907/09  &  \ciii]  &  0.89  &  0.047  &  0.024  &  1.11 & \\
         &  2325-29  &   \cii]  &  0.16  &  0.047  &  0.024  &  0.19 & could be 2321/31 [\oiii] \\
 SMP~45  &  1548/50  &   \civ   &  0.14  &  0.162  &  0.000  &  0.22 & \\
         &  1640     &  \heii   &  0.25  &  0.162  &  0.000  &  0.39 & Balmer-$\alpha$, G140L \\
         &  1907/09  & \ciii]   &  0.79  &  0.162  &  0.000  &  1.28 & low S/N \\
 SMP~46  &  1548/50  &   \civ   &  1.01  &  0.069  &  0.116  &  2.05  & Voigt profile \\
         &  1640     &  \heii   &  0.51  &  0.069  &  0.116  &  0.97  & Balmer-$\alpha$, G140L\\
         &  1658-1666 &  \oiii] &  0.08  &  0.069  &  0.116  &  0.15  & \\
         &  1907/09  &  \ciii]  &  3.39  &  0.069  &  0.116  &  6.14  & \\
   	 &  2325-29  &   \cii]  &  1.19  &  0.069  &  0.116  &  2.04  & \\
   	 &  2423/25  &  [\neiv] &  0.10  &  0.069  &  0.116  &  0.15  & deblended from [\oii] \\
   	 &  2470     &   [\oii] &  0.13  &  0.069  &  0.116  &  0.19  & deblended from [\neiv] \\
   	 &  3133     &  \oiii   &  0.24  &  0.069  &  0.116  &  0.31  & low S/N, Bowen fluorescence \\
 SMP~48  &  1239/43 &  \nv       &  0.01 & 0.147 &   0.000 &   0.026 &   \\
	 &   1548/50  &   \civ   &  0.02  &  0.147  &  0.000  &  0.04  & P-Cygni, Voigt profile \\
         &  1658-1666 &  \oiii] &  0.04  &  0.147  &  0.000  &  0.06  & Voigt profile \\
	 &  1907/09  &  \ciii]  &  1.13  &  0.147  &  0.000  &  1.76  & Voigt profile \\
   	 &  2325-29  &   \cii]  &  0.29  &  0.147  &  0.000  &  0.45  & Voigt profile \\
	 &  2470     &   [\oii] &  0.06  &  0.147  &  0.000  &  0.09  & \\
	 &  2837     &   \oiii &  0.02  &  0.147  &  0.000  &  0.02  & Bowen flourescence \\
 SMP~59 &  1483/87  &   \niv]  &  0.22  &  0.097  &  0.000  &  0.29 & \\
   	&  1548/50  &   \civ   &  0.19  &  0.097  &  0.000  &  0.25 & \\
 	&  1640     &  \heii   &  0.68  &  0.097  &  0.000  &  0.88 & Balmer-$\alpha$, Voigt profile, G140L\\
   	&  1907/09  &  \ciii]  &  0.50  &  0.097  &  0.000  &  0.67 & low S/N spectrum \\
   	&  2306     &   \heii  &  0.10  &  0.097  &  0.000  &  0.14 & low S/N, Paschen-$\epsilon$, deblended from [\neiv] \\
   	&  2423/25  &  [\neiv] &  0.36  &  0.097  &  0.000  &  0.46 & low S/N, deblended from \heii \\
 SMP~71 &  1335/36  &   \cii   &  0.18  &  0.062  &  0.182  &  0.60 & marginal feature, smoothed with box=3\\
   	&  1397/1407 &  \oiv]  &  0.12  &  0.062  &  0.182  &  0.37 & Voigt profile\\
   	&  1483/87  &   \niv]  &  0.04  &  0.062  &  0.182  &  0.12 & \\
   	&  1520     &  $\dots$   &  0.01  &  0.062  &  0.182  &  0.02 & not identified\\
   	&  1548/50  &   \civ   &  3.33  &  0.062  &  0.182  &  8.75 & \\
	&  1575     &   [\nev]   &  0.04  &  0.062  &  0.182  &  0.09 & deblended from [\neiv] \\
   	&  1602     &  [\neiv] &  0.03  &  0.062  &  0.182  &  0.06 & deblended from [\nev] \\
   	&  1640     &  \heii   &  0.95  &  0.062  &  0.182  &  2.32 & Balmer-$\alpha$, Voigt function, deblended from \oiii, G140L\\
   	&  1640     &  \heii   &  1.27  &  0.062  &  0.182  &  3.09 & Balmer-$\alpha$, Voigt function, G230L \\
   	&  1658-1666 &  \oiii] &  0.10  &  0.062  &  0.182  &  0.25 & deblended from \heii, G140L \\
   	&  1907/09  &  \ciii]  &  4.21  &  0.062  &  0.182  &  9.29 & \\
   	&  2306     &   \heii  &  0.10  &  0.062  &  0.182  &  0.21 & marginal, Paschen-$\epsilon$, deblended from \cii] \\
   	&  2325-29  &   \cii]  &  0.52  &  0.062  &  0.182  &  1.06 & Voigt profile, deblended from \heii \\
   	&  2423/25  &  [\neiv] &  0.21  &  0.062  &  0.182  &  0.38 & \\
   	&  2470     &   [\oii] &  0.05  &  0.062  &  0.182  &  0.08 & \\
   	&  2511     &  \heii   &  0.02  &  0.062  &  0.182  &  0.03 & marginal, Paschen-${\gamma}$ \\
   	&  2733     &   \heii  &  0.06  &  0.062  &  0.182  &  0.09 & Paschen-${\beta}$ \\
   	&  2784     &  [\mgv]? &  0.04  &  0.062  &  0.182  &  0.06 & Possibly 2786 [\arv] \\
   	&  2837/38  &   \cii   &  0.05  &  0.062  &  0.182  &  0.08 & Bowen fluorescence \\
   	&  3023     &  \oiii   &  0.02  &  0.062  &  0.182  &  0.03 & marginal, Bowen fluor., deblended from 3043/47 \oiii\\
   	&  3043/47  &  \oiii   &  0.08  &  0.062  &  0.182  &  0.10 & Bowen fluorescence,deblended from 3023 \oiii \\
   	&  3133     &  \oiii   &  0.42  &  0.062  &  0.182  &  0.57 & Bowen fluorescence \\
 SMP~72 &  1397/1407 &  \oiv] &  0.61  &  0.191  &  0.000  &  1.15  & diffuse spectrum \\
  	&  1548/50  &   \civ   &  4.37  &  0.191  &  0.000  &  7.54  & \\
	&  1640     &  \heii   &  3.45  &  0.191  &  0.000  &  5.81  & Balmer-$\alpha$, G140L \\
  	&  1907/09  &  \ciii]  &  4.80  &  0.191  &  0.000  &  8.55  & \\
  	&  2423/25  &  [\neiv] &  1.09  &  0.191  &  0.000  &  1.75  & \\
 SMP~79 &  1175/76 &  \ciii    &  0.16  &  0.221  &  0.000  &  0.48  & \\
 	&  1239/43 &   \nv     &  0.07  &  0.221  &  0.000  &  0.19  & marginally P-Cygni \\
  	&  1335/36 &   \cii    &  0.06  &  0.221  &  0.000  &  0.13  & \\
  	&  1371    &    \ov    &  0.06  &  0.221  &  0.000  &  0.12  & marginally P-Cygni \\
  	&  1548/50 &   \civ    &  0.90  &  0.221  &  0.000  &  1.69  & Voigt profile \\
  	&  1575    &   [\nev]  &  0.04  &  0.221  &  0.000  &  0.07  & \\
  	&  1640    &  \heii    &  0.39  &  0.221  &  0.000  &  0.72  & Balmer-$\alpha$, Voigt profile, G140L \\
	&  1640    &  \heii    &  0.33  &  0.221  &  0.000  &  0.61  & Balmer-$\alpha$, G230L \\
  	&  1658-1666 &  \oiii] &  0.07  &  0.221  &  0.000  &  0.12  & \\
  	&  1907/09 &  \ciii]   &  4.50  &  0.221  &  0.000  &  8.77  & Voigt profile \\
  	&  2297    &   \ciii   &  0.05  &  0.221  &  0.000  &  0.11  & deblended from \cii] \\
  	&  2325-29 &   \cii]   &  0.62  &  0.221  &  0.000  &  1.20  & deblended from \ciii \\
  	&  2423/25 &  [\neiv]  &  0.01  &  0.221  &  0.000  &  0.02  & \\
  	&  2470    &  [\oii]   &  0.06  &  0.221  &  0.000  &  0.09  & \\
  	&  2733    &   \heii   &  0.02  &  0.221  &  0.000  &  0.03  & Paschen-${\beta}$ \\
  	&  2784/2929 &  \mgv   &  0.01  &  0.221  &  0.000  &  0.02  & \\
  	&  2837/38 &   \cii    &  0.04  &  0.221  &  0.000  &  0.05 &  P-Cygni \\
  	&  3043/47 &  \oiii    &  0.03  &  0.221  &  0.000  &  0.04  & Bowen fluorescence \\
  	&  3133    &  \oiii    &  0.15  &  0.221  &  0.000  &  0.19  & Bowen fluorescence  \\
 SMP~80  &  1239/43 &    \nv   &  0.02  &  0.044  &  0.046  &  0.04  & P-Cygni \\
  	 &  1548/50 &   \civ   &  0.06  &  0.044  &  0.046  &  0.08  & low S/N, P-Cygni, Voigt profile \\
   	 &  1907/09 &  \ciii]  &  0.23  &  0.044  &  0.046  &  0.30  & \\
        &  2325-29  &   \cii]  &  0.08  &  0.044  &  0.046  &  0.05  & marginal feature \\
 SMP~81 &  1239/43  &    \nv   &  0.04  &  0.221  &  0.056  &  0.14  & P-Cygni\\
  	&  1483/87  &   \niv]  &  0.01  &  0.221  &  0.056  &  0.02  & marginal feature \\
 	&  1548/50  &   \civ   &  0.03  &  0.221  &  0.056  &  0.06  & \\
 	&  1658-1666 &  \oiii] &  0.03  &  0.221  &  0.056  &  0.06  & two components deblended, G140L \\
 	&  1658-1666 &  \oiii] &  0.06  &  0.221  &  0.056  &  0.14  & two components deblended, G140L \\
 	&  1658-1666 &  \oiii] &  0.07  &  0.221  &  0.056  &  0.16  & primary spatial component, G230L \\
 	&  1658-1666 &  \oiii] &  0.08  &  0.221  &  0.056  &  0.18  & two spatial components summed, G230L \\
	&  1747-1754 &  \niii] &  0.03  &  0.221  &  0.056  &  0.06  & primary spatial component \\
	&  1747-1754 &  \niii] &  0.04  &  0.221  &  0.056  &  0.08  & two spatial components \\
	&  1892     & \siiii]  &  0.04  &  0.221  &  0.056  &  0.10  & primary spatial component, deblended from \ciii] \\
	&  1892     & \siiii]  &  0.03  &  0.221  &  0.056  &  0.07  & two spatial components, deblended from \ciii] \\
	&  1907/09  &  \ciii]  &  0.18  &  0.221  &  0.056  &  0.42  & primary spatial component, deblended from \siiii] \\
	&  1907/09  &  \ciii]  &  0.26  &  0.221  &  0.056  &  0.62  & two spatial components, deblended from \siiii] \\
	&  2321/31  &  [\oiii] &  0.05  &  0.221  &  0.056  &  0.10  & primary spatial component \\
	&  2321/31  &  [\oiii] &  0.05  &  0.221  &  0.056  &  0.12  & two spatial components \\
	&  2470     &  [\oii]  &  0.04  &  0.221  &  0.056  &  0.076  & primary spatial component \\
	&  2470     &  [\oii]  &  0.04  &  0.221  &  0.056  &  0.077  & two spatial components \\
 SMP~93 &  1483/87  &   \niv]  &  0.15  &  0.056  &  0.000  &  0.17  & diffuse, low S/N spectrum \\
  	&  1548/50  &   \civ   &  0.18  &  0.056  &  0.000  &  0.21  & marginal feature \\
  	&  1640     &  \heii   &  0.52  &  0.056  &  0.000  &  0.60  & Balmer-$\alpha$, G140L\\
  	&  1907/09  &   \ciii] &  0.49  &  0.056  &  0.000  &  0.58  & low S/N\\
  	&  2325-29     &   \cii  &  0.28  &  0.056  &  0.000  &  0.33  & low S/N\\
 SMP~95 &  1335/36  &   \cii   &  0.18  &  0.096  &  0.031  &  0.31  & \\
        &  1397-1407 &  \oiv] &  0.06  &  0.096  &  0.031  &  0.09  & \\
 	&  1548/50  &   \civ   &  1.84  &  0.096  &  0.031  &  2.76  & \\
	&  1640     &  \heii   &  1.40  &  0.096  &  0.031  &  2.05  & Balmer-$\alpha$, G140L\\
  	&  1640     &  \heii   &  1.90  &  0.096  &  0.031  &  2.78  & Balmer-$\alpha$, G230L \\
  	&  1907/09  &  \ciii]  &  4.96  &  0.096  &  0.031  &  7.33  & \\
  	&  2325-29  &   \cii]  &  1.69  &  0.096  &  0.031  &  2.46  & \\
  	&  2423/25  &  [\neiv] &  0.11  &  0.096  &  0.031  &  0.15  & \\
  	&  2796/2803 &  \mgii &  0.09  &  0.096  &  0.031  &  0.11  & \\
SMP~102 &  1342     &   \oiv   &  0.04  &  0.162  &  0.000  &  0.07  & marginal feature\\
	&  1548/50  &   \civ   &  4.86  &  0.162  &  0.000  &  7.72  & Voigt profile\\
	&  1640     &  \heii   &  2.59  &  0.162  &  0.000  &  4.03  & Balmer-$\alpha$, G140L \\
	&  1907/09  &  \ciii]  &  6.16  &  0.162  &  0.000  &  10.0  & \\
	&  2321/31  &  [\oiii] &  0.47  &  0.162  &  0.000  &  0.77  & \\
	&  2423/25  &  [\neiv] &  0.34  &  0.162  &  0.000  &  0.51  & \\
	&  2511     &  \heii   &  0.09  &  0.162  &  0.000  &  0.13  & Paschen-${\gamma}$ \\
	&  2733     &  \heii   &  0.23  &  0.162  &  0.000  &  0.30  & Paschen-${\beta}$ \\
	&  3043/47  &  \oiii   &  0.15  &  0.162  &  0.000  &  0.18  & low S/N, Bowen fluorescence\\
	&  3133     &  \oiii   &  0.39  &  0.162  &  0.000  &  0.47  & low S/N, Bowen fluorescence  \\

\enddata
\end{deluxetable}

\begin{deluxetable}{lrrr}

\tablewidth{0pt}
\tablecaption {Velocities from P-Cygni profiles \label{Velocity}}

\tablehead{ \colhead{Name} & \colhead{$\lambda$} & \colhead{ID} & 
\colhead{v$_{\infty}$} \\
\colhead{} & \colhead{[\AA~]} & \colhead{} & \colhead{[km s$^{-1}$}]}

\startdata

 SMP~18 & 1239/43 & \nv	&   1670 \\ 
        & 1548/50 & \civ&   1570 \\ 
 SMP~25 & 1239/43 & \nv	&   1830 \\ 
        & 1548/50 & \civ&   1900 \\ 
 SMP~27 & 1239/43 & \nv	&   1840 \\ 
        & 1548/50 & \civ&   1770 \\ 
 SMP~34 & 1239/43 & \nv	&   (1680)\tablenotemark{a}\\ 
        & 1548/50 & \civ&   1920\\ 
 SMP~48 & 1239/43 & \nv	&   (2200)\tablenotemark{a}\\ 
        & 1548/50 & \civ&     1320\\
 SMP~79 & 1371    &  \ov&   (3200)\tablenotemark{a}\\ 
        & 2837/38 & \cii&   4200\\ 
 SMP~80 & 1239/43 & \nv &    3000\\ 
        & 1548/50 & \civ&    2800\\
 SMP~81 & 1239/43 & \nv &    (3042)\tablenotemark{a}\\ 

\enddata
\tablenotetext{a}{very uncertain measurement, low S/N}
\end{deluxetable}

\begin{deluxetable}{lrrrrrrr}
\tabletypesize{\small}
\tablewidth{0pt}

\tablecaption {Plasma Diagnostics and Nebular Excitation Class \label{Diags}}

\tablehead{ \colhead{Name} & \colhead{} & \colhead{N$_{\rm e}$} & \colhead{} & \colhead{} & \colhead{T$_{\rm e}$} & \colhead{} & \colhead{EC}\\ 
\colhead{} & \colhead{low}& \colhead{medium} & \colhead{high} & \colhead{low} &\colhead{medium} & \colhead{high} & \colhead{}\\}

\startdata
   J~41 &   500 &   500 &   500 & 10000 & 10000 & 10000 & \nodata \\
  SMP~4 &  5000 &  5000 &  5000 & 11800 & 11800 & 11800 & \nodata \\ 
  SMP~9 &  3000 &  3000 &  3000 & 14800 & 14800 & 14800 & 6-7     \\
 SMP~10 &  2800 &  2800 &  2800 & 17900 & 17900 & 17900 & 6-7     \\    
 SMP~16 &   600 &   600 &   600 & 12000 & 19800 & 19800 & 8       \\
 SMP~18 &   400 &   400 &   400 & 11600 & 11600 & 11600 & 2-6     \\
 SMP~19 &  3000 &  4000 &  4000 & 12000 & 13000 & 13500 & 7       \\
 SMP~25 & 16600 & 16600 & 16600 & 10900 & 13000 & 13000 & \nodata \\     
 SMP~27 &   500 &   500 &   500 & 11500 & 17200 & 17200 & 3-6     \\
 SMP~28 &  2000 &  2000 &  2000 & 10000 & 10000 & 10000 & 6-7     \\   
 SMP~30 &   400 &   400 &   400 & 12300 & 15700 & 15700 & 8 \\
 SMP~34 &  5000 &  5000 &  5000 & 10000 & 10000 & 10000 & 2 \\    
 SMP~45 &  1000 &  1000 &  1000 & 16300 & 16300 & 16300 & 6 \\    
 SMP~46 &  2500 &  2500 &  2500 & 11800 & 11800 & 11800 & 6 \\   
 SMP~48 &  1900 &  1900 &  1900 &  8800 & 11300 & 11300 & 3-4 \\    
 SMP~59 &   200 &   200 &   200 & 12400 & 15000 & 15000 & 7-9?\\  
 SMP~71 &  7500 & 10000 & 10000 & 11500 & 12800 & 12800 & 6-7 \\     
 SMP~72 &   100 &   100 &   100 & 16000 & 16000 & 16000 & 7 \\
 SMP~79 &  3100 &  3100 &  3100 & 13600 & 12900 & 12900 & 5\\   
 SMP~80 &  4500 &  8700 &  8700 & 12500 &  9800 & 10000 & 2-4 \\
 SMP~81 & 10000 & 20000 & 20000 & 15000 & 15000 & 15000 & 5 \\     
 SMP~93 &   300 &   300 &   300 & 11000 & 17100 & 17100 & 8 \\
 SMP~95 &  1000 &  1000 &  1000 & 11300 & 12700 & 12700 & 5-7 \\  
SMP~102 &  1000 &  1000 &  1000 & 13700 & 13700 & 13700 & 7 \\   
\enddata
\end{deluxetable}

\begin{deluxetable}{lrcc}

\tablewidth{0pt}
\tablecaption {Emission Lines Used for Ionic Abundances \label{LinesUsed}}

\tablehead{ 
\colhead{Ion} & \colhead{Spectrum} & \colhead{Wavelengths} &  \colhead{Ionization} \\
\colhead{} & \colhead{} & \colhead{(\AA)} & \colhead{Zone} 
}

\startdata
 C$^{+}$  &  \ion{C}{2}] & 2626+28     & Low \\ 
 C$^{2+}$ &  \ion{C}{3}] & 1907+09     & Med \\ 
 C$^{3+}$ &  \ion{C}{4}\tablenotemark{a} & 1548+1550 & Med \\ 
 N$^{2+}$ &  \ion{N}{3}] & 1749+52     & Med \\ 
 N$^{3+}$ & [\ion{N}{4}] & 1483+87     & Med \\ 
 N$^{4+}$ &  \ion{N}{5}\tablenotemark{a} & 1239+43 & High \\ 
 O$^{3+}$ &  \ion{O}{3}] & 1400+01+05+07 & High \\ 
Ne$^{3+}$ &  \ion{Ne}{4}] & 2423+25    & High \\ 
\enddata
\tablenotetext{a}{Abundance not derived in {\bf nebular} software}

\end{deluxetable}


\begin{deluxetable}{lccccccccr}
\tabletypesize{\scriptsize}
\tablecaption{Ionic Abundances \label{Abund}}
\tablehead{
\colhead{Name} & \colhead{C$^+$/H$^+$} & \colhead{C$^{2+}$/H$^+$} & \colhead{C$^{3+}$/H$^+$} & \colhead{C/H}&
 \colhead{N$^{2+}$/H$^+$} & \colhead{N$^{3+}$/H$^+$} & \colhead{N$^{4+}$/H$^+$} & \colhead{O$^{3+}$} & \colhead{Ne$^{3+}$/H$^+$} \\ 
\colhead{(1)} & \colhead{(2)} & \colhead{(3)} & \colhead{(4)} & \colhead{(5)} & 
\colhead{(6)} & \colhead{(7)} & \colhead{(8)} & \colhead{(9)} & \colhead{(10)}\\} 

\startdata
   J~41  &   \nodata   &   9.80e-05  &   \nodata   &   9.80e-05  &   \nodata   &   \nodata   &   \nodata   &   \nodata   &   \nodata    \\
  SMP~04  &   \nodata   &   2.27e-04  &   2.27e-04  &   4.54e-04  &   \nodata   &   \nodata   &   2.10e-04  &   \nodata   &   \nodata    \\
  SMP~09  &   5.42e-05  &   1.52e-04  &   6.59e-05  &   2.72e-04  &   \nodata   &   \nodata   &   \nodata   &   4.90e-05  &   5.44e-06   \\
  SMP~10  &   \nodata   &   2.02e-05  &   2.09e-06  &   2.23e-05  &   \nodata   &   \nodata   &   3.09e-06  &   \nodata   &   \nodata    \\
  SMP~16  &   1.30e-05  &   6.96e-06  &   1.46e-06  &   2.14e-05  &   1.50e-05  &   1.25e-05  &   5.74e-06  &   2.19e-05  &   1.12e-05   \\
  SMP~18  &   1.42e-05  &   1.79e-04  &   4.41e-05  &   2.38e-04  &   \nodata   &   \nodata   &   5.45e-04  &   \nodata   &   \nodata    \\
  SMP~19  &   9.80e-05  &   4.61e-05  &   1.47e-04  &   2.91e-04  &   \nodata   &   3.09e-05  &   \nodata   &   1.46e-04  &   2.15e-05   \\
  SMP~25  &   4.91e-05  &   9.76e-05  &   4.62e-05  &   1.93e-04  &   \nodata   &   \nodata   &   3.15e-04  &   \nodata   &   \nodata    \\
  SMP~27  &   3.65e-05  &   2.85e-05  &   4.59e-06  &   6.96e-05  &   \nodata   &   \nodata   &   4.09e-05  &   \nodata   &   \nodata    \\
  SMP~28  &   \nodata   &   1.83e-05  &   8.29e-05  &   1.01e-04  &   3.52e-03  &   1.75e-03  &   1.87e-02  &   \nodata   &   5.53e-05   \\
  SMP~30  &   \nodata   &   5.11e-05  &   4.59e-06  &   5.57e-05  &   2.80e-04  &   6.15e-05  &   \nodata   &   6.51e-05  &   1.91e-05   \\
  SMP~34  &   3.76e-05  &   5.50e-05  &   4.14e-05  &   1.34e-04  &   \nodata   &   \nodata   &   2.90e-04  &   \nodata   &   \nodata    \\
  SMP~45  &   \nodata   &   3.67e-05  &   1.38e-06  &   3.81e-05  &   \nodata   &   \nodata   &   \nodata   &   \nodata   &   \nodata    \\
  SMP~46  &   1.46e-04  &   3.03e-04  &   1.29e-04  &   5.78e-04  &   \nodata   &   \nodata   &   \nodata   &   \nodata   &   1.13e-05   \\
  SMP~48  &   2.18e-04  &   3.13e-05  &   3.62e-06  &   2.53e-04  &   \nodata   &   \nodata   &   1.95e-05  &   \nodata   &   \nodata    \\
  SMP~59  &   \nodata   &   1.19e-05  &   2.66e-06  &   1.45e-05  &   \nodata   &   1.75e-05  &   \nodata   &   \nodata   &   1.03e-05   \\
  SMP~71  &   8.80e-05  &   4.19e-04  &   2.88e-04  &   7.94e-04  &   \nodata   &   2.22e-05  &   \nodata   &   2.62e-04  &   2.01e-05   \\
  SMP~72  &   \nodata   &   1.08e-04  &   5.30e-05  &   1.61e-04  &   \nodata   &   \nodata   &   \nodata   &   1.92e-04  &   2.98e-05   \\
  SMP~79  &   4.02e-05  &   3.74e-04  &   5.23e-05  &   4.67e-04  &   \nodata   &   \nodata   &   3.73e-05  &   \nodata   &   9.49e-07   \\
  SMP~80  &   5.20e-06  &   5.37e-06  &   2.21e-05  &   3.27e-05  &   \nodata   &   \nodata   &   1.17e-04  &   \nodata   &   \nodata    \\
  SMP~81  &   2.51e-06  &   1.12e-05  &   6.38e-07  &   1.44e-05  &   7.46e-06  &   2.69e-06  &   7.65e-06  &   \nodata   &   \nodata    \\
  SMP~93  &   3.57e-05  &   5.26e-06  &   9.87e-07  &   4.19e-05  &   \nodata   &   4.61e-06  &   \nodata   &   \nodata   &   \nodata    \\
  SMP~95  &   2.26e-04  &   3.44e-04  &   9.64e-05  &   6.67e-04  &   \nodata   &   \nodata   &   \nodata   &   \nodata   &   7.49e-06   \\
 SMP~102  &   \nodata   &   2.96e-04  &   1.53e-04  &   4.50e-04  &   \nodata   &   \nodata   &   \nodata   &   \nodata   &   1.75e-05   \\
\enddata
\end{deluxetable}

\clearpage

\begin{deluxetable}{lrrrr}
\tablewidth{10truecm}
\tablecaption {Average C/H abundances \label{AvgAbund}}

\tablehead{\colhead{} & \colhead{median} & \colhead{mean} & \colhead{sigma} & \colhead{N}\\}

\startdata
R & 3.16e-04 & 3.23e-04 & 3.59e-04 &8\\
E & 3.39e-04 & 5.11e-04 & 4.88e-04 &8\\
BC& 2.91e-04 & 3.19e-04 & 2.06e-04 &14\\
B & 3.81e-05 & 3.58e-05 & 1.93e-05 &8\\
P & 1.01e-04 & 6.17e-05 & 5.60e-05 &2\\

\enddata
\end{deluxetable}

\newpage
\plotone{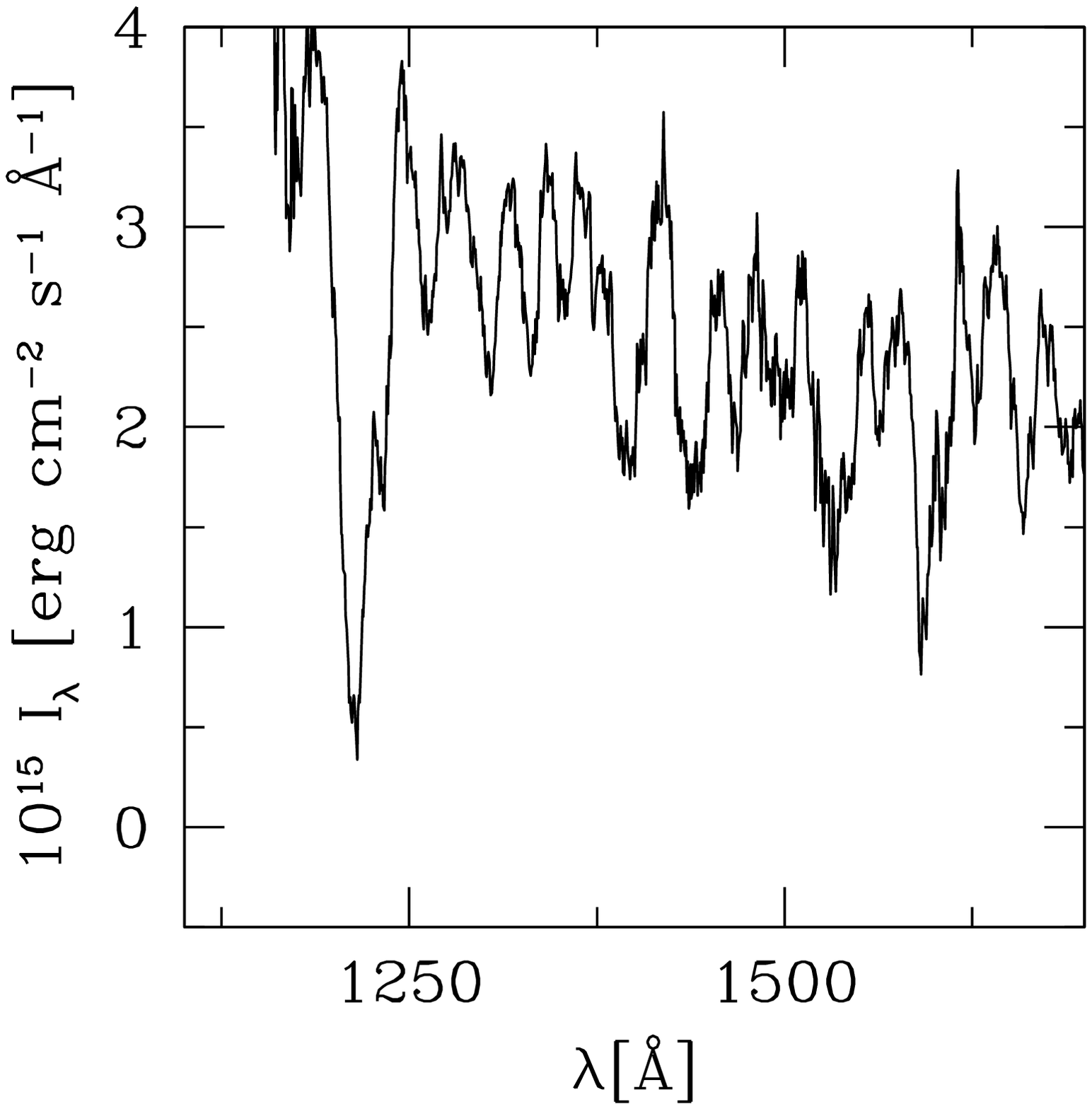}
\newpage
\plotone{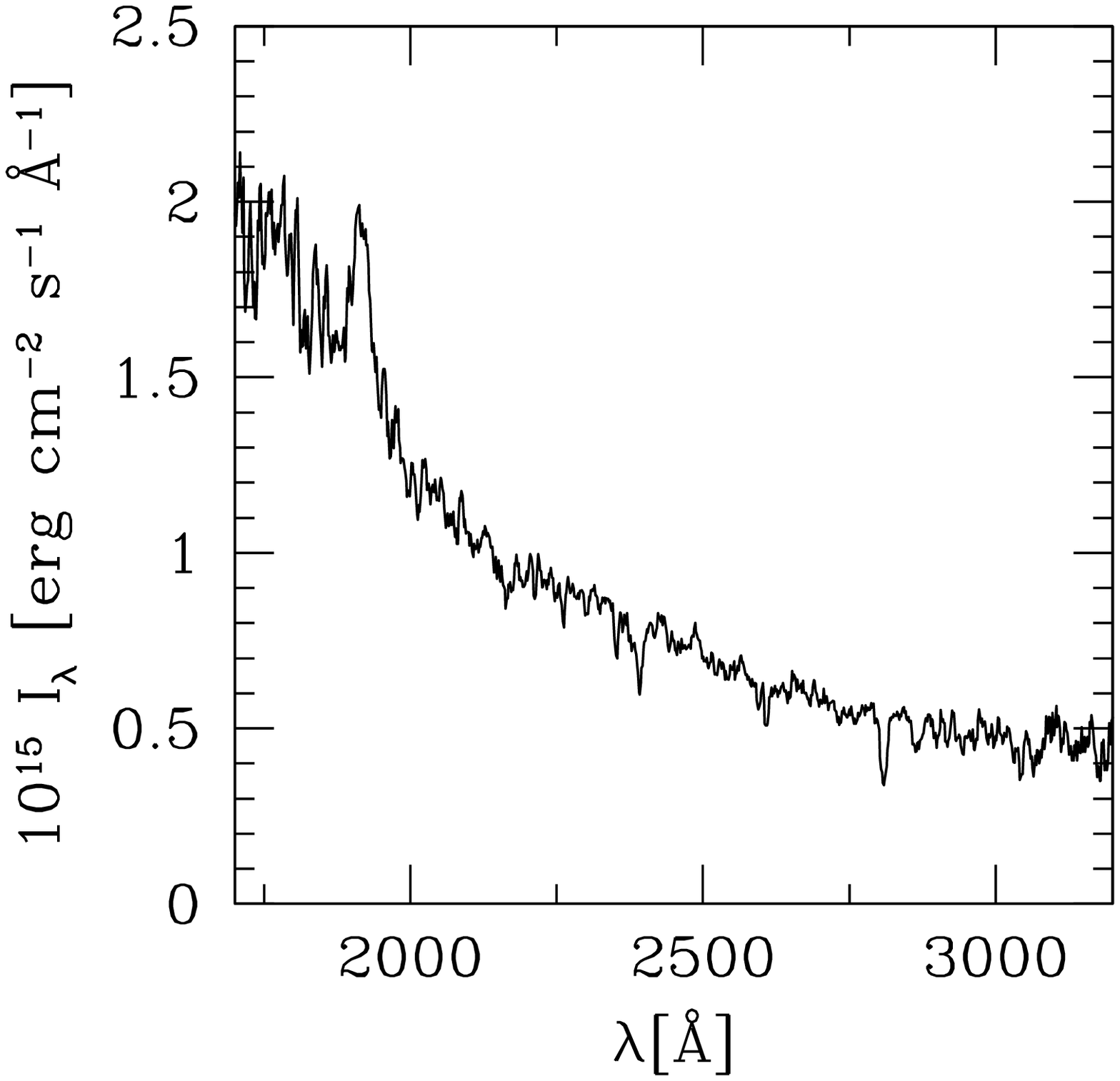}
\newpage
\plotone{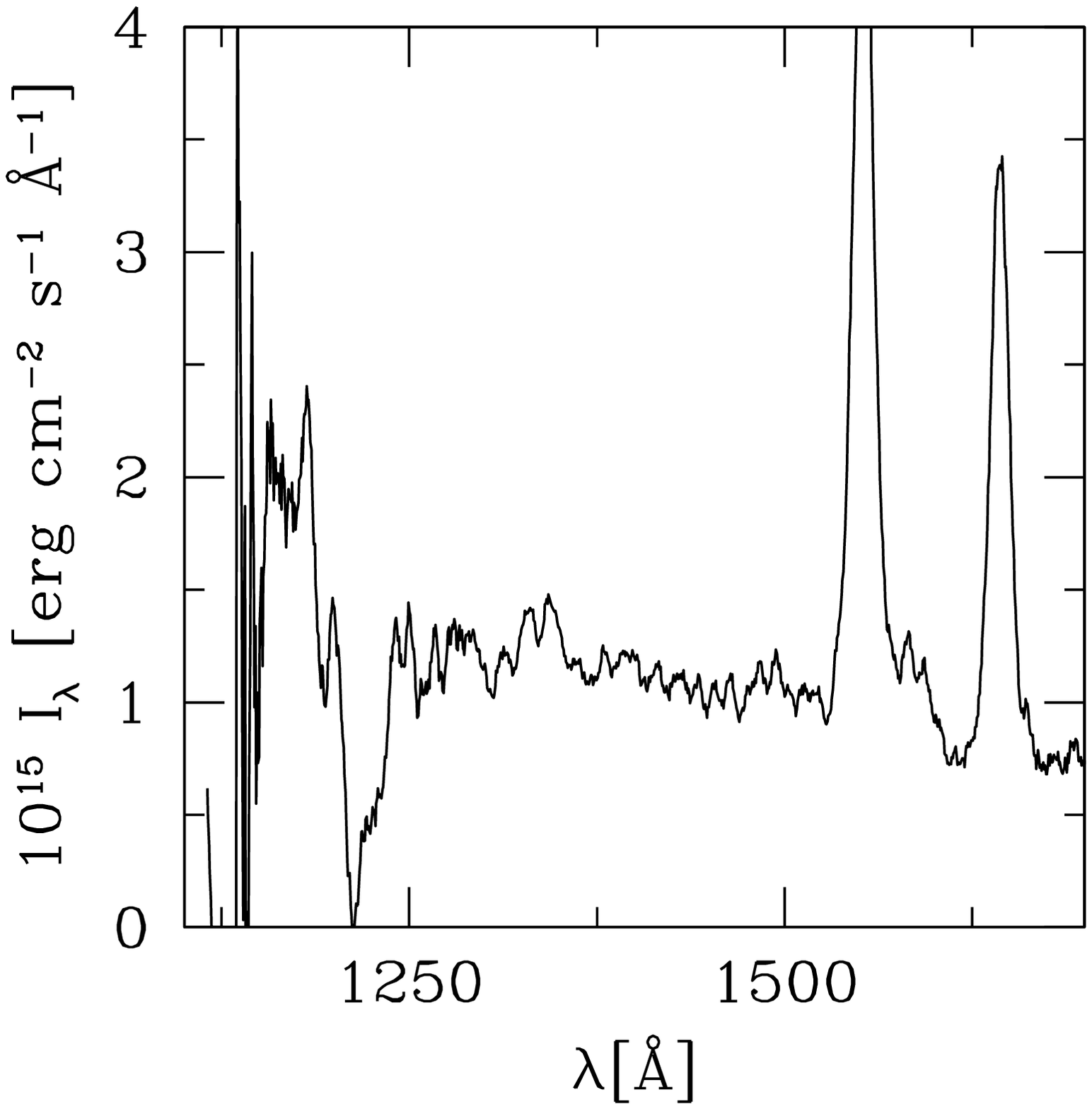}
\newpage
\plotone{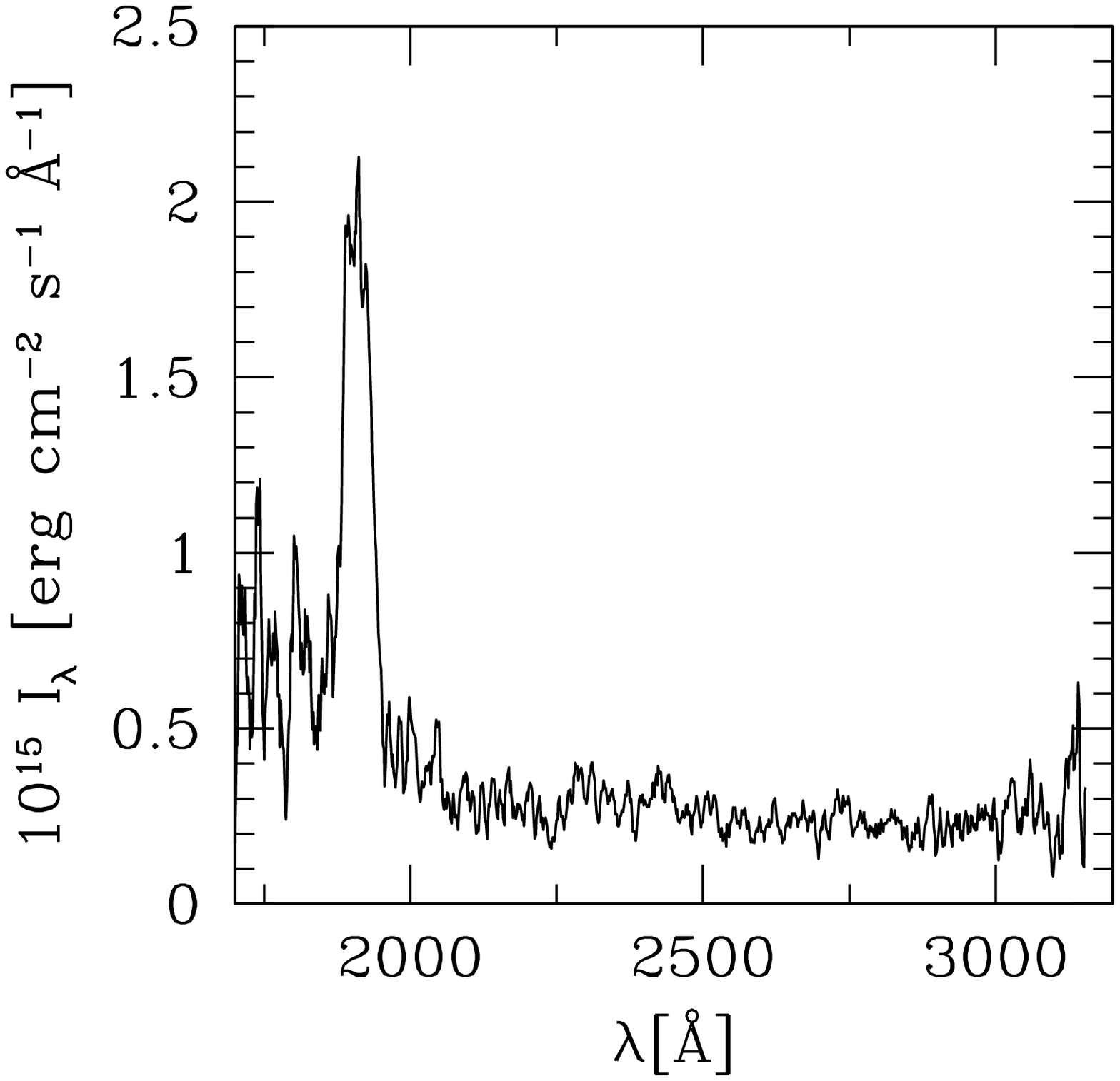}
\newpage
\plotone{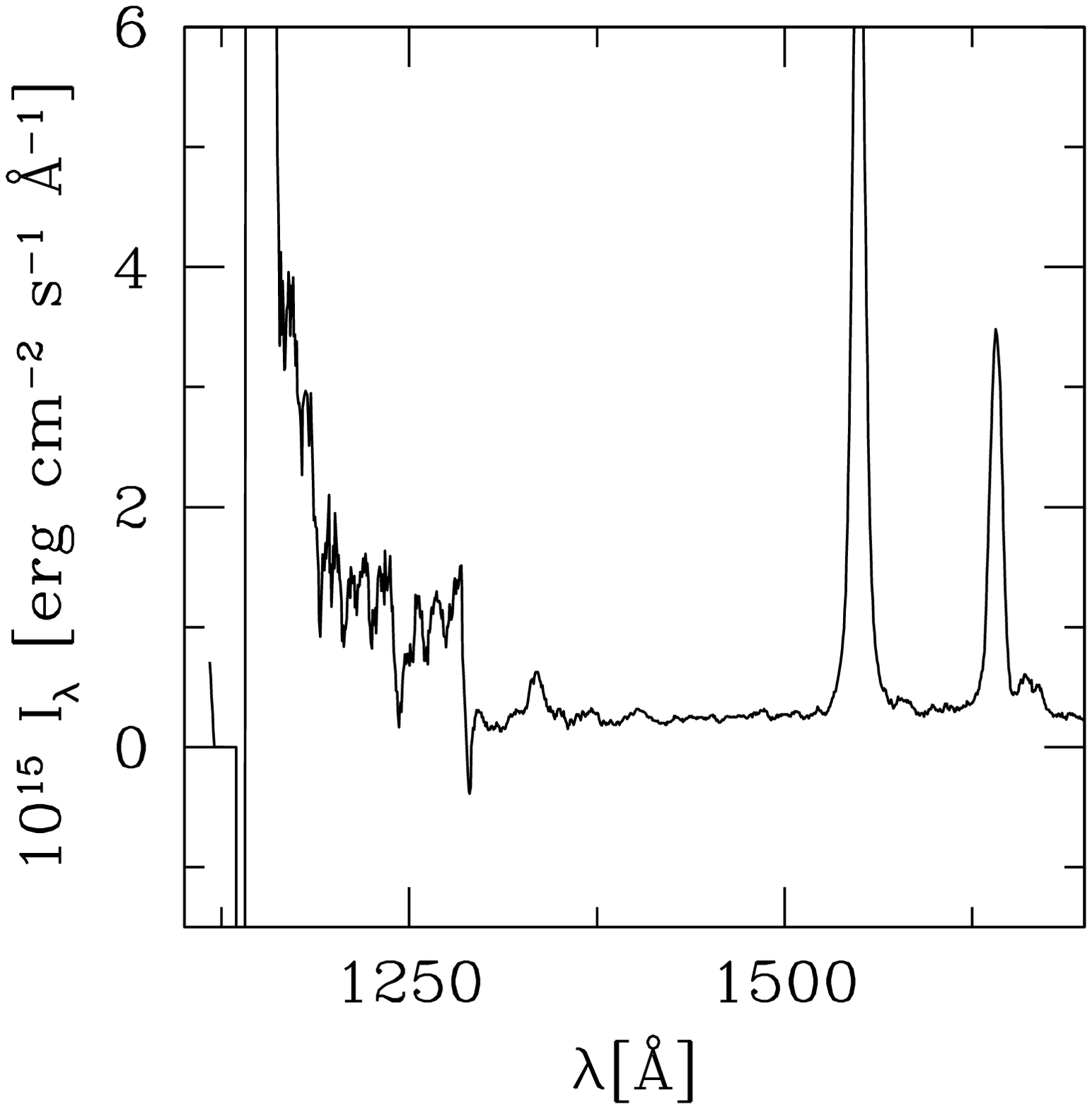}
\newpage
\plotone{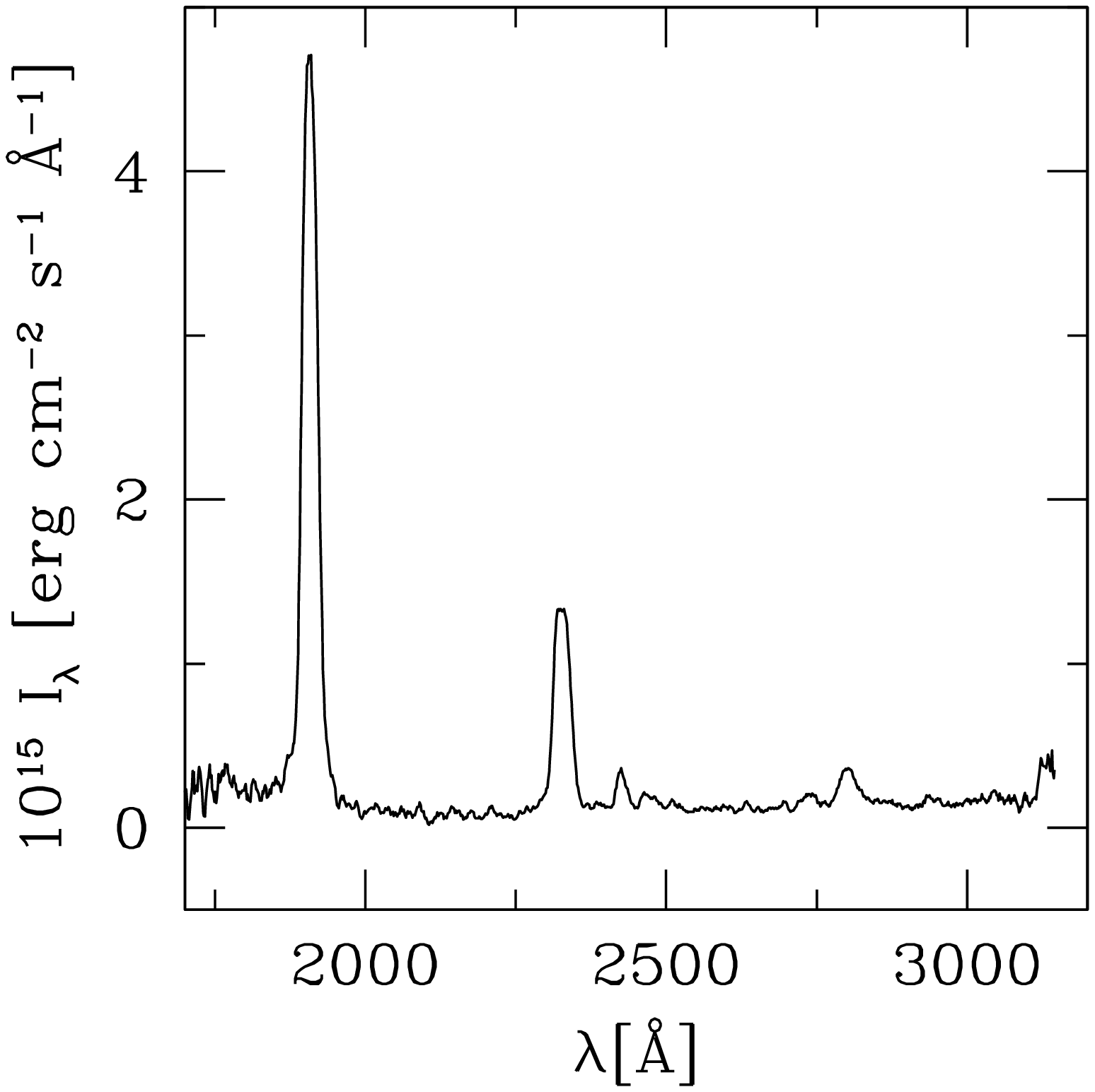}
\newpage
\plotone{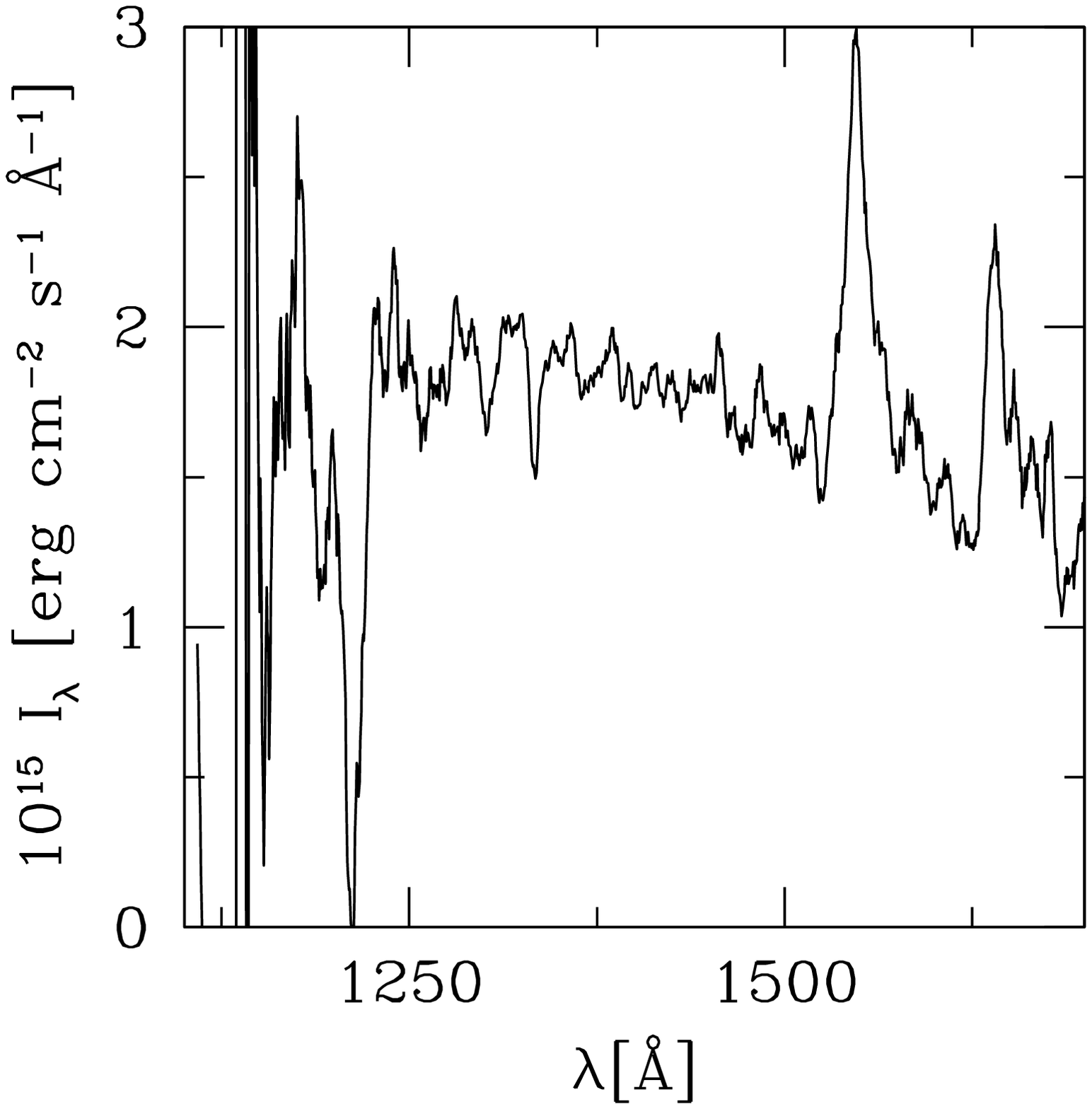}
\newpage
\plotone{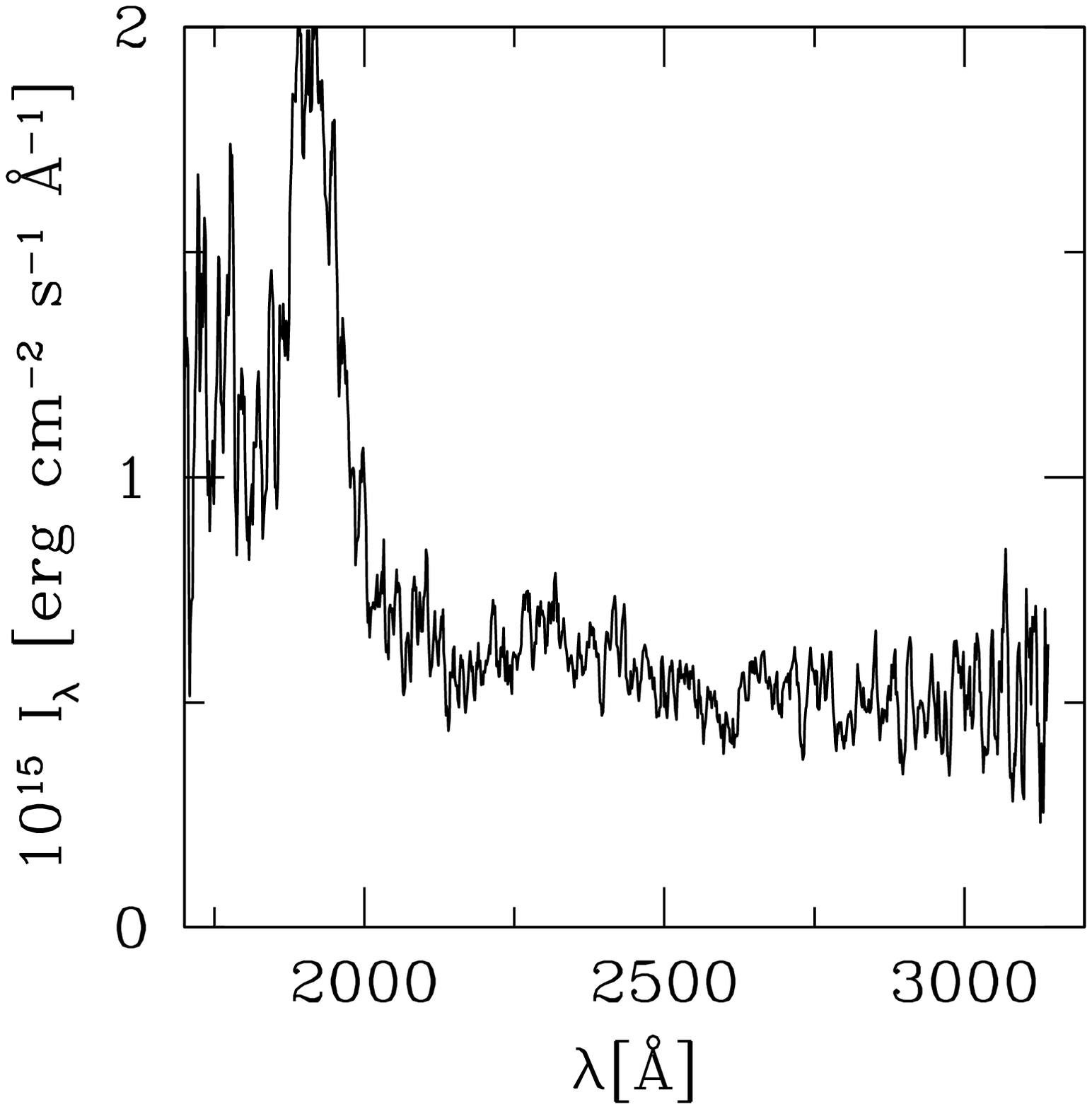}
\newpage
\plotone{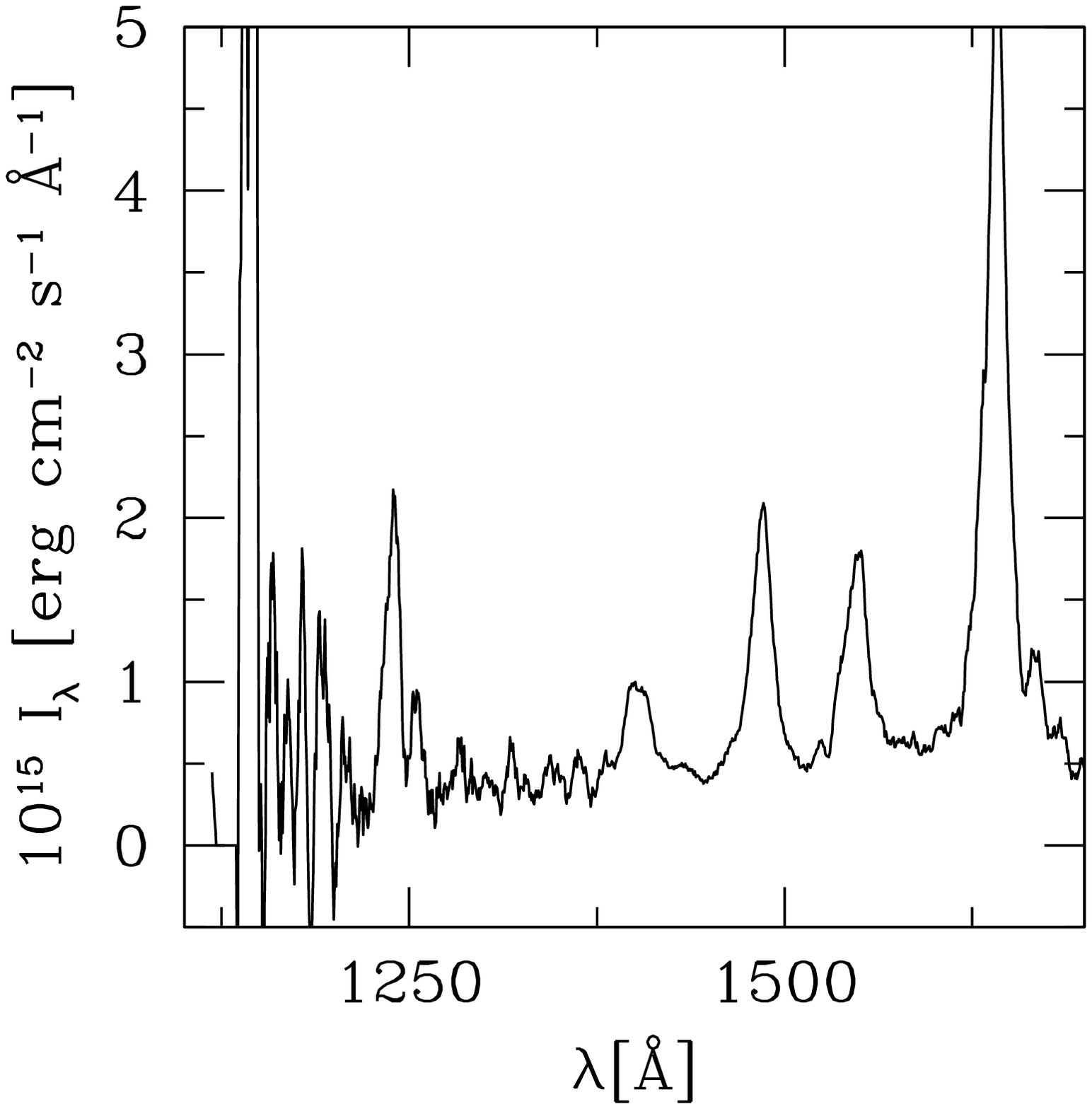}
\newpage
\plotone{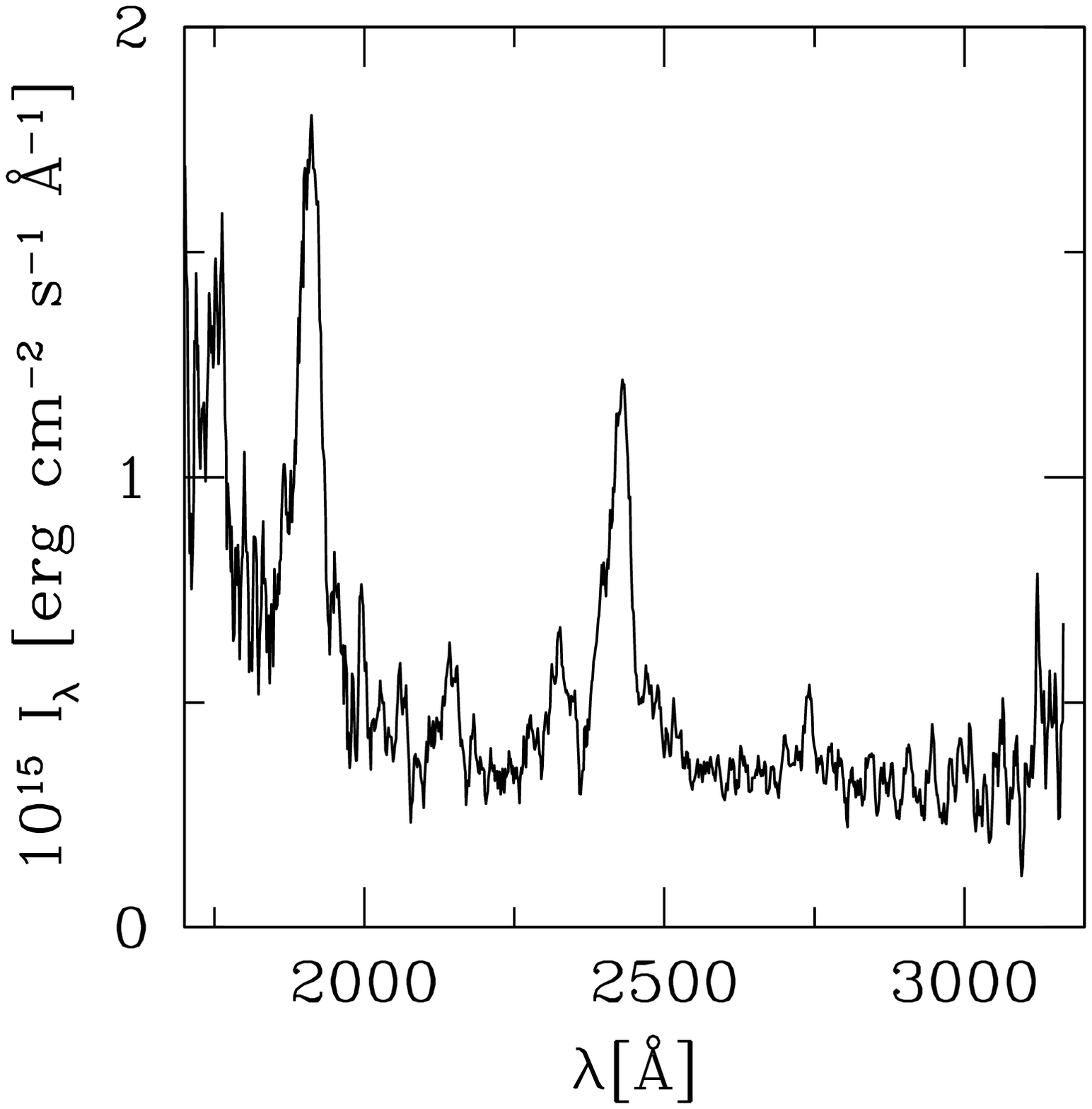}
\newpage
\plotone{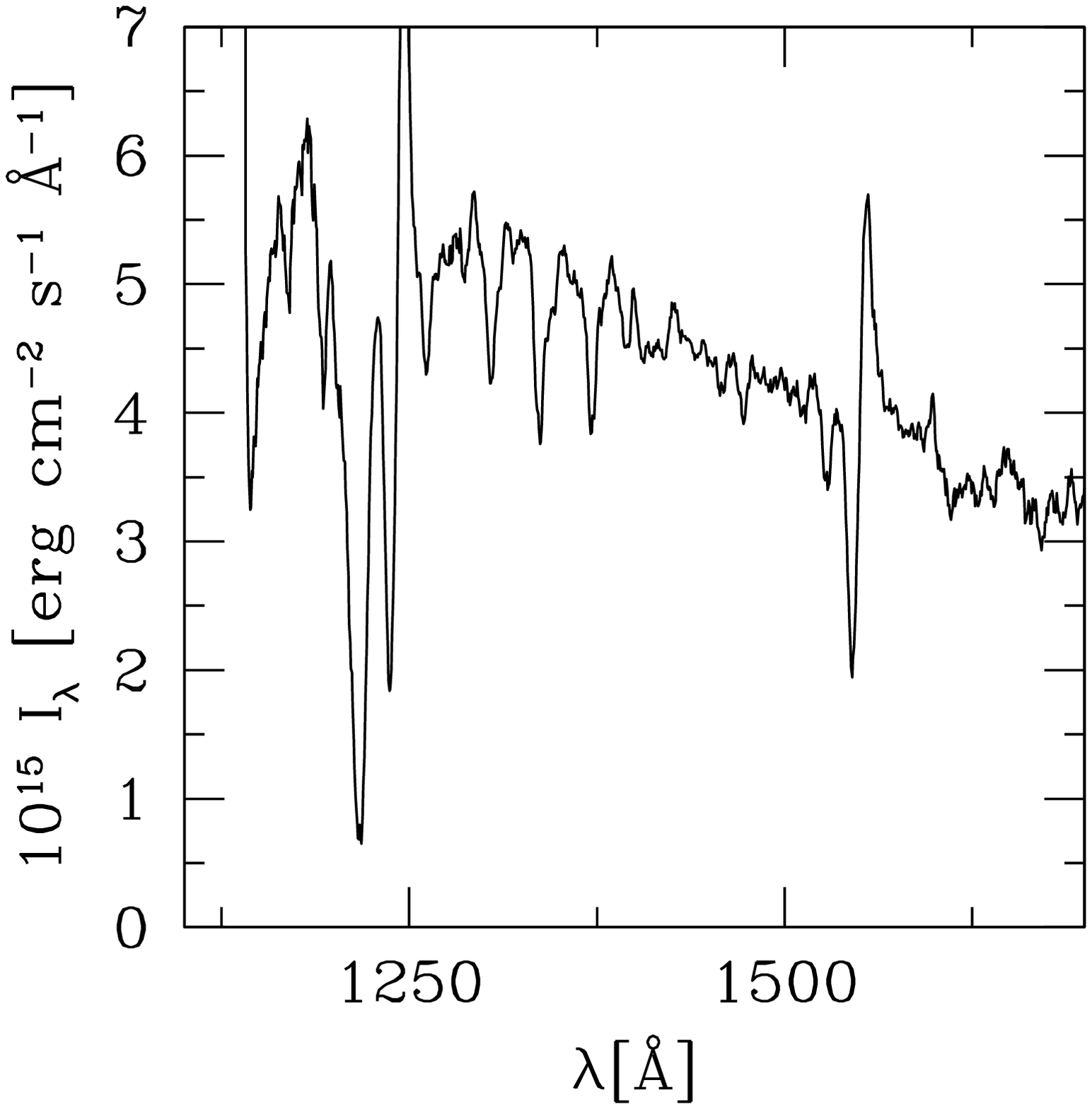}
\newpage
\plotone{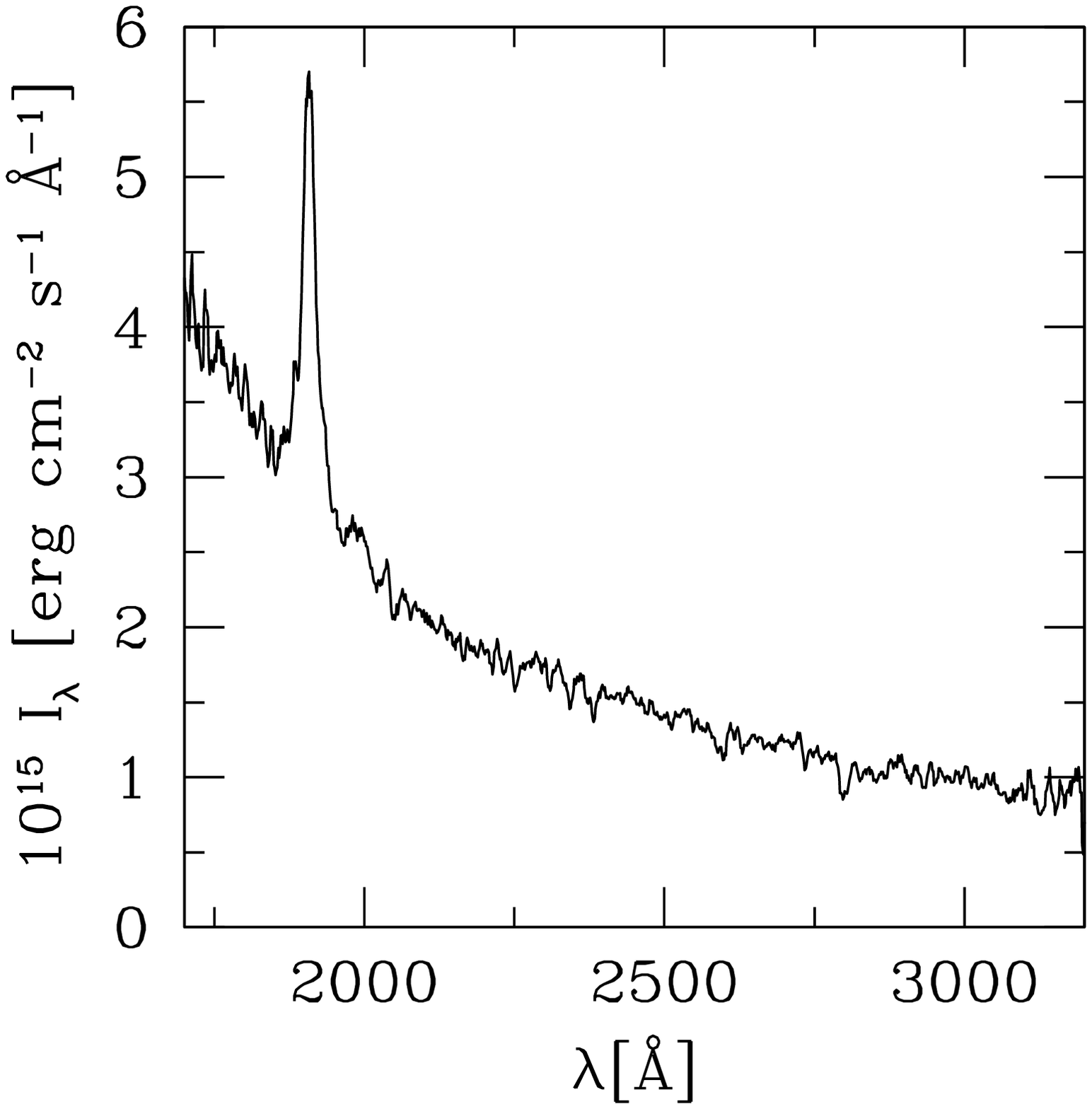}
\newpage
\plotone{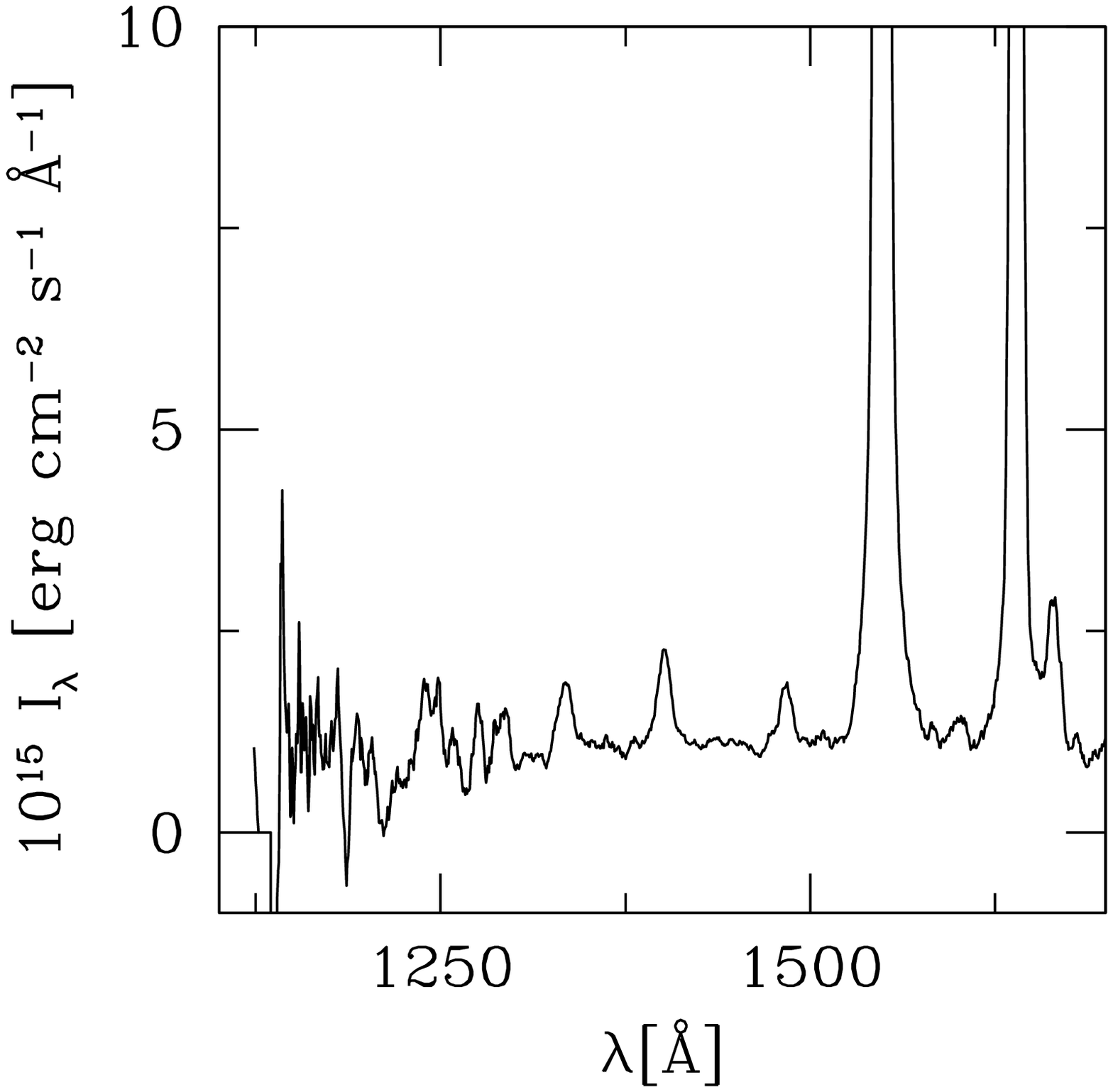}
\newpage
\plotone{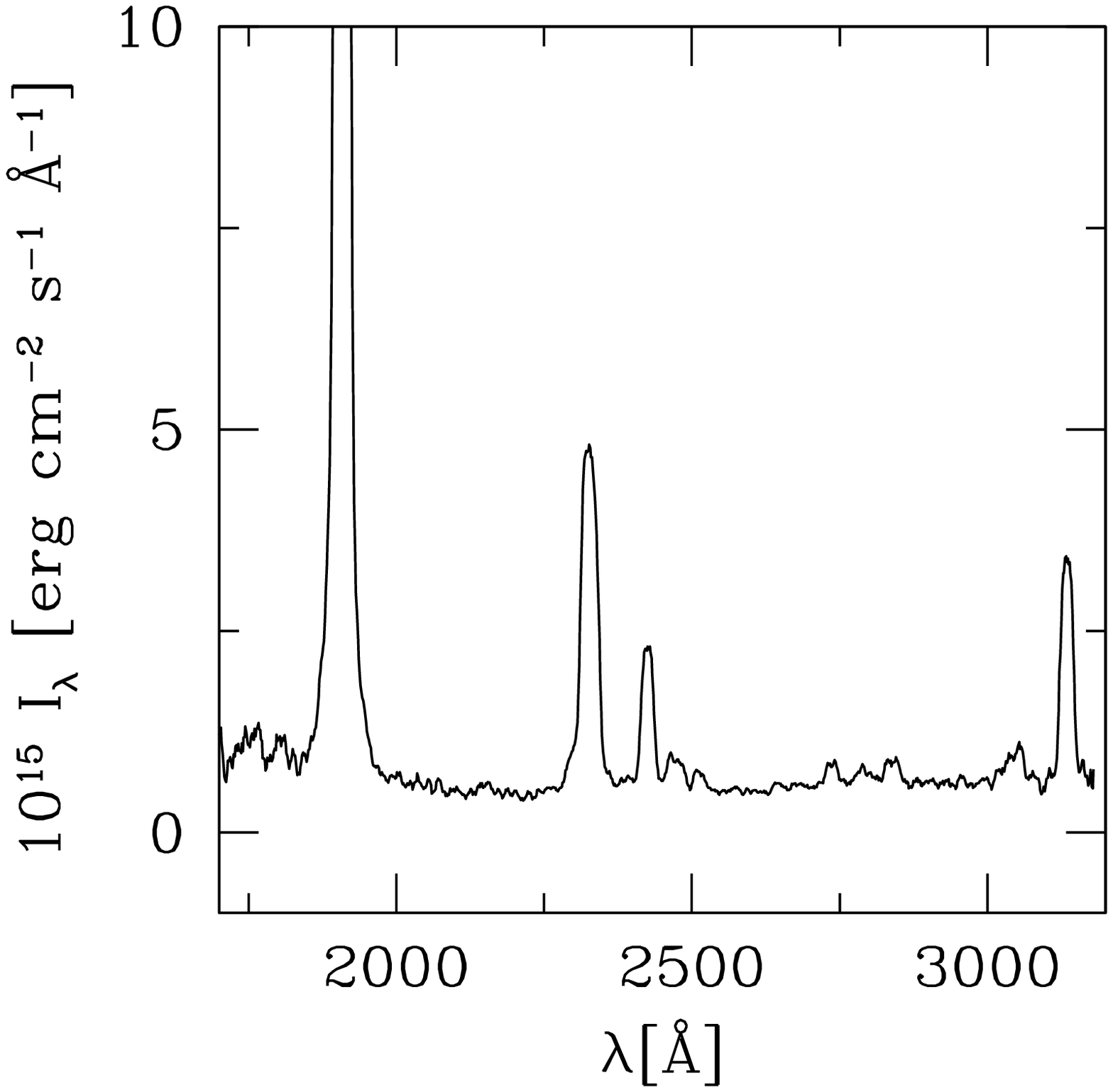}
\newpage
\plotone{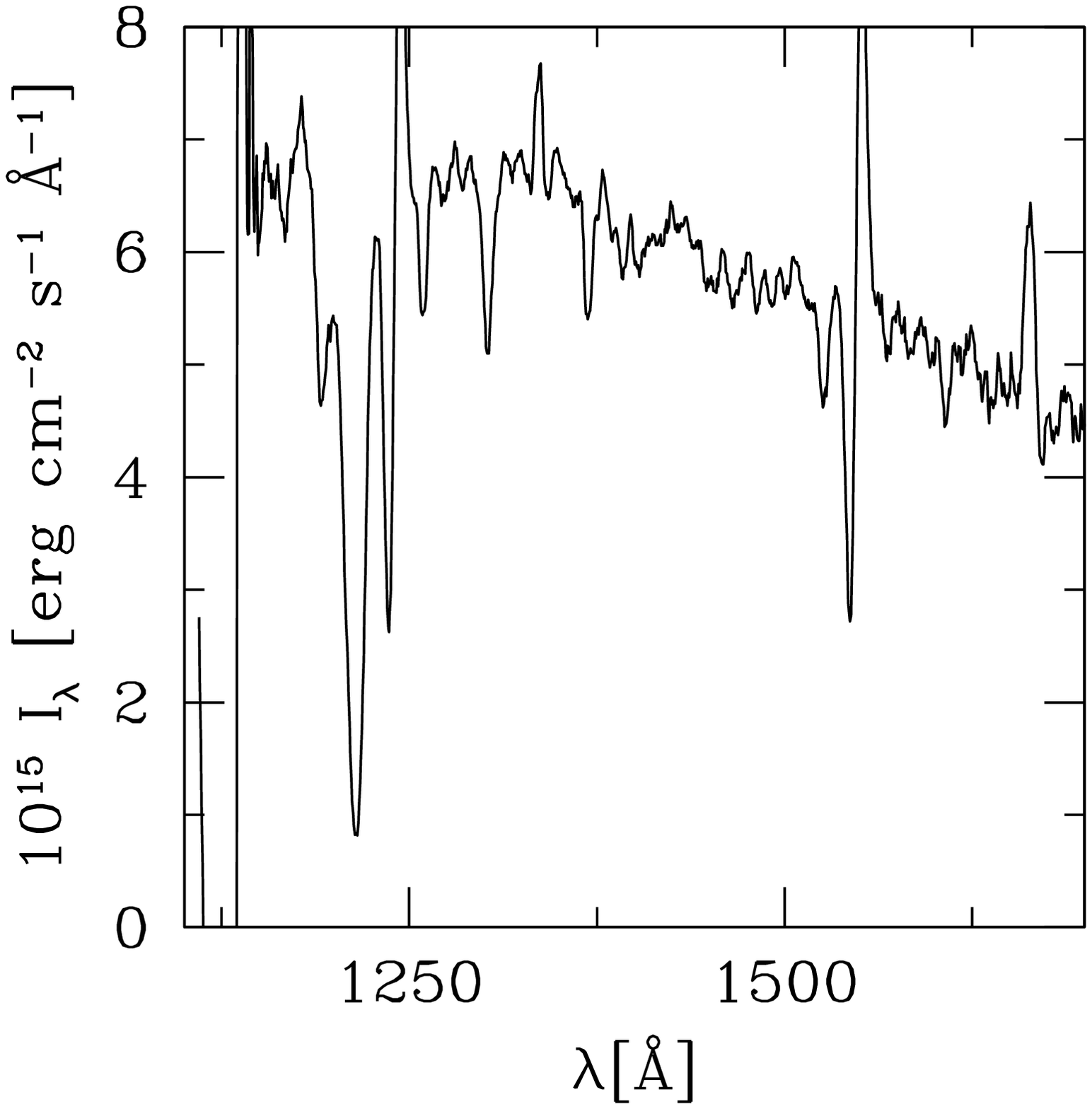}
\newpage
\plotone{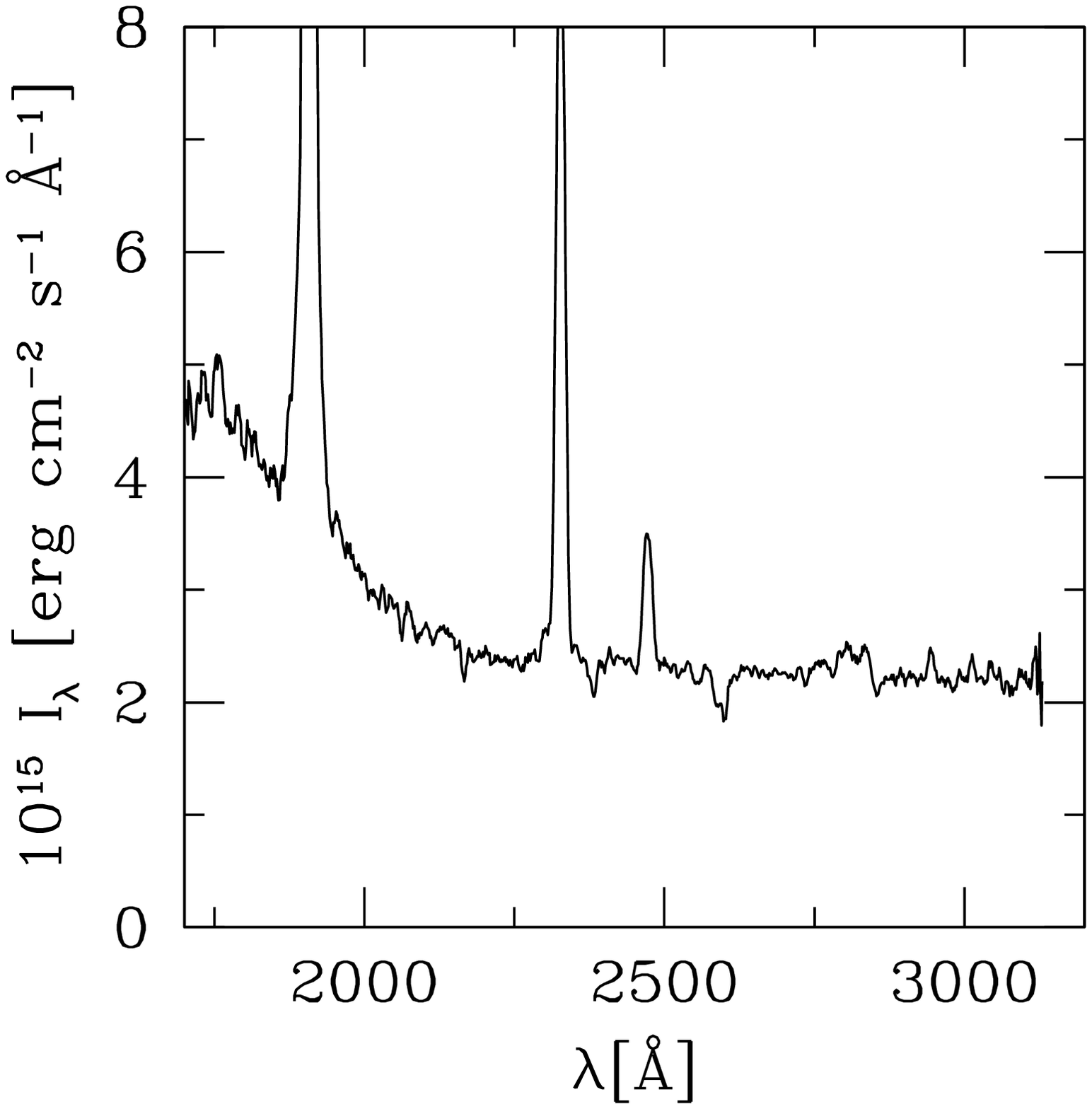}
\newpage
\plotone{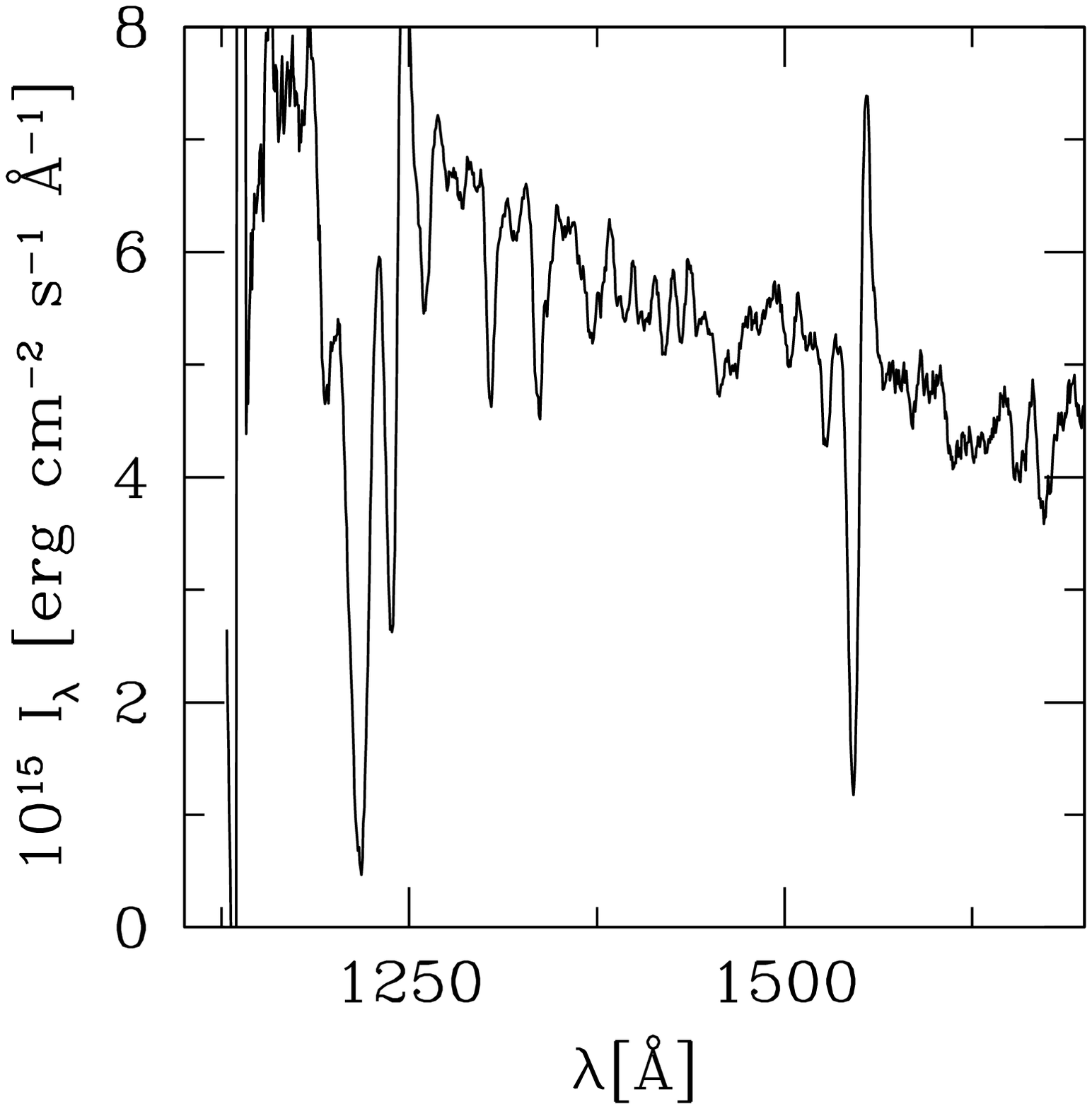}
\newpage
\plotone{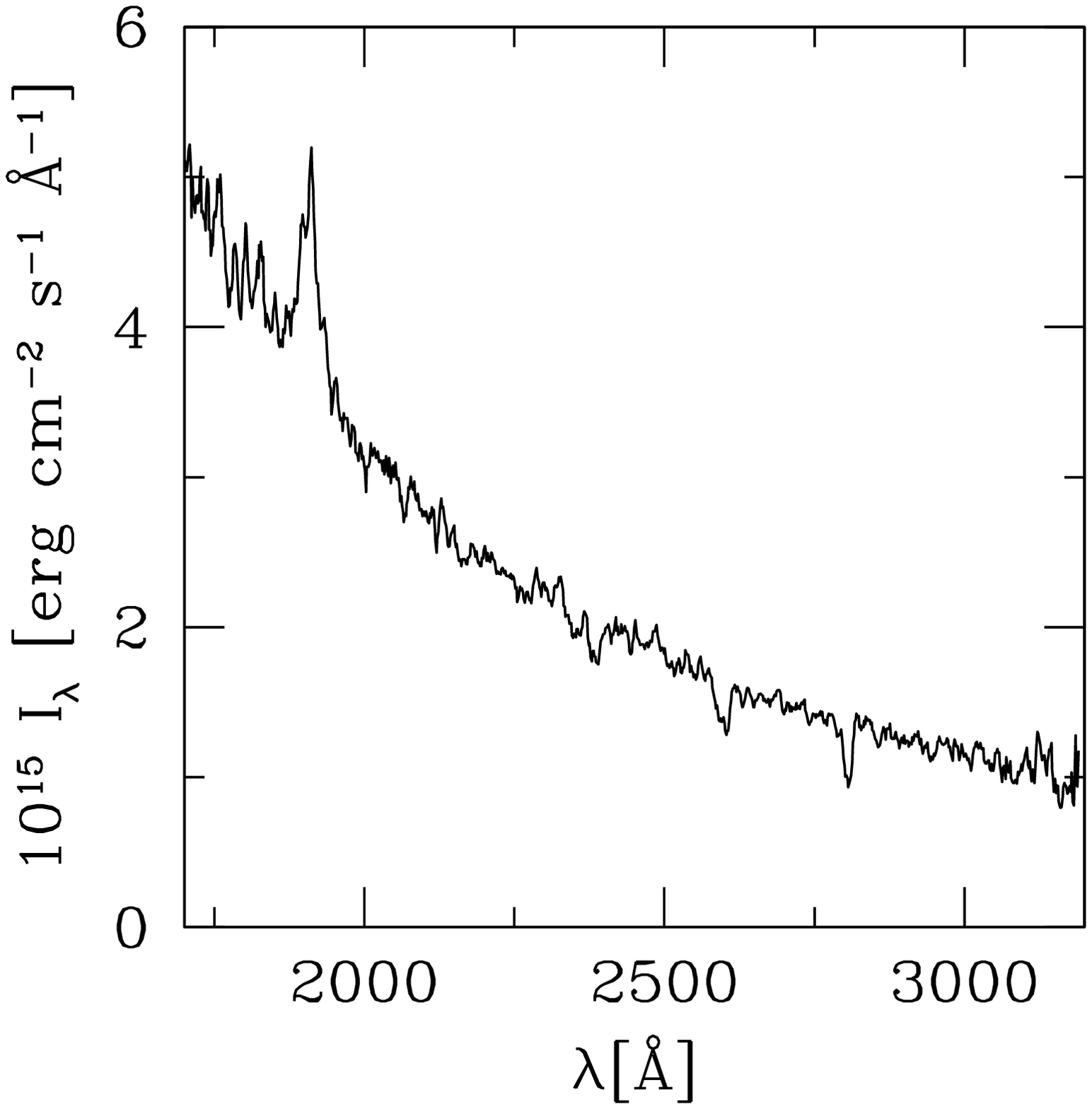}
\newpage
\plotone{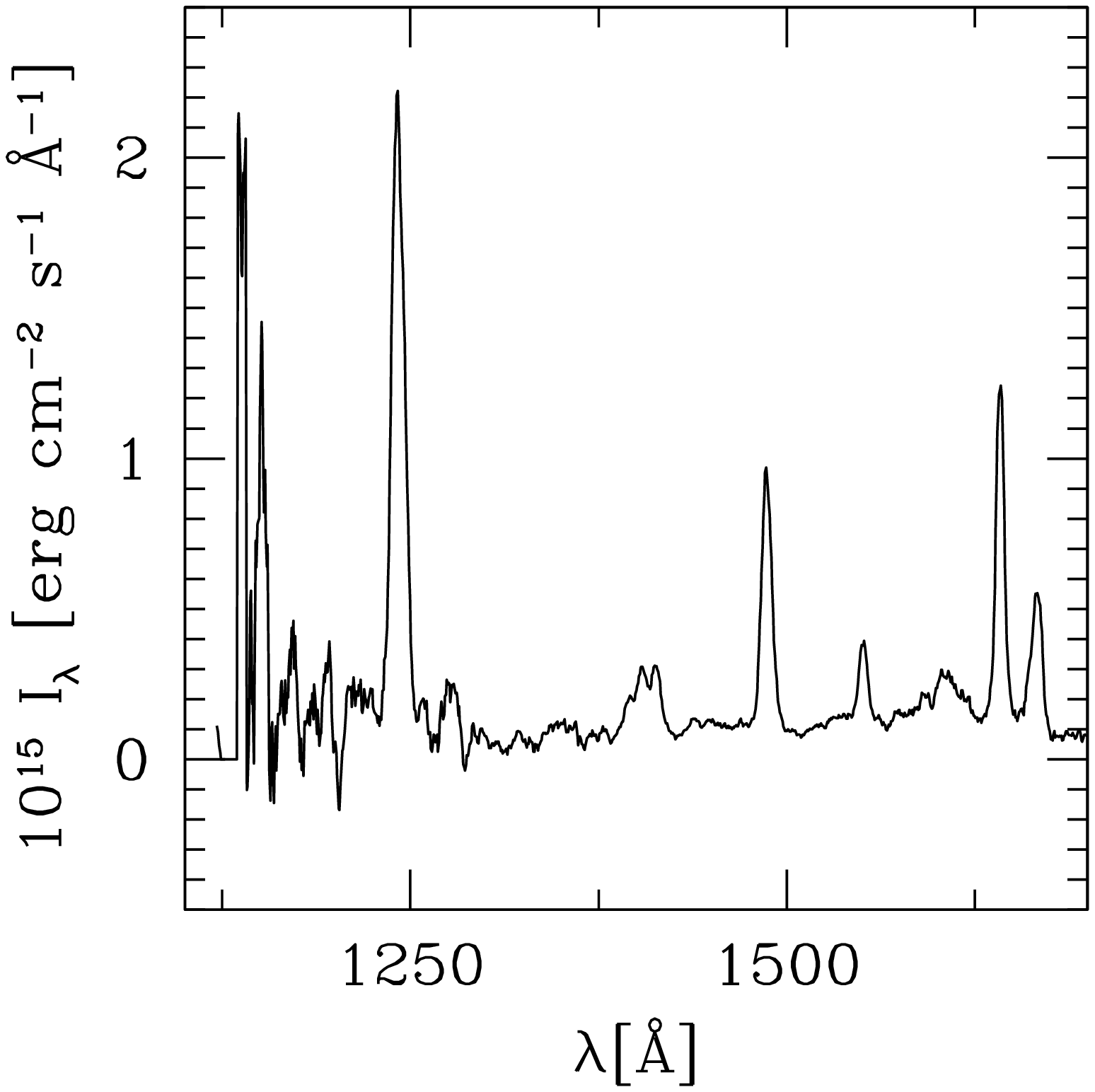}
\newpage
\plotone{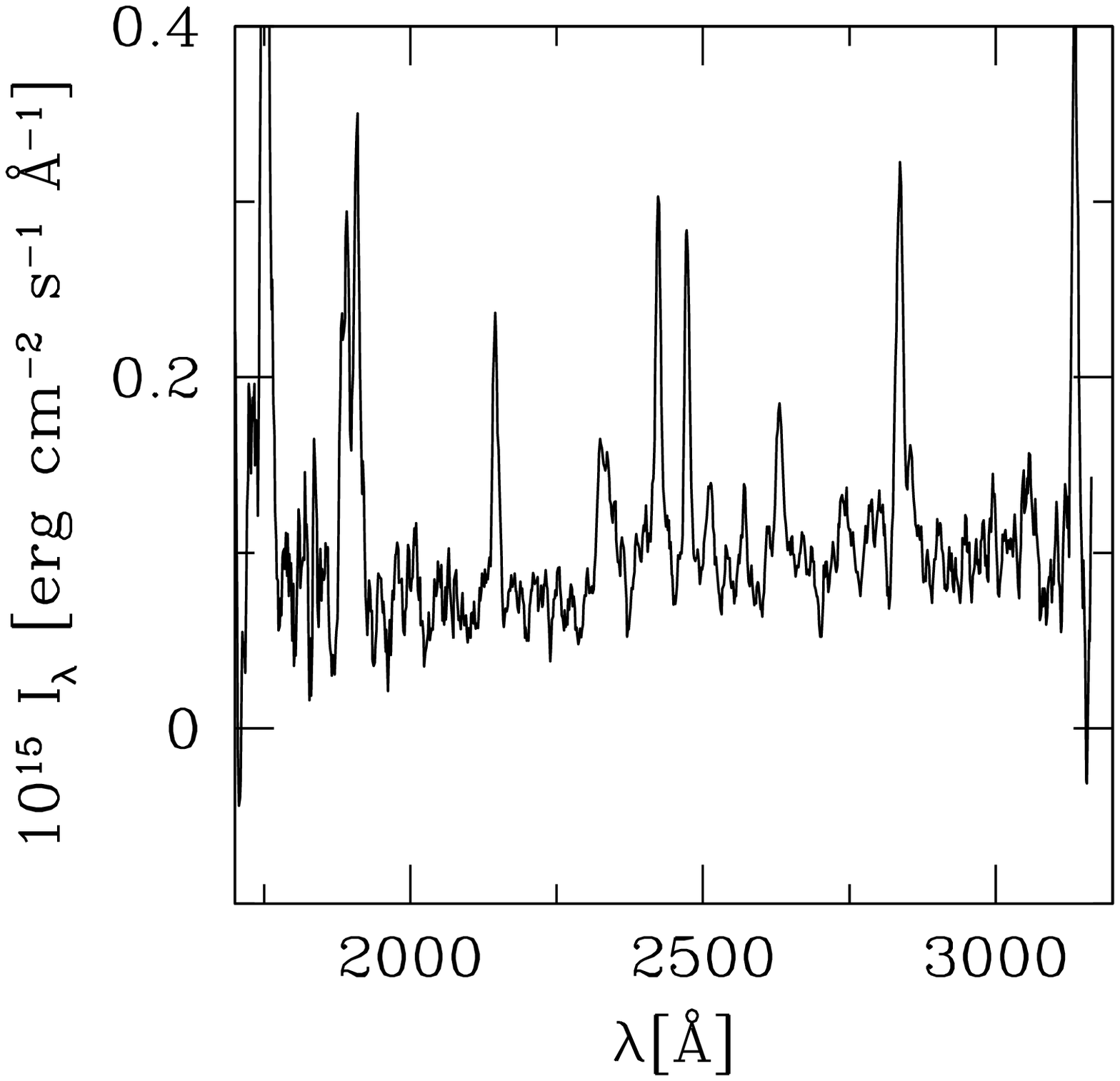}
\newpage
\plotone{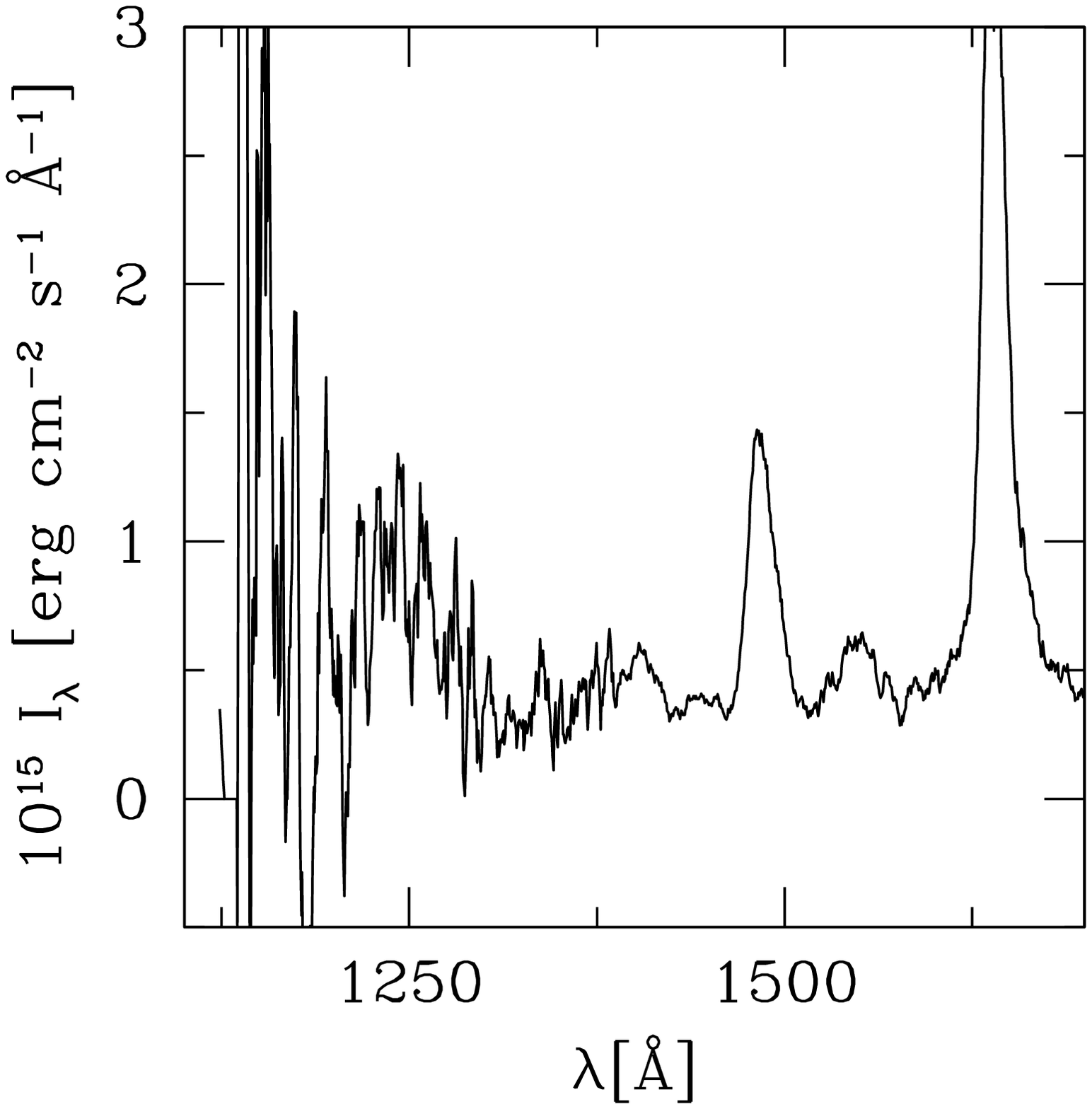}
\newpage
\plotone{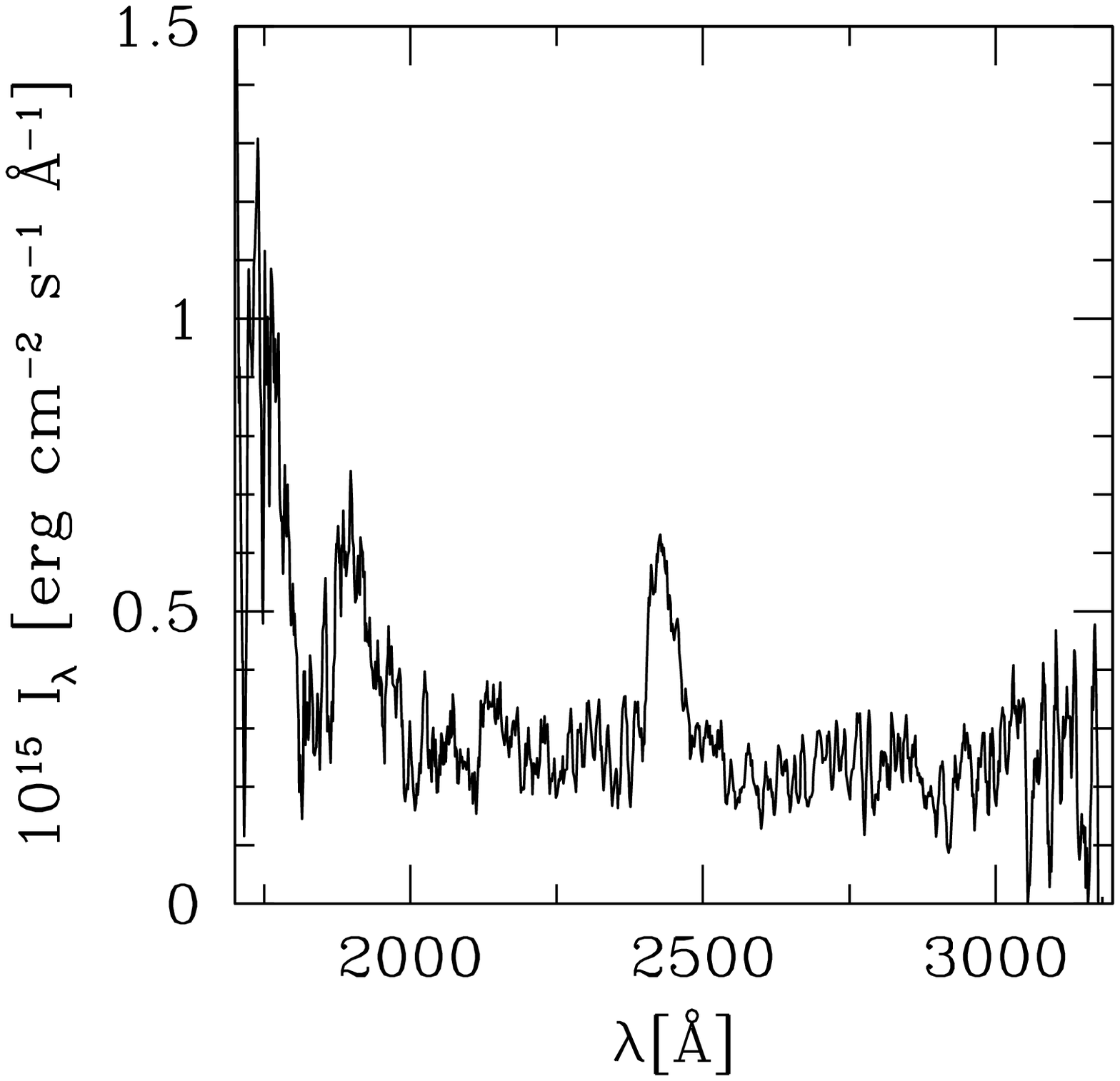}
\newpage
\plotone{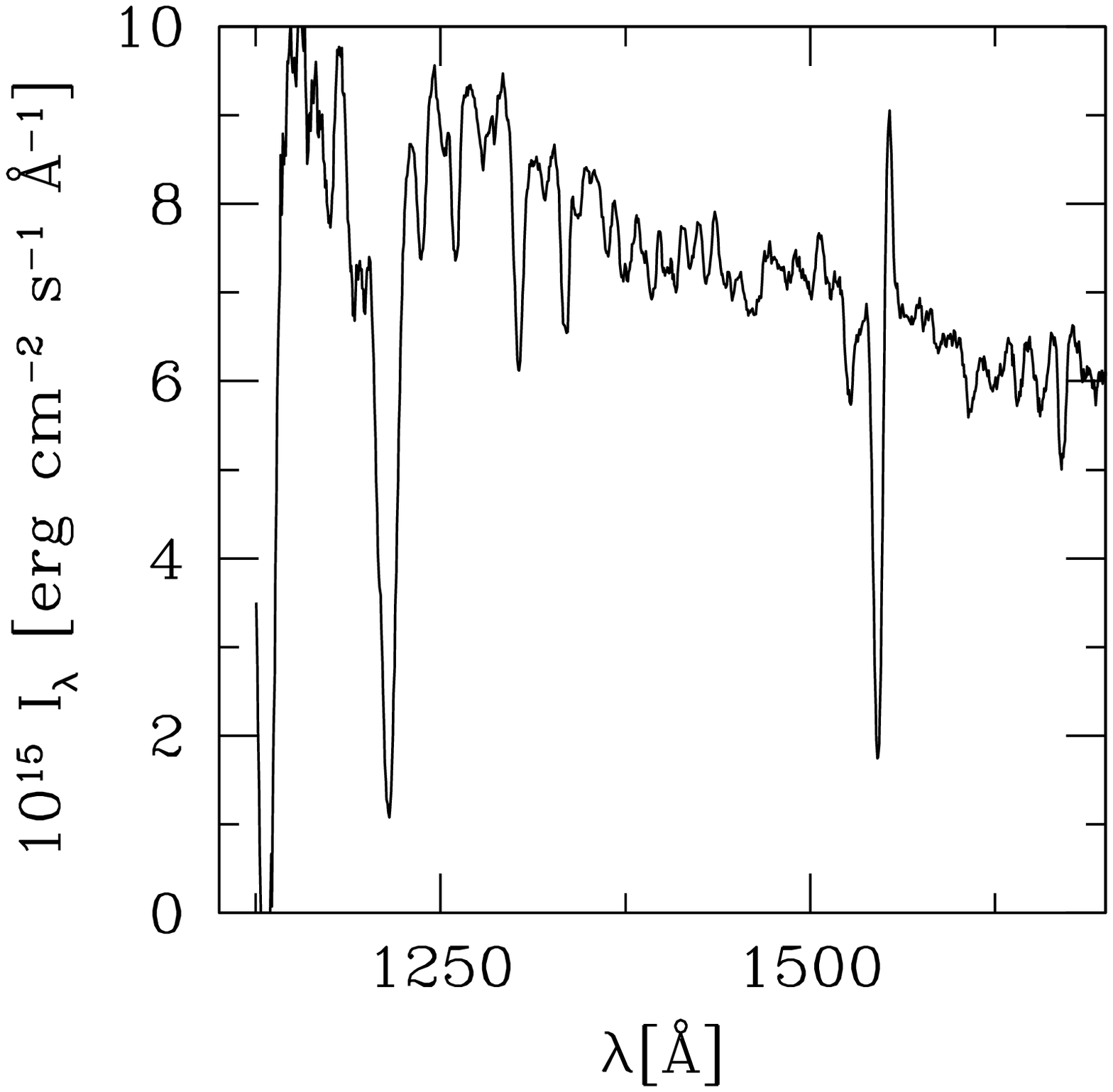}
\newpage
\plotone{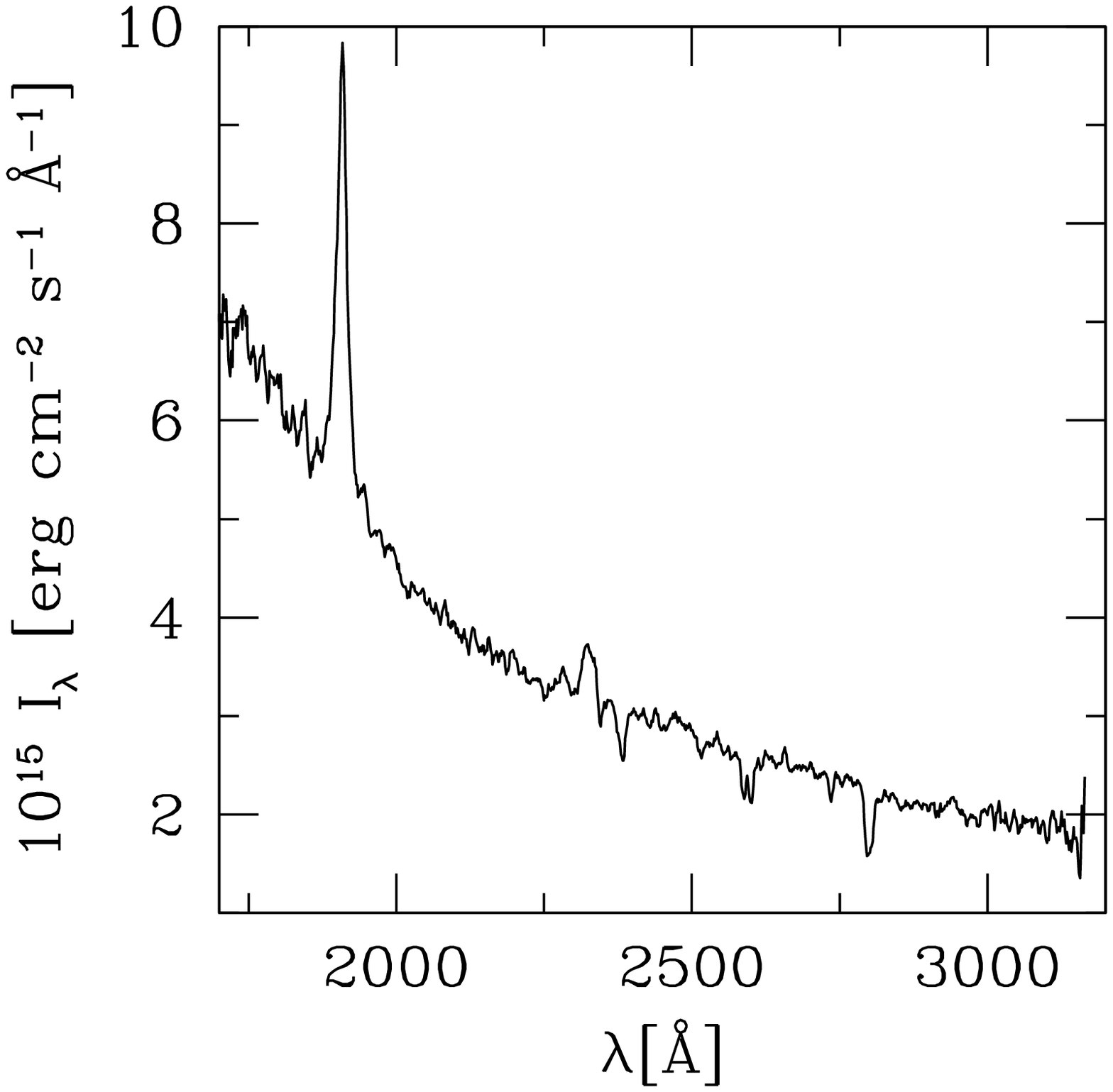}
\newpage
\plotone{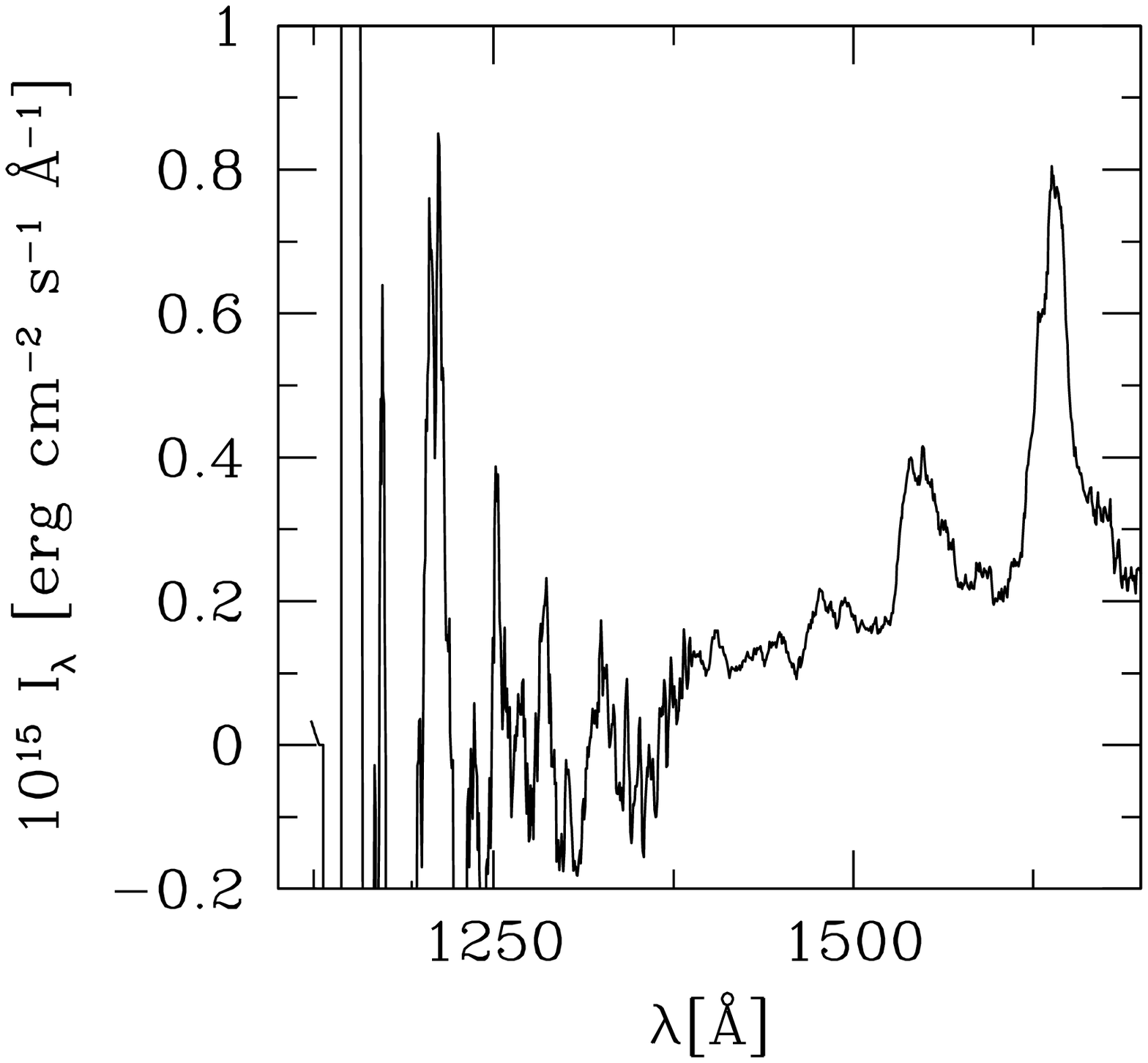}
\newpage
\plotone{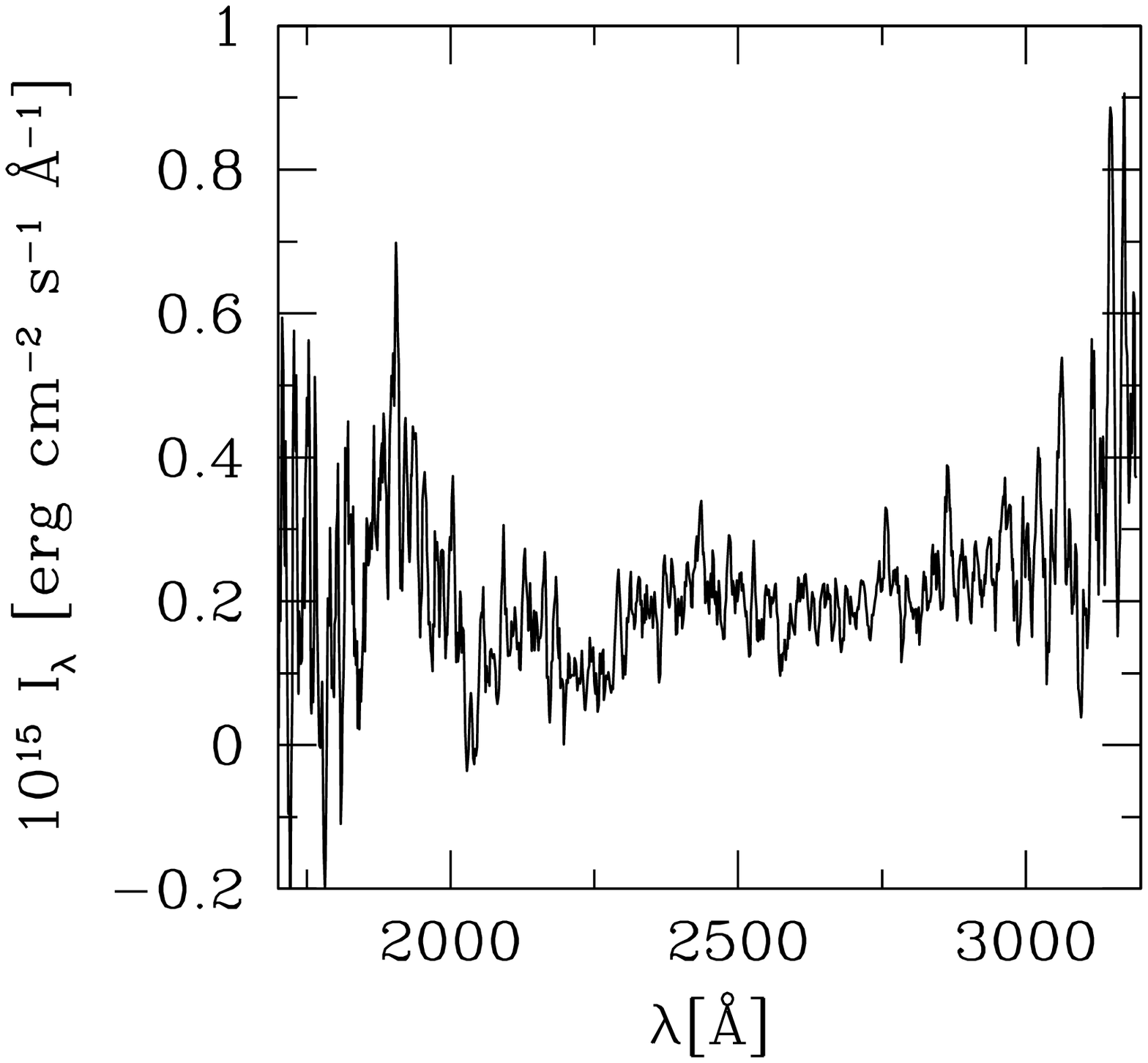}
\newpage
\plotone{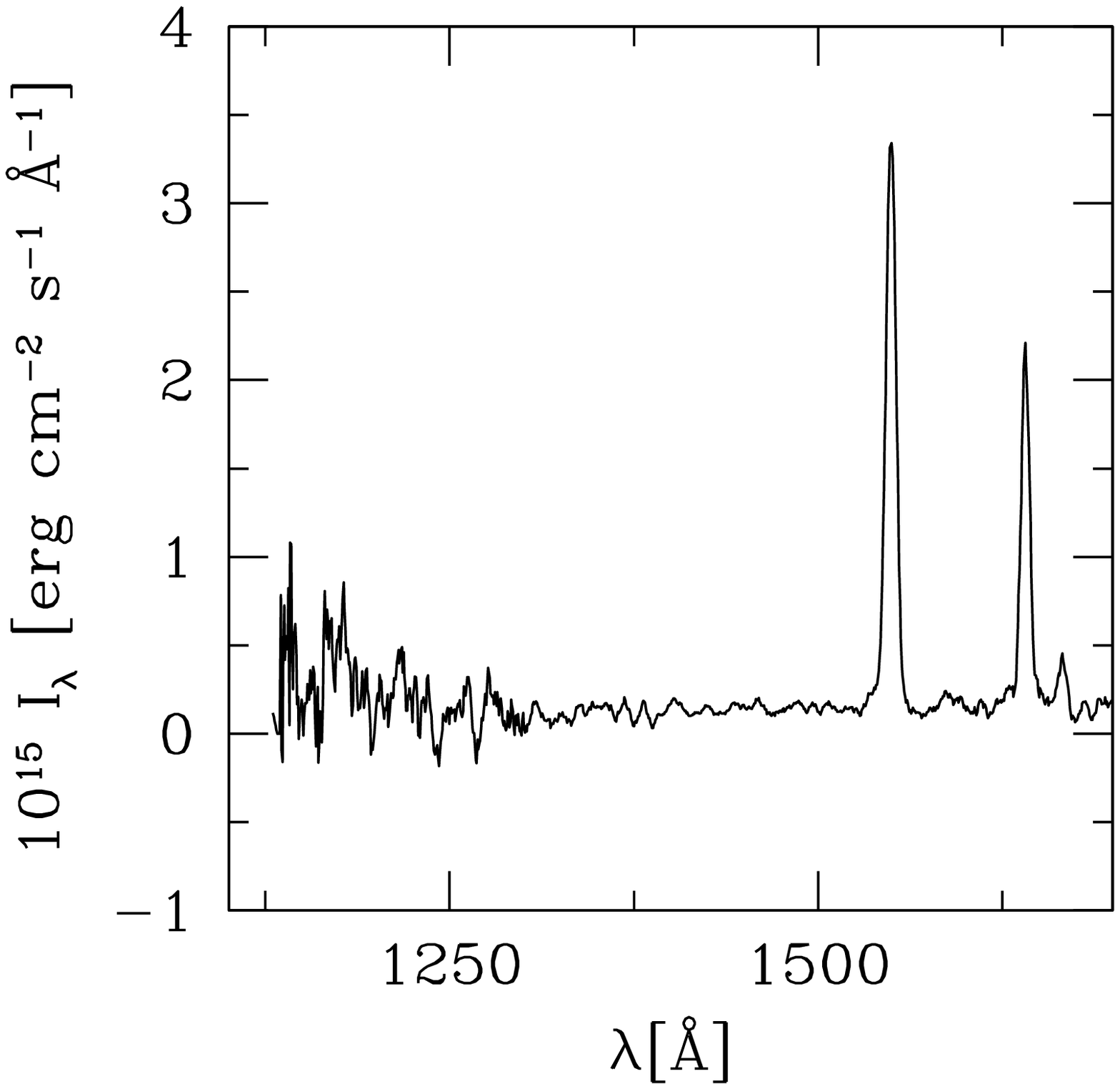}
\newpage
\plotone{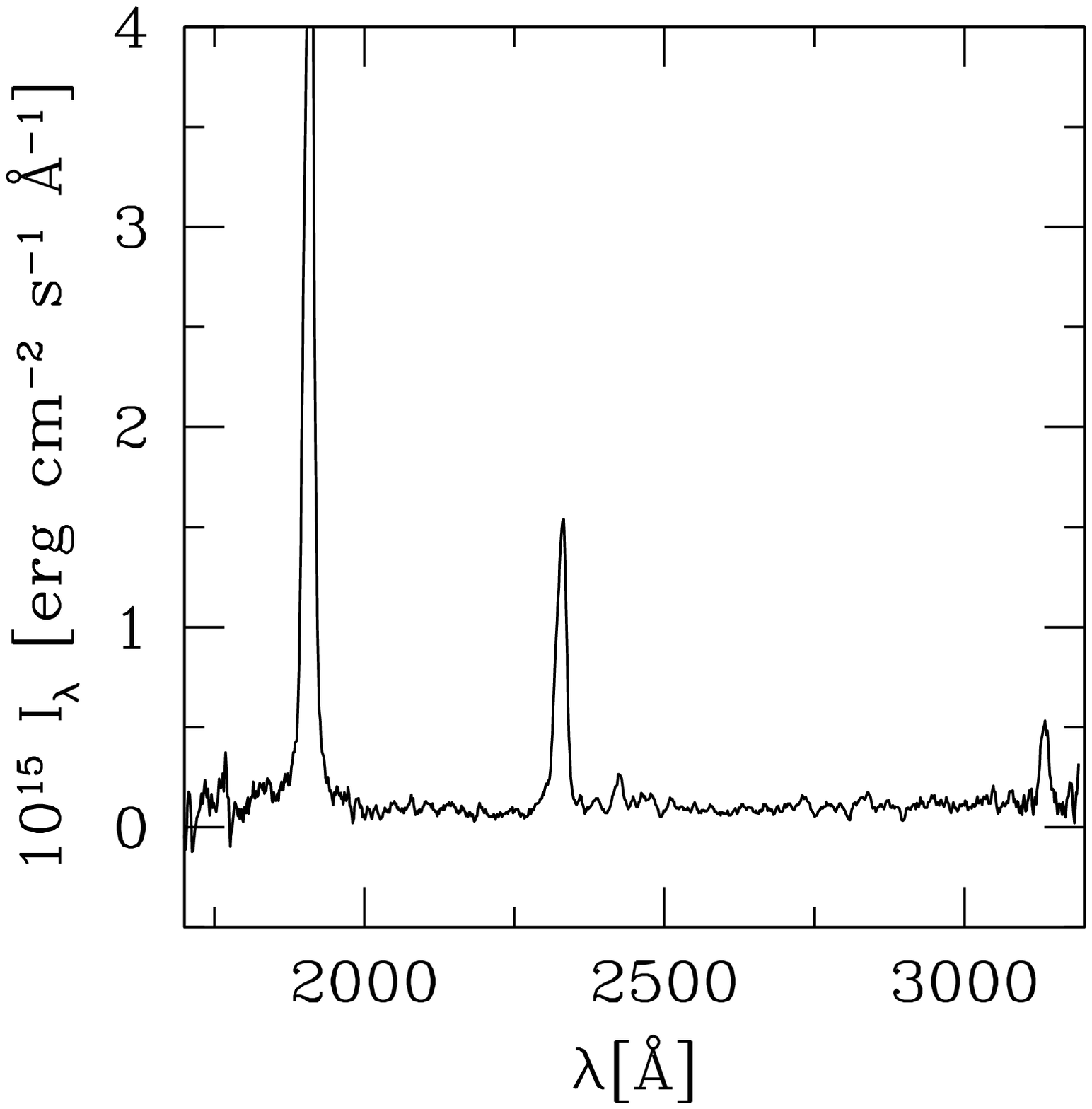}
\newpage
\plotone{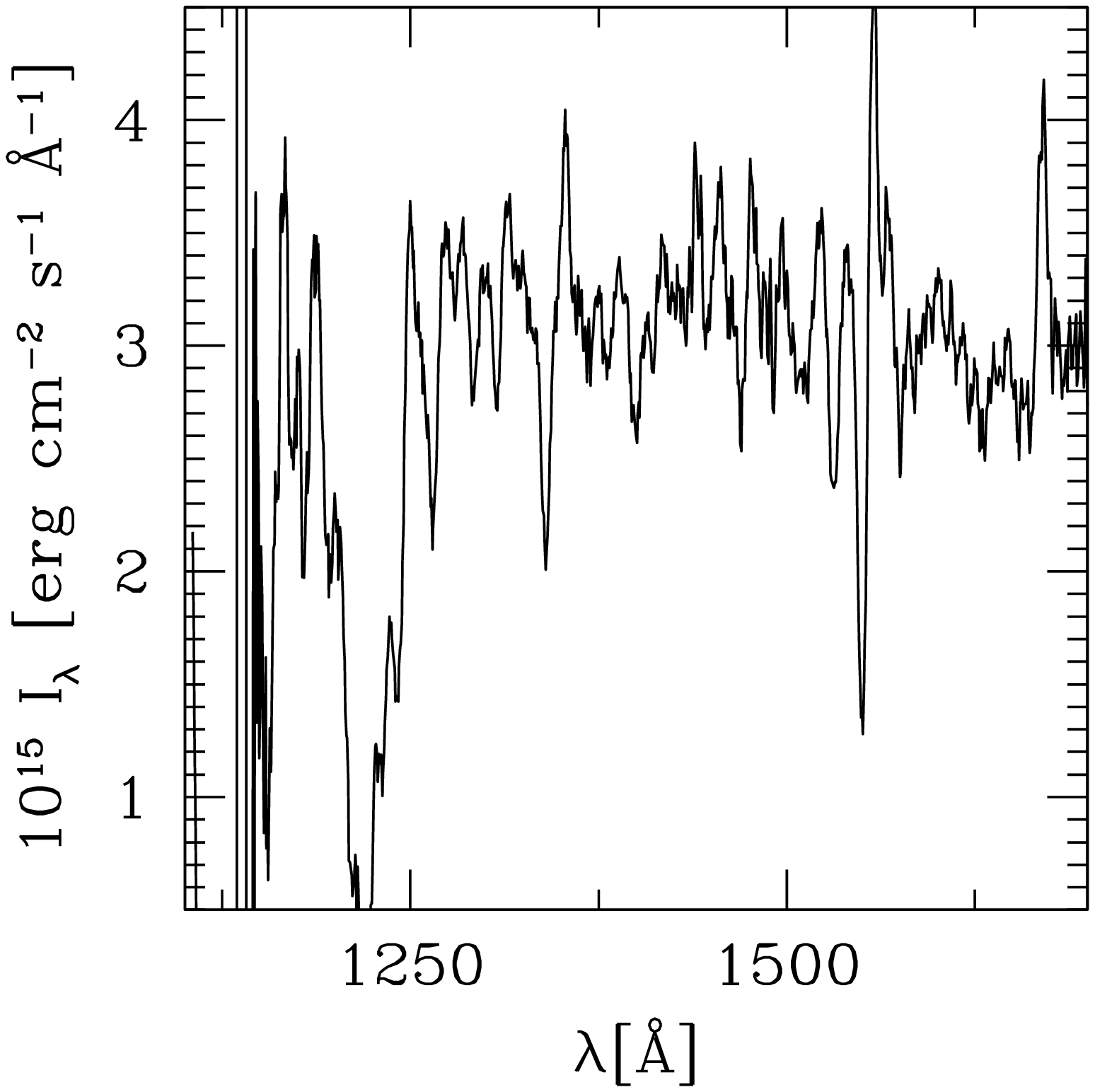}
\newpage
\plotone{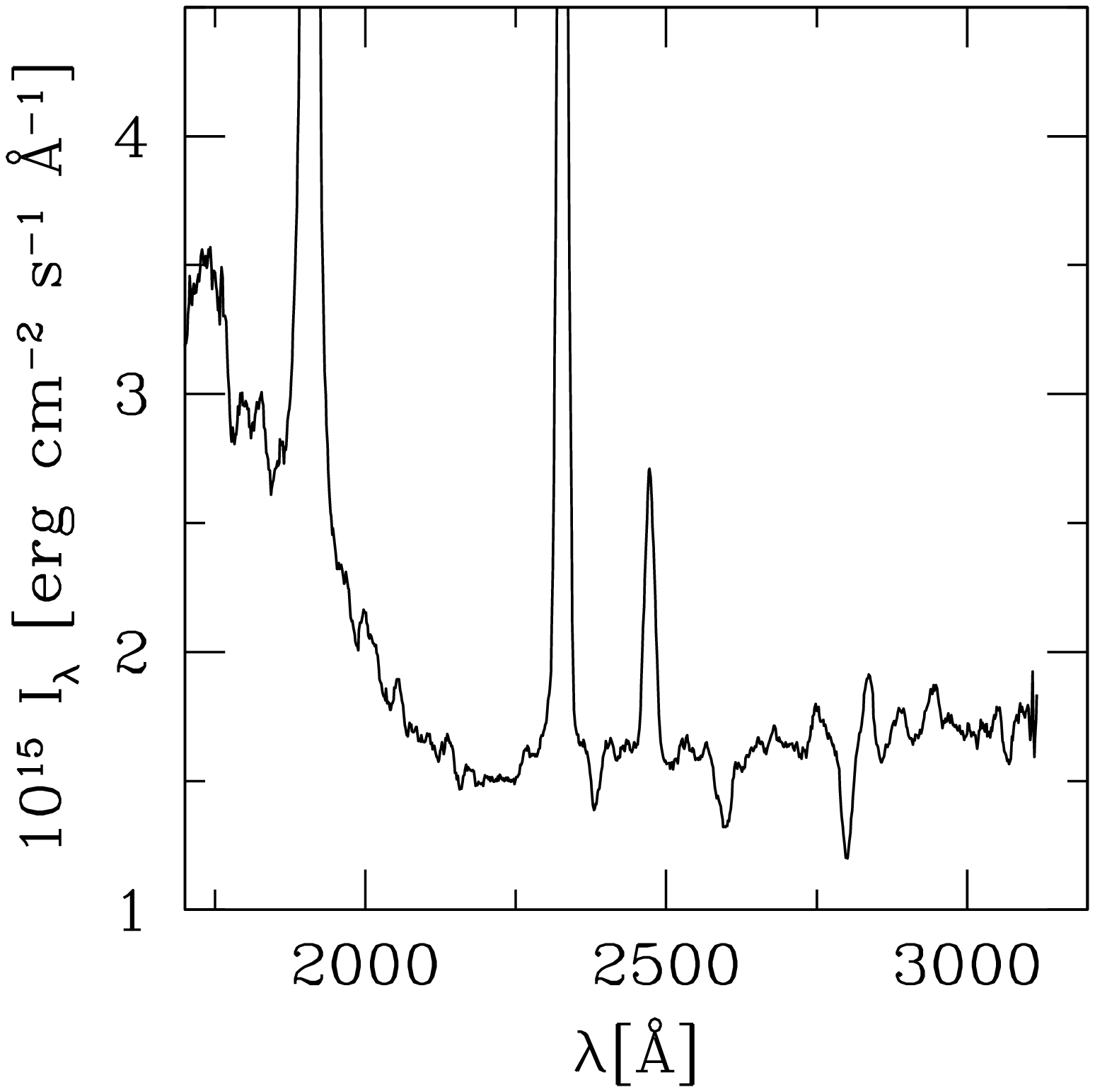}
\newpage
\plotone{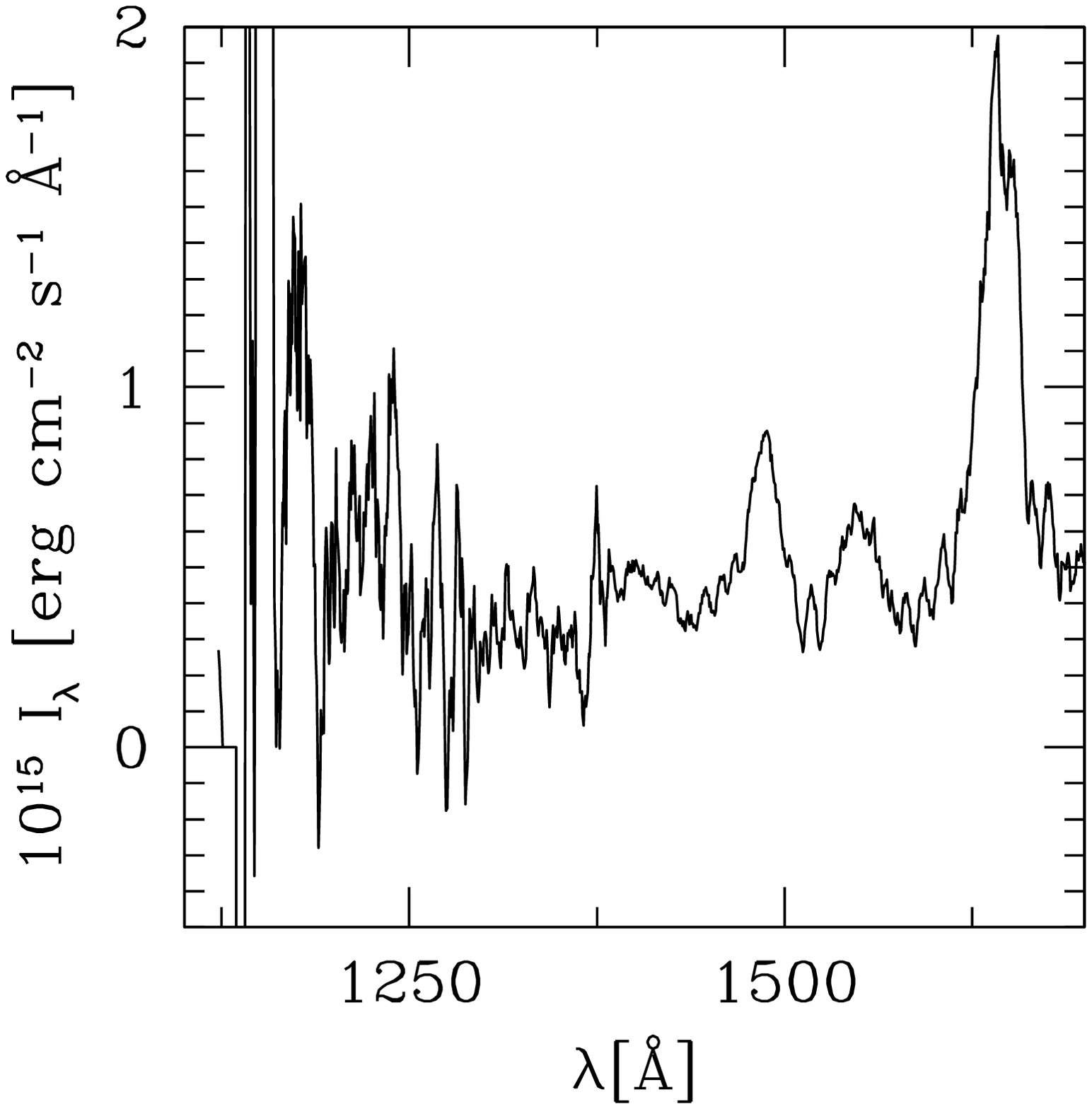}
\newpage
\plotone{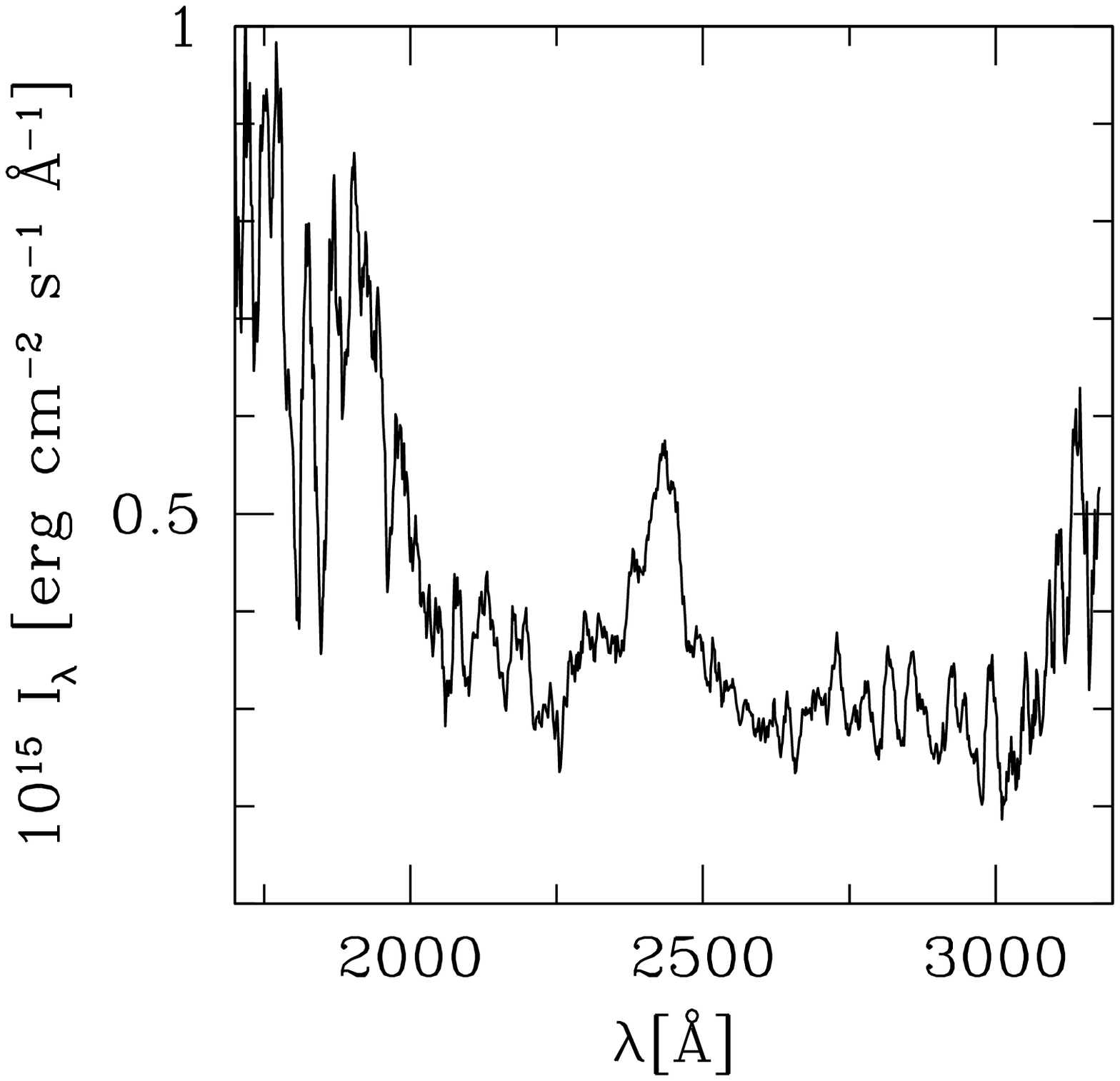}
\newpage
\plotone{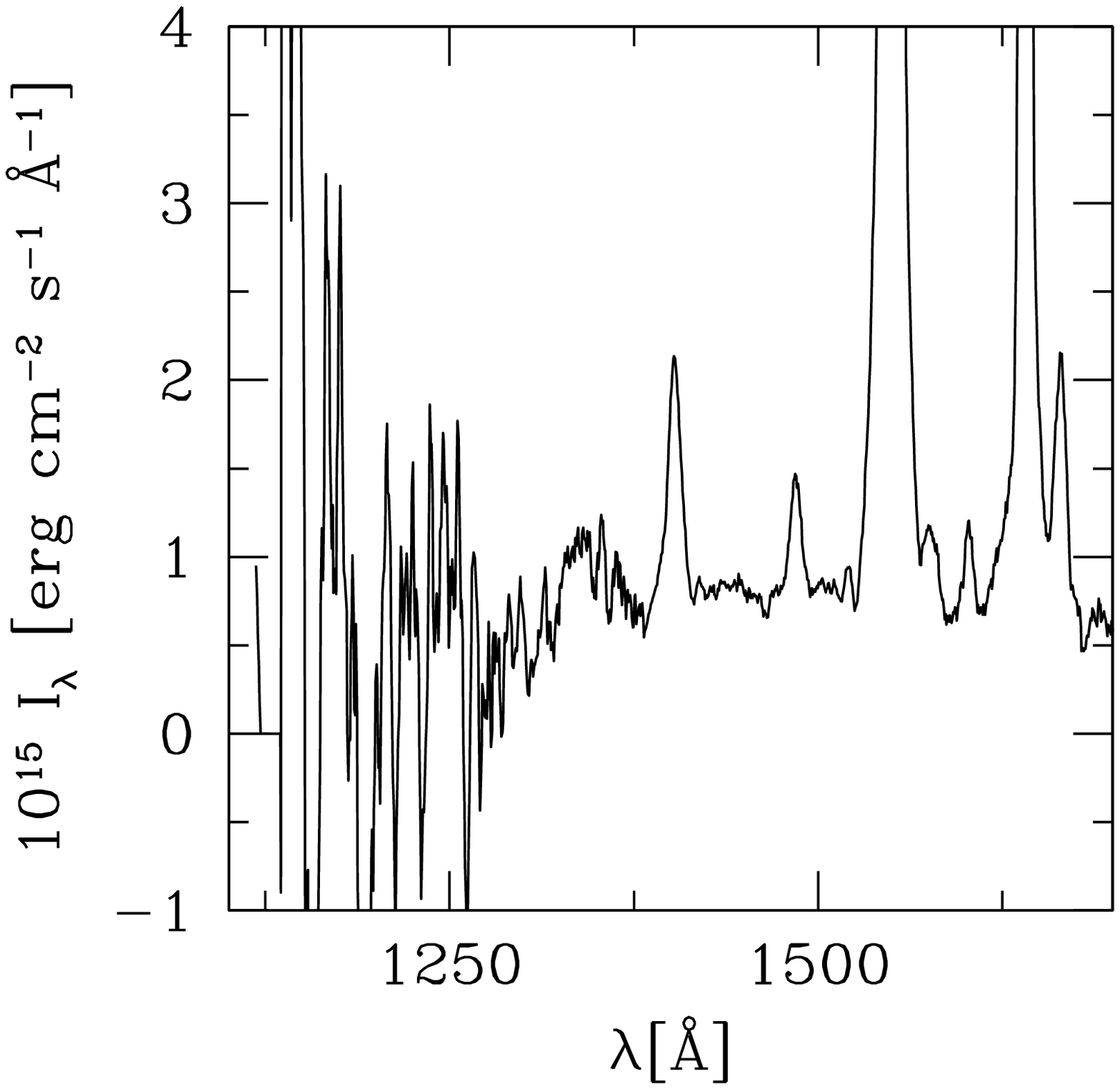}
\newpage
\plotone{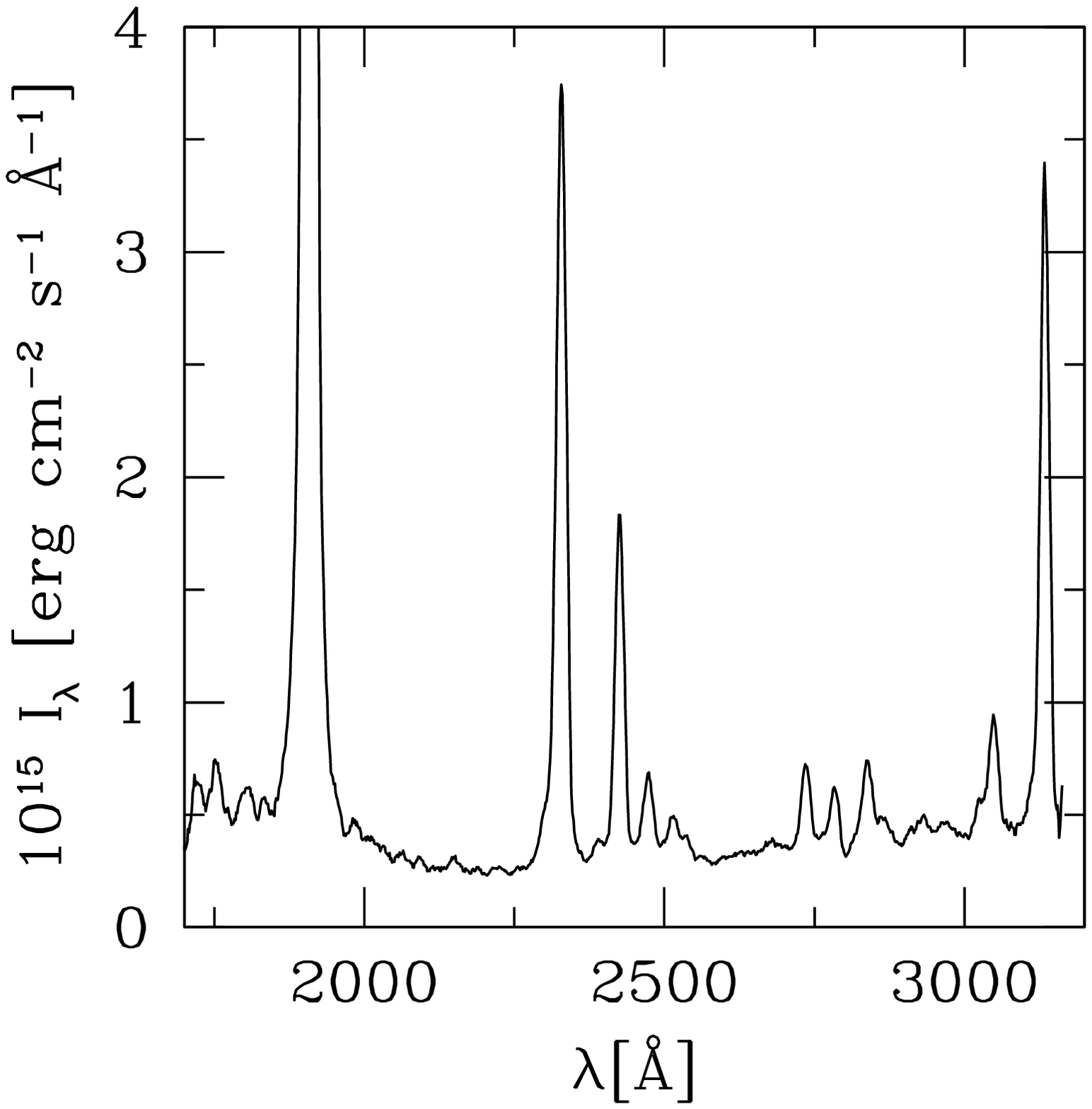}
\newpage
\plotone{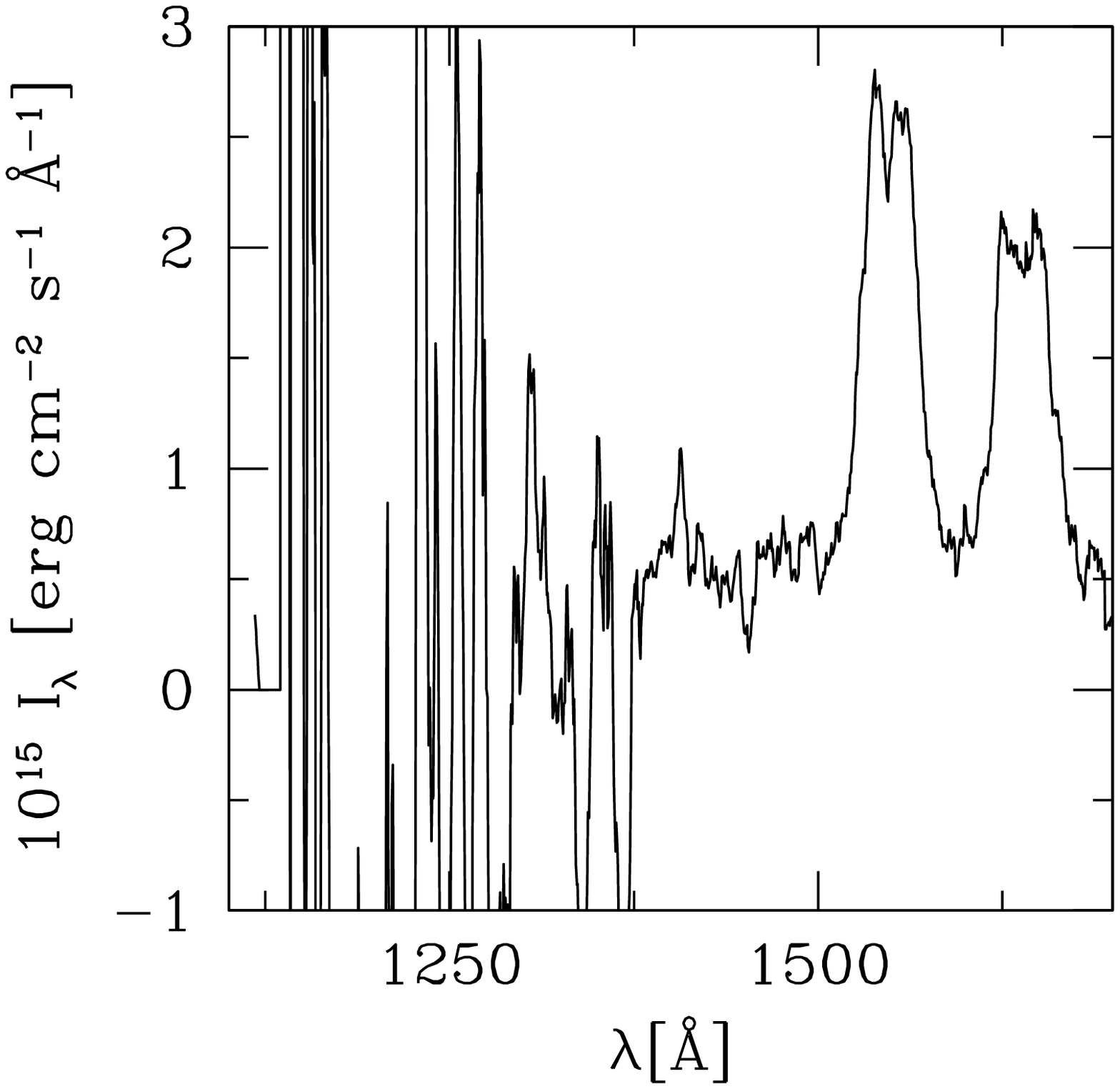}
\newpage
\plotone{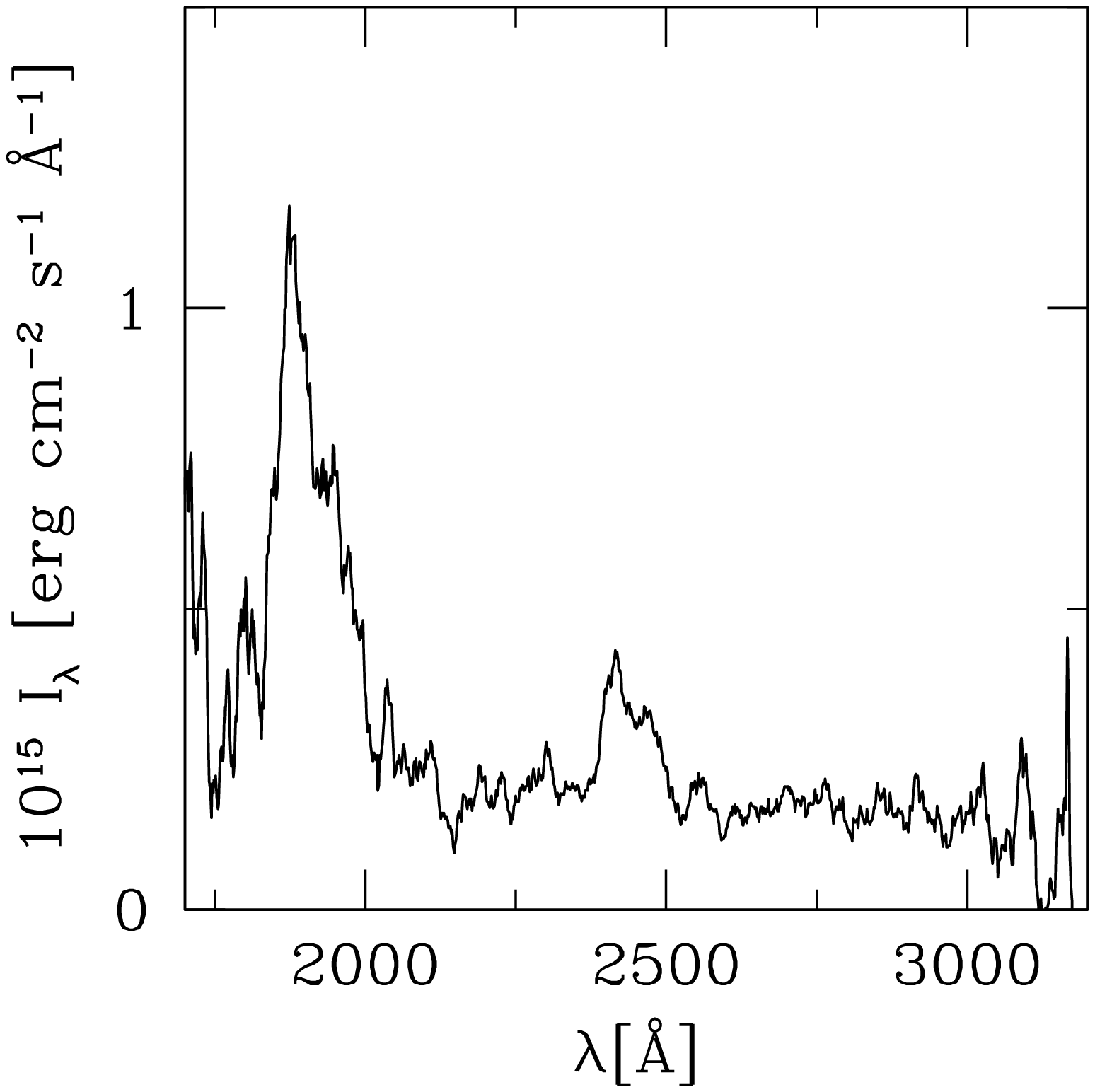}
\newpage
\plotone{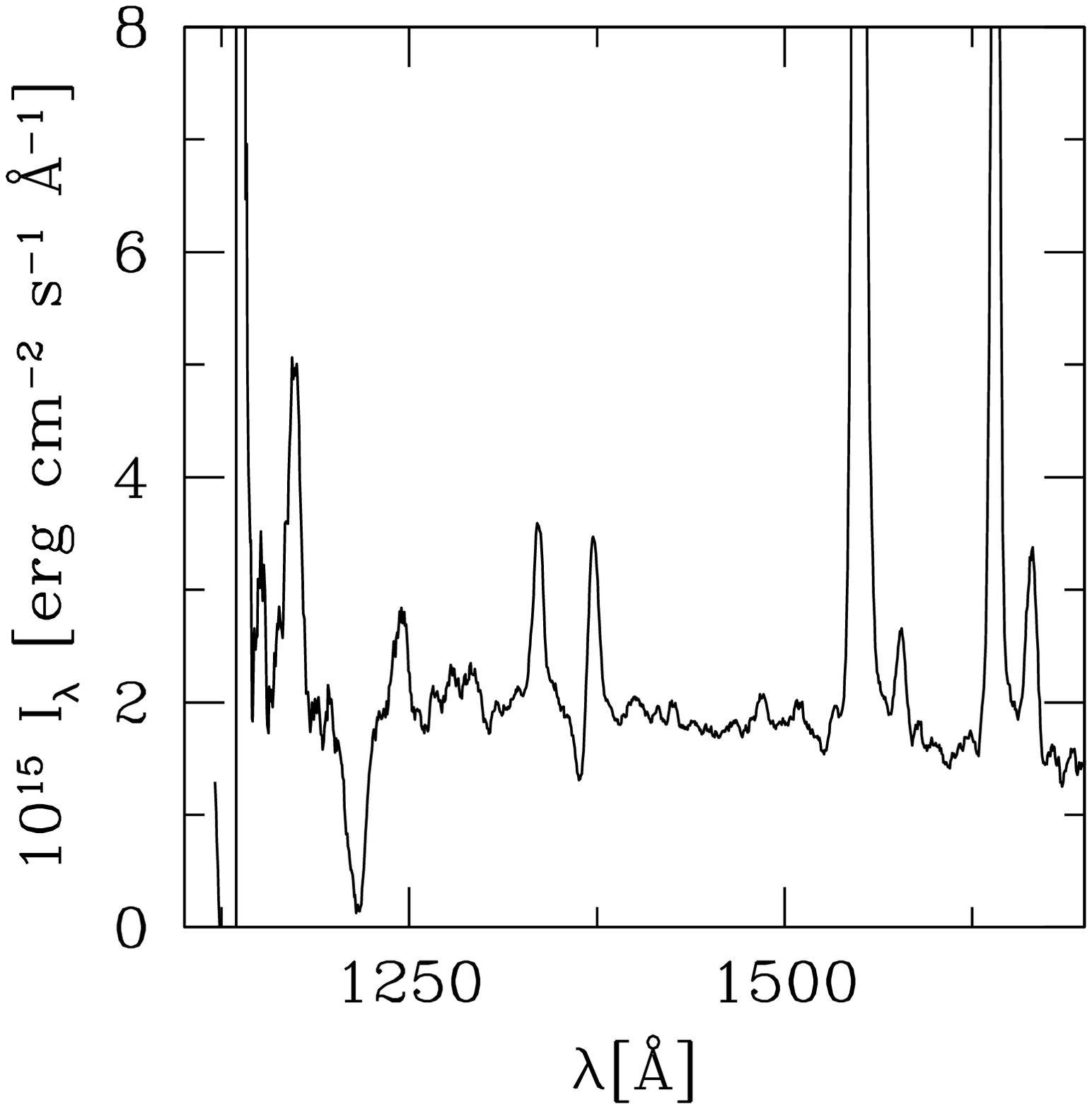}
\newpage
\plotone{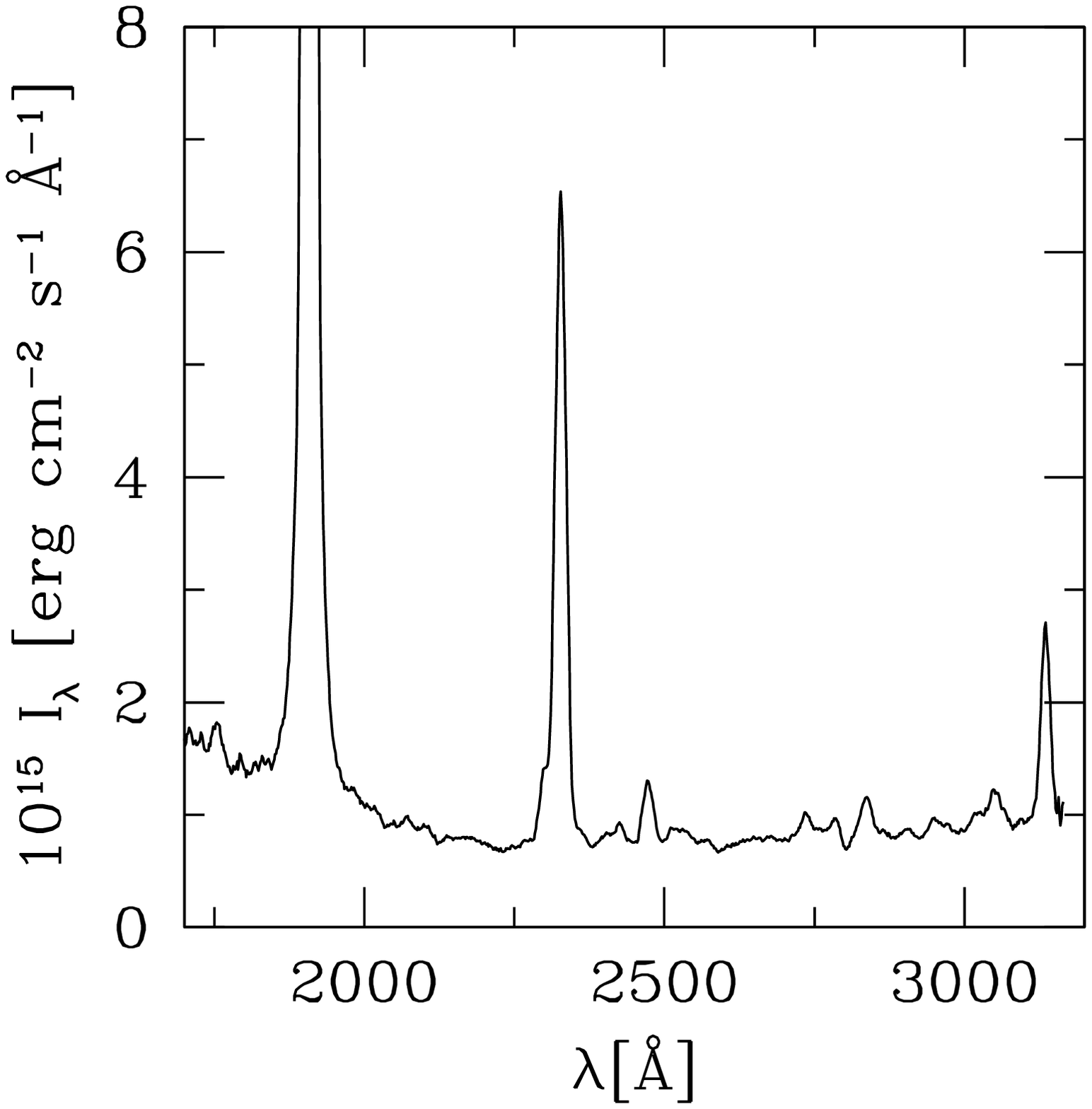}
\newpage
\plotone{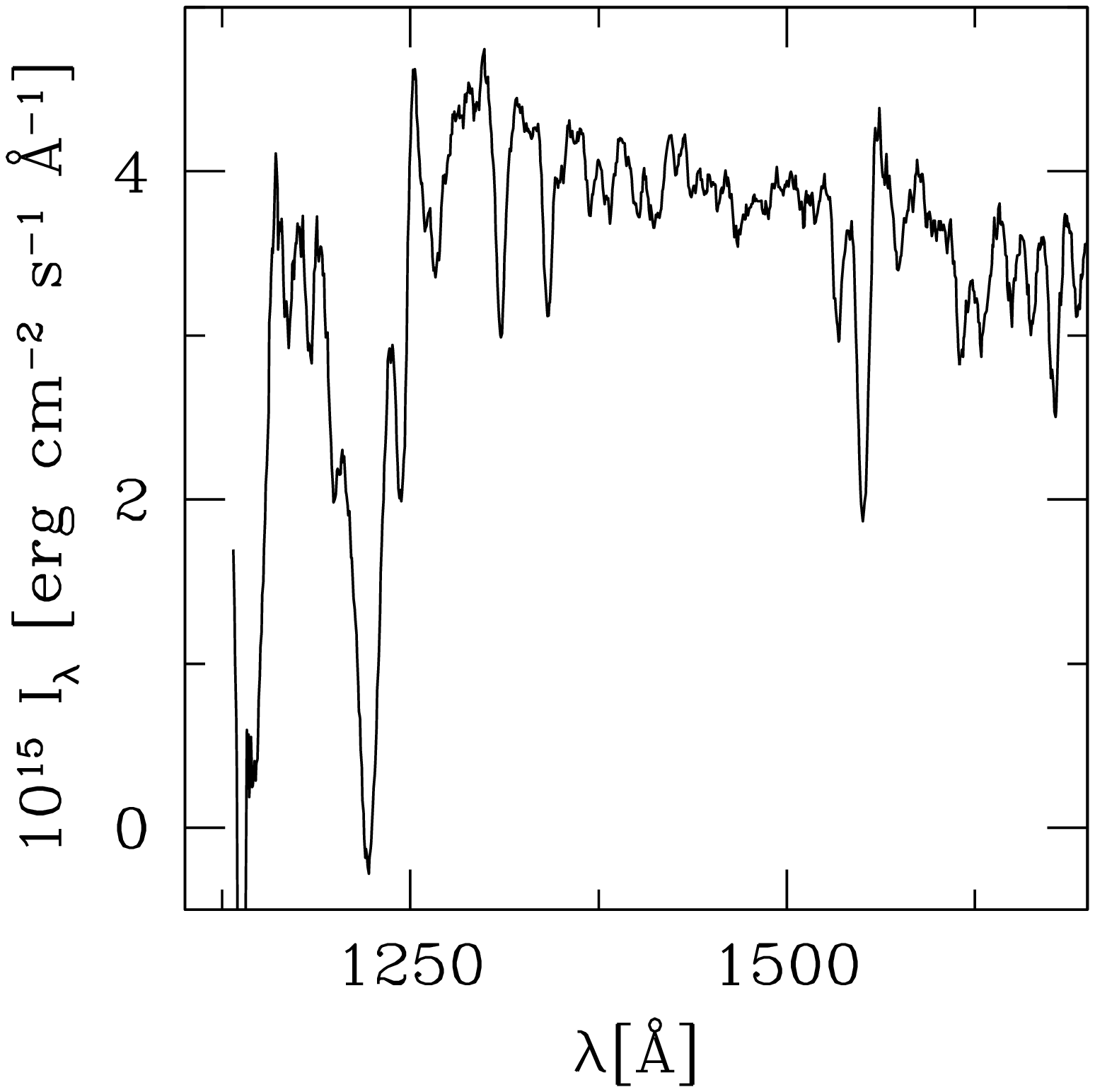}
\newpage
\plotone{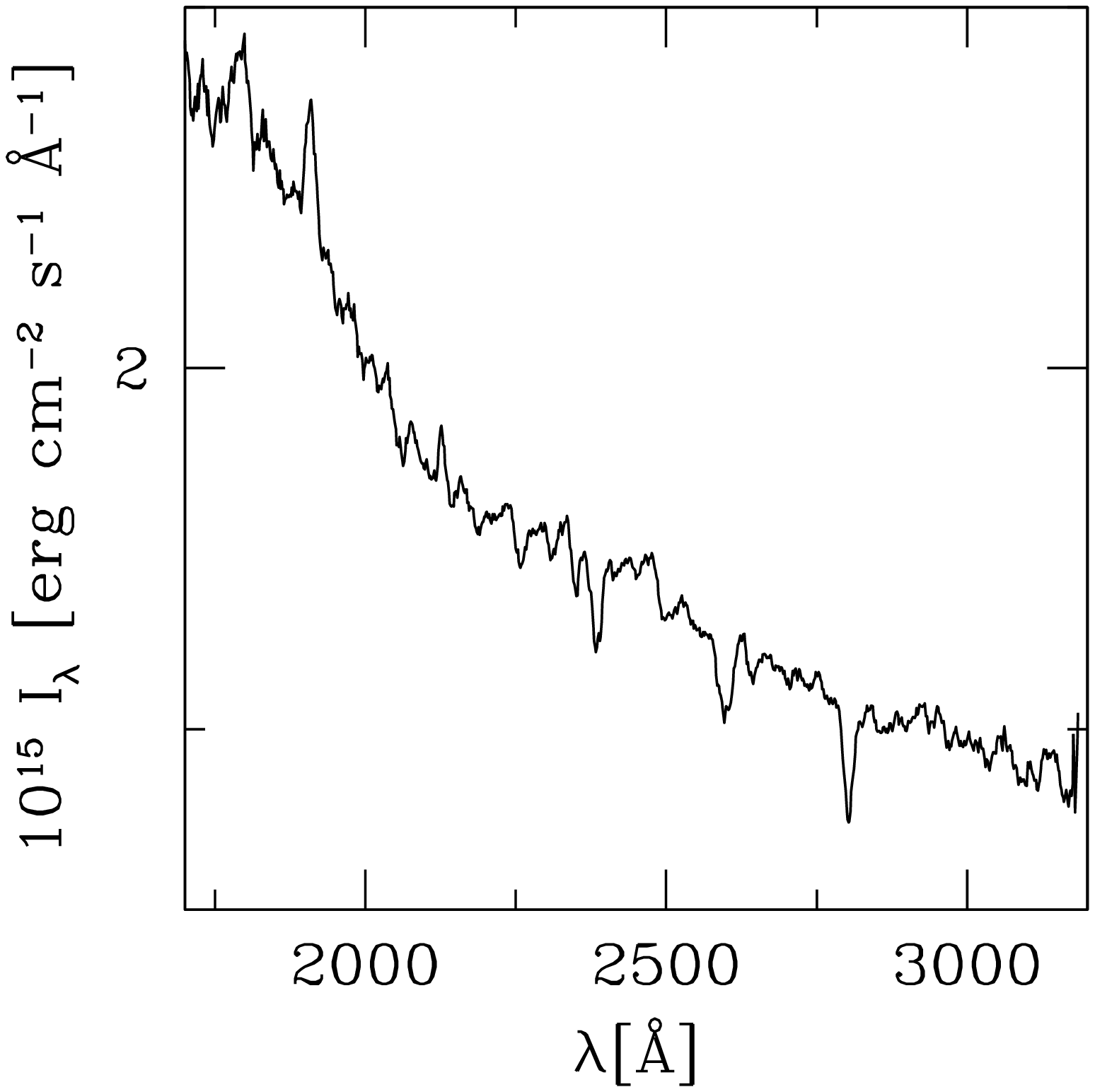}
\newpage
\plotone{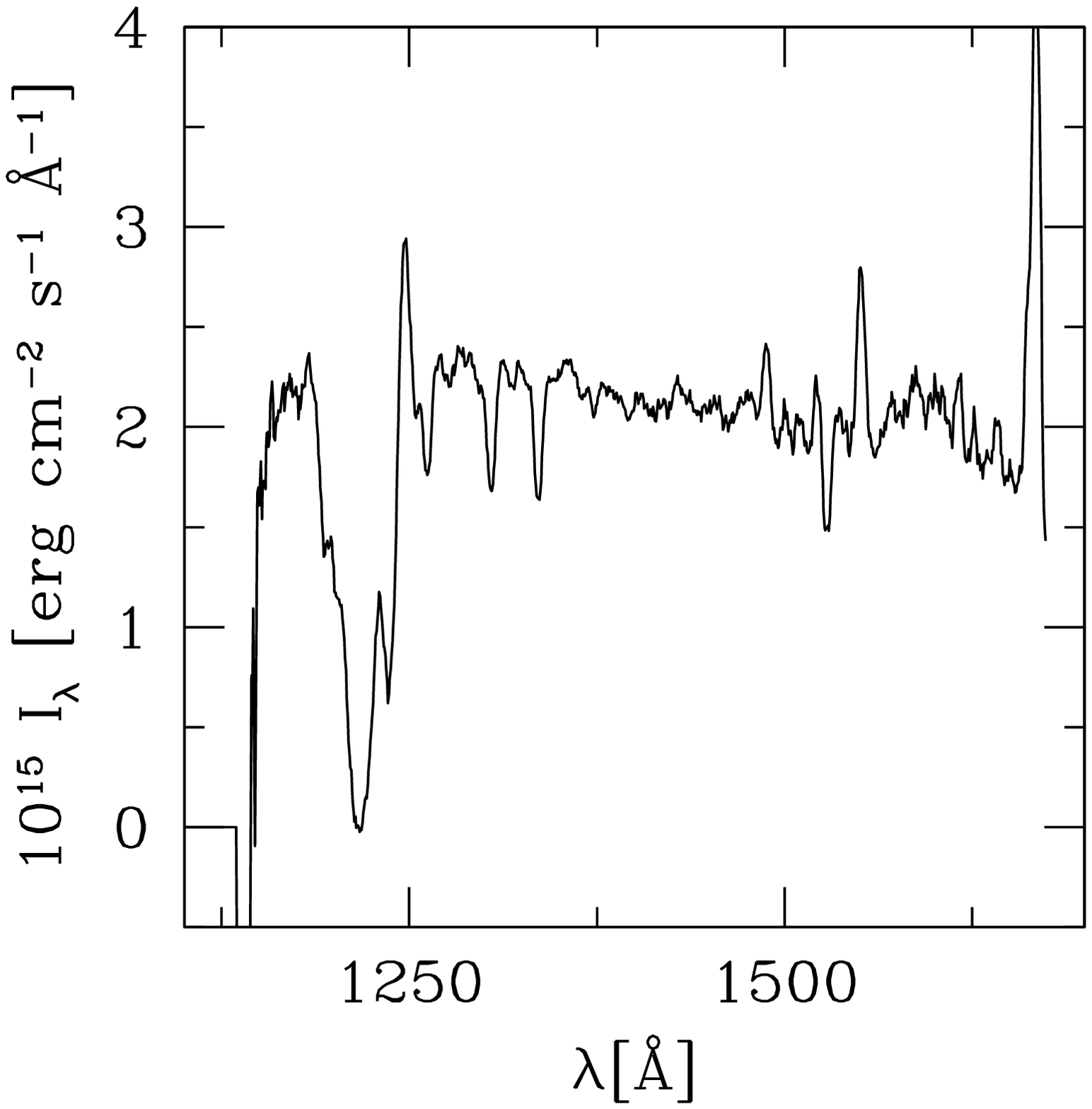}
\newpage
\plotone{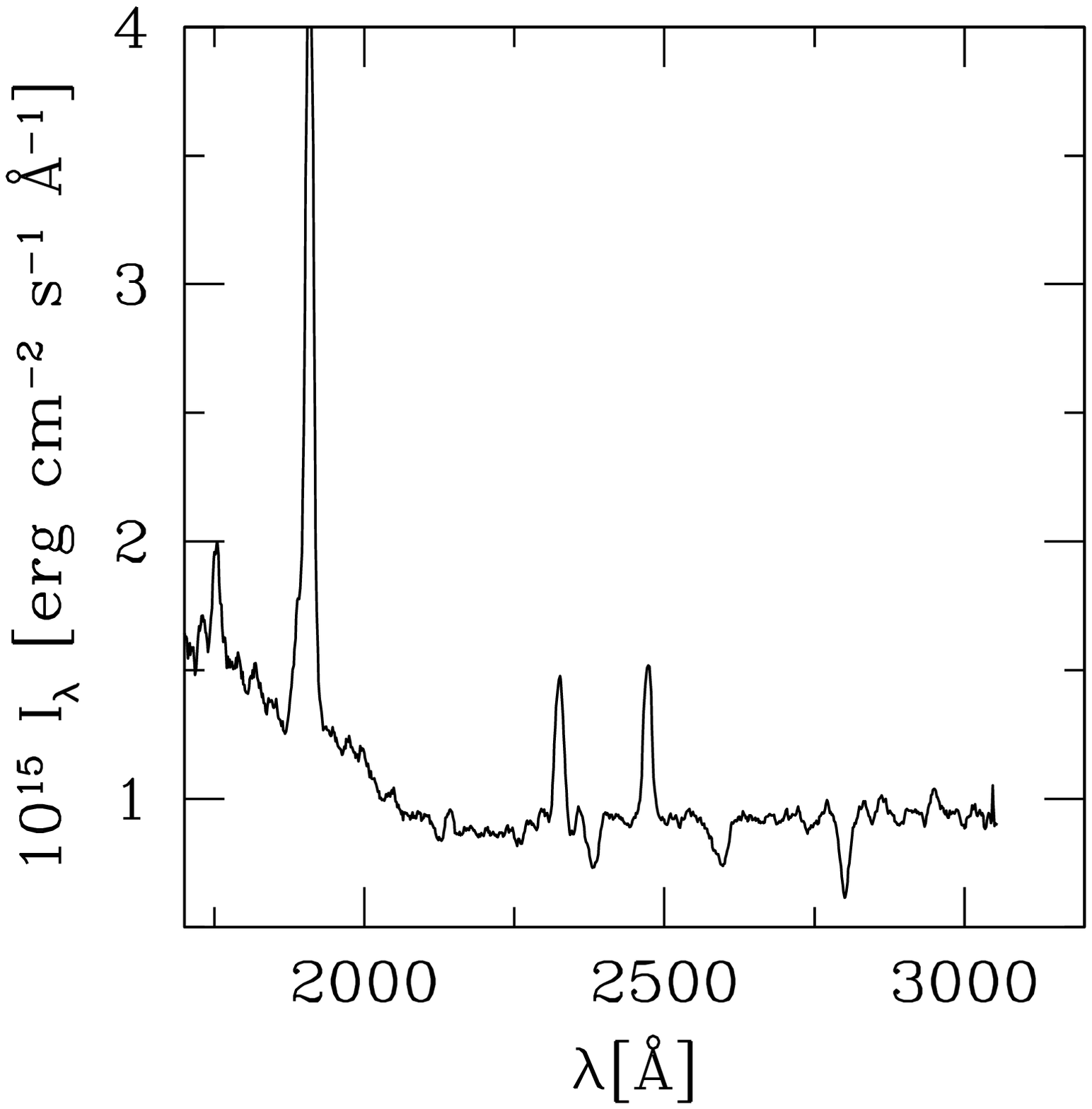}
\newpage
\plotone{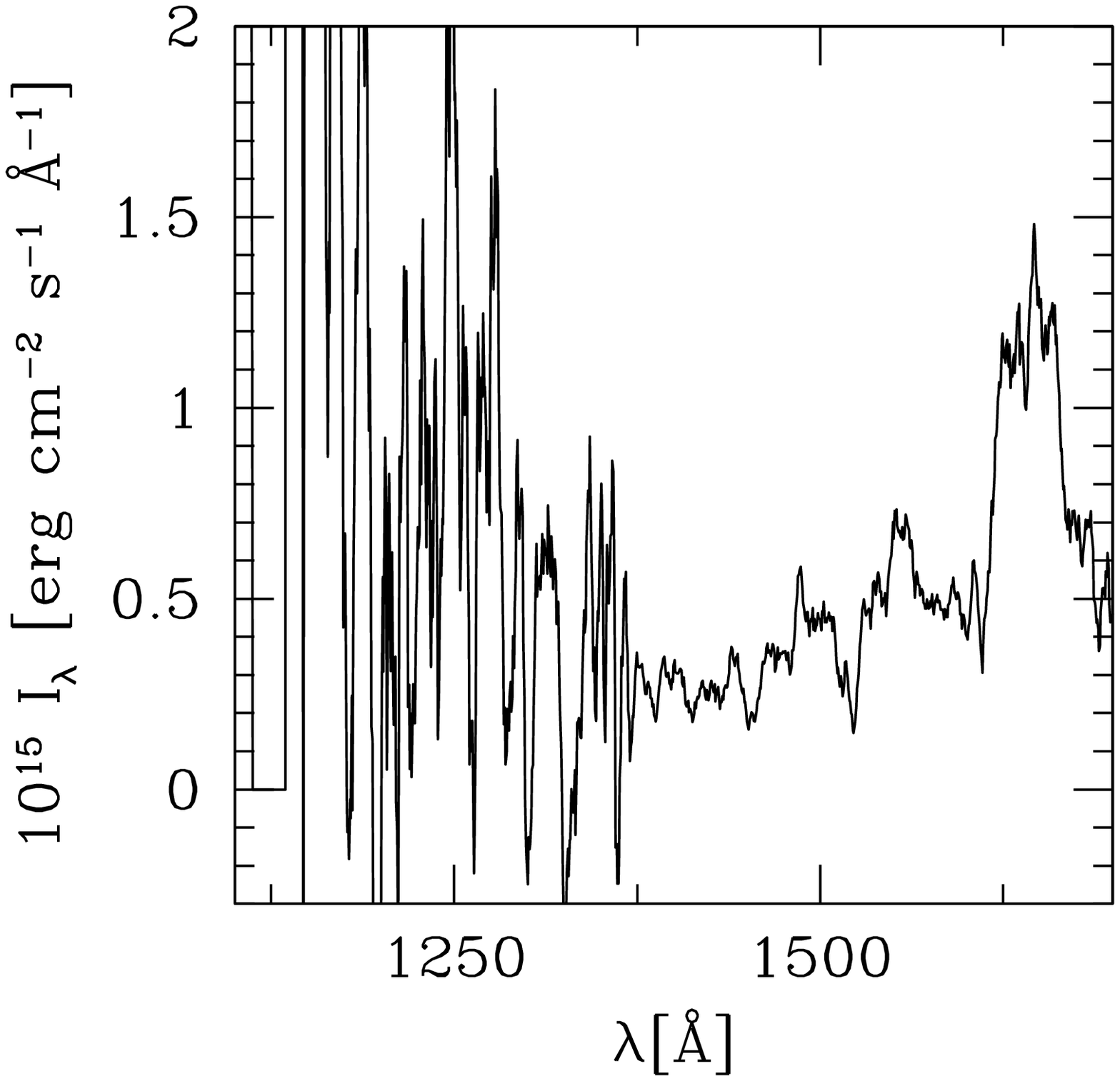}
\newpage
\plotone{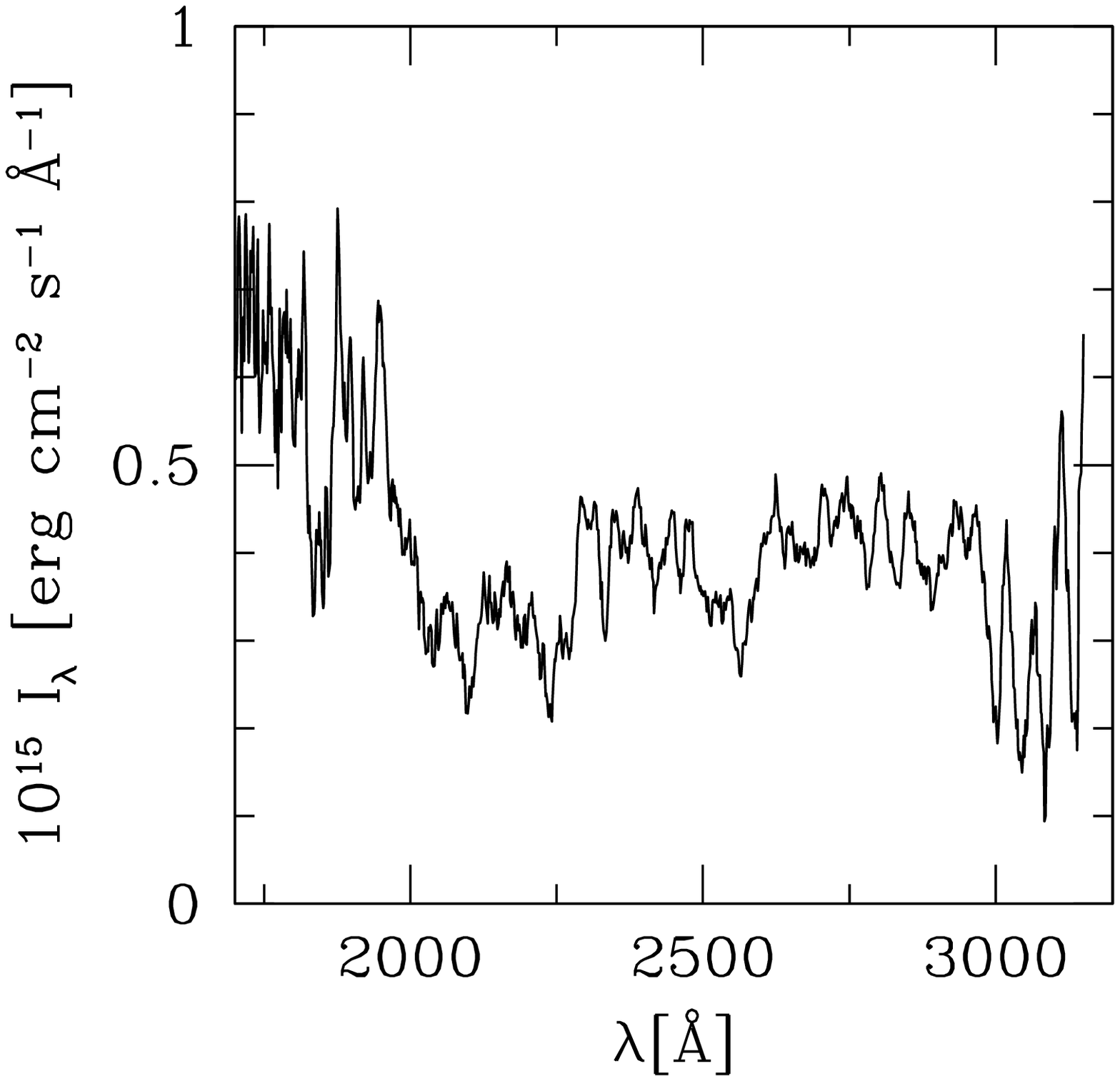}
\newpage
\plotone{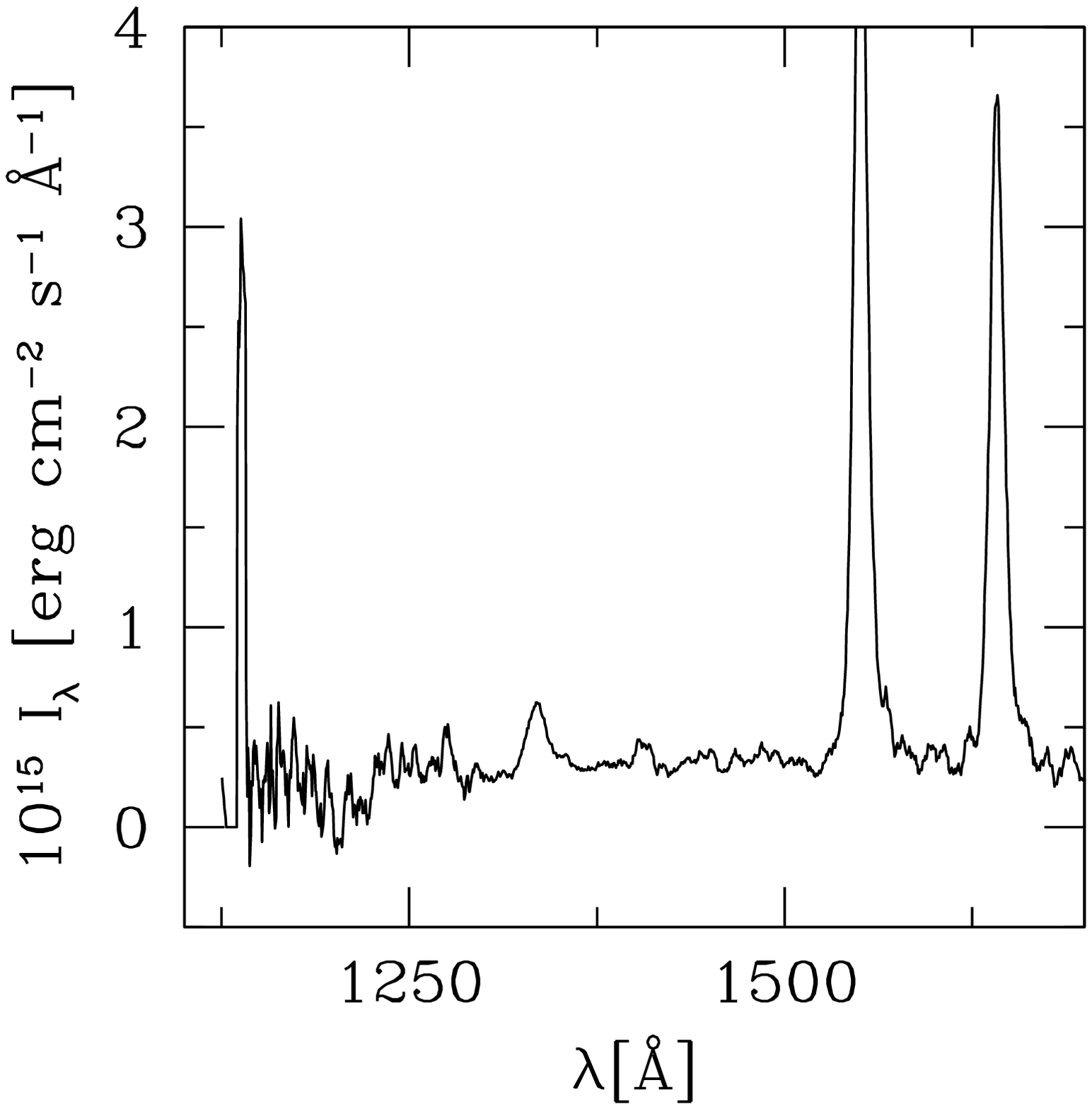}
\newpage
\plotone{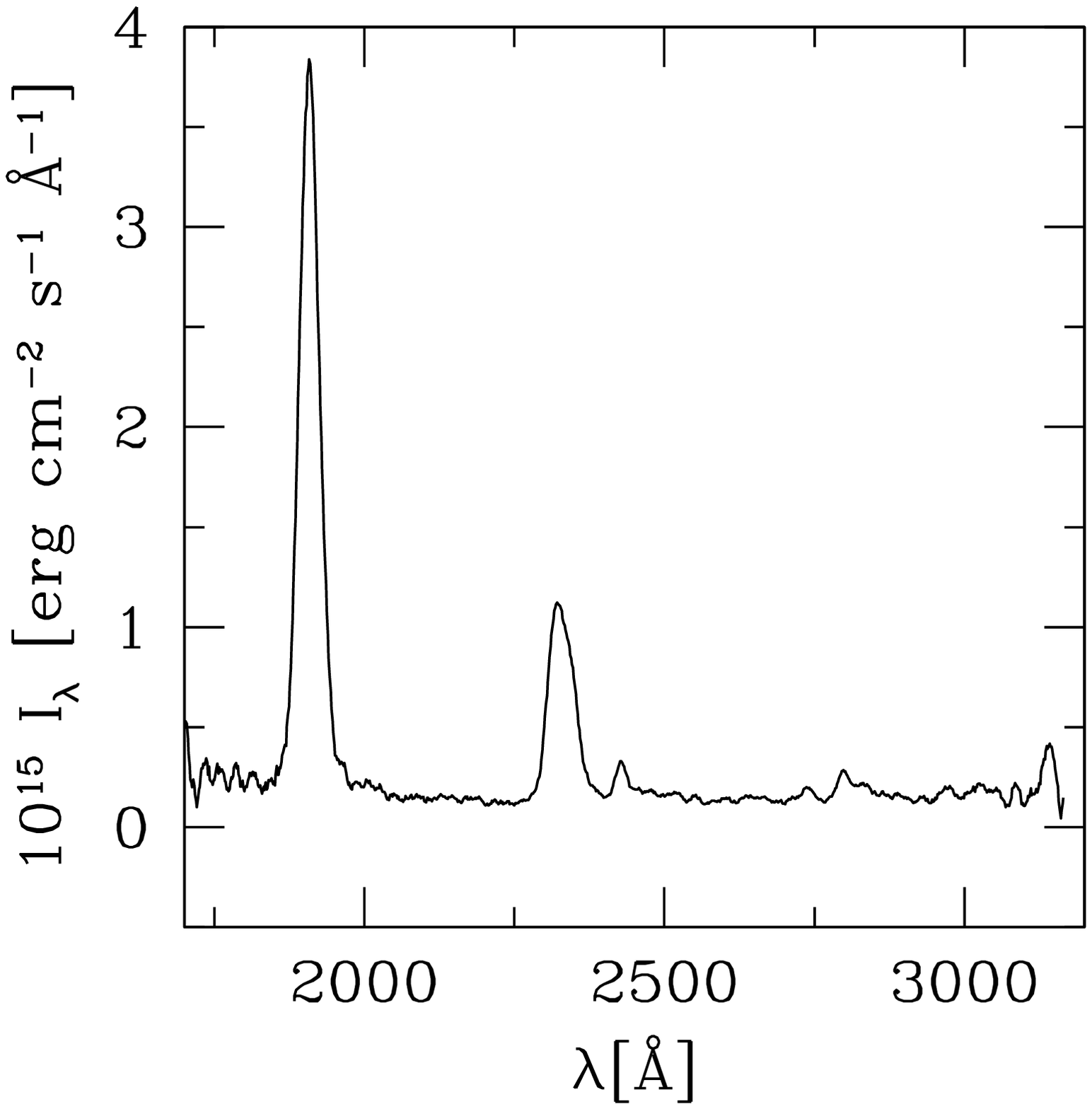}
\newpage
\plotone{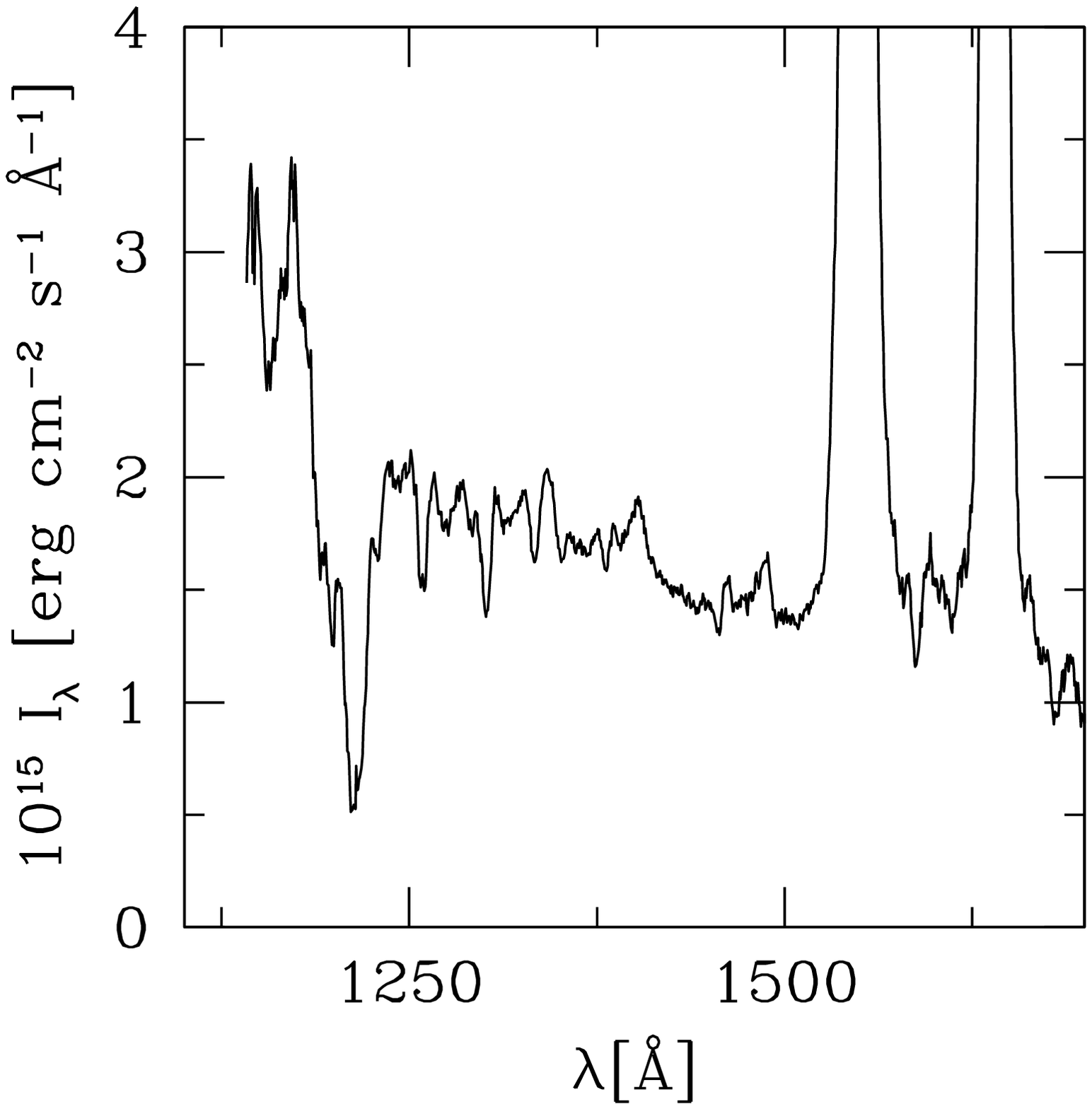}
\newpage
\plotone{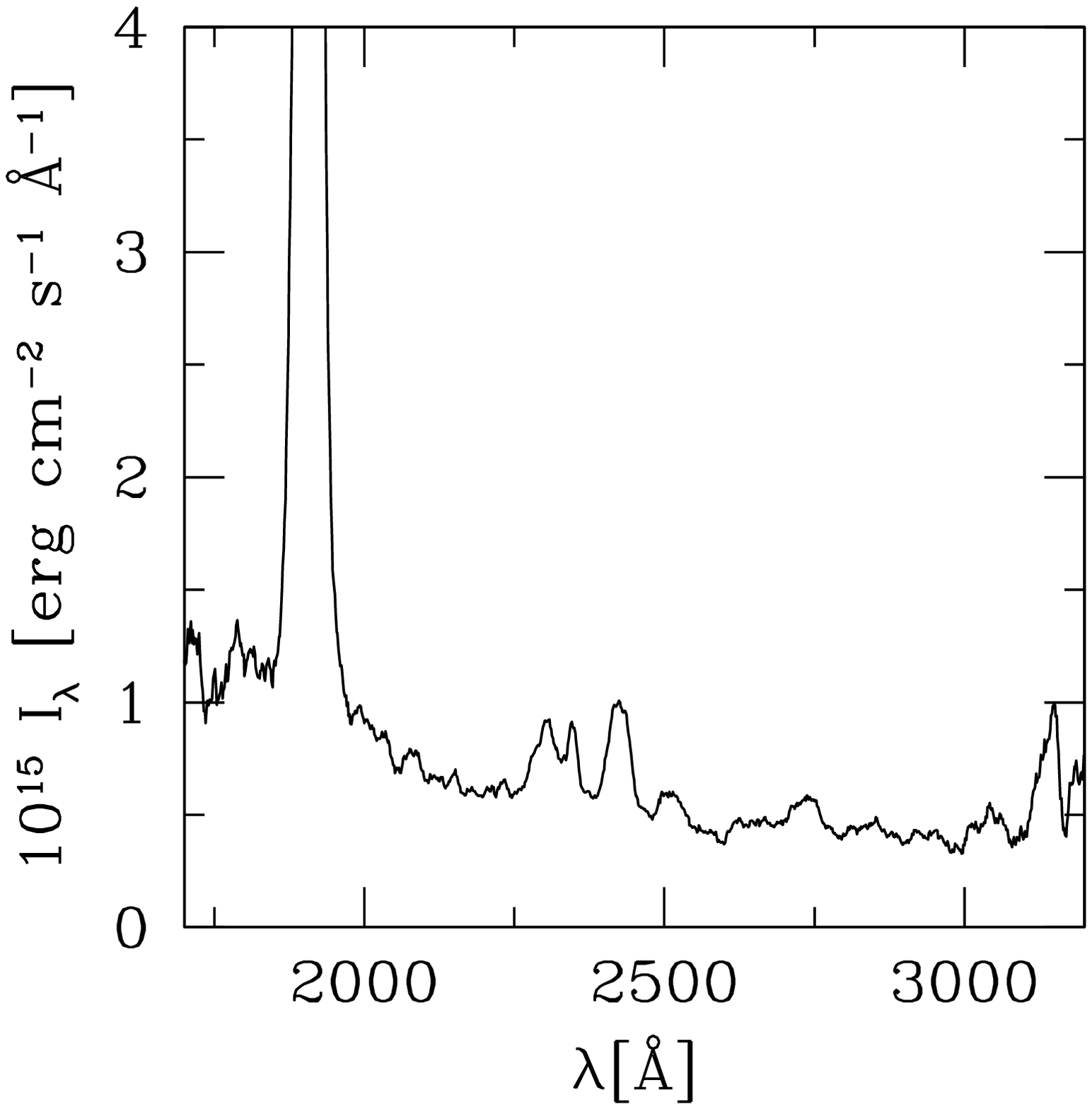}
\newpage
\plotone{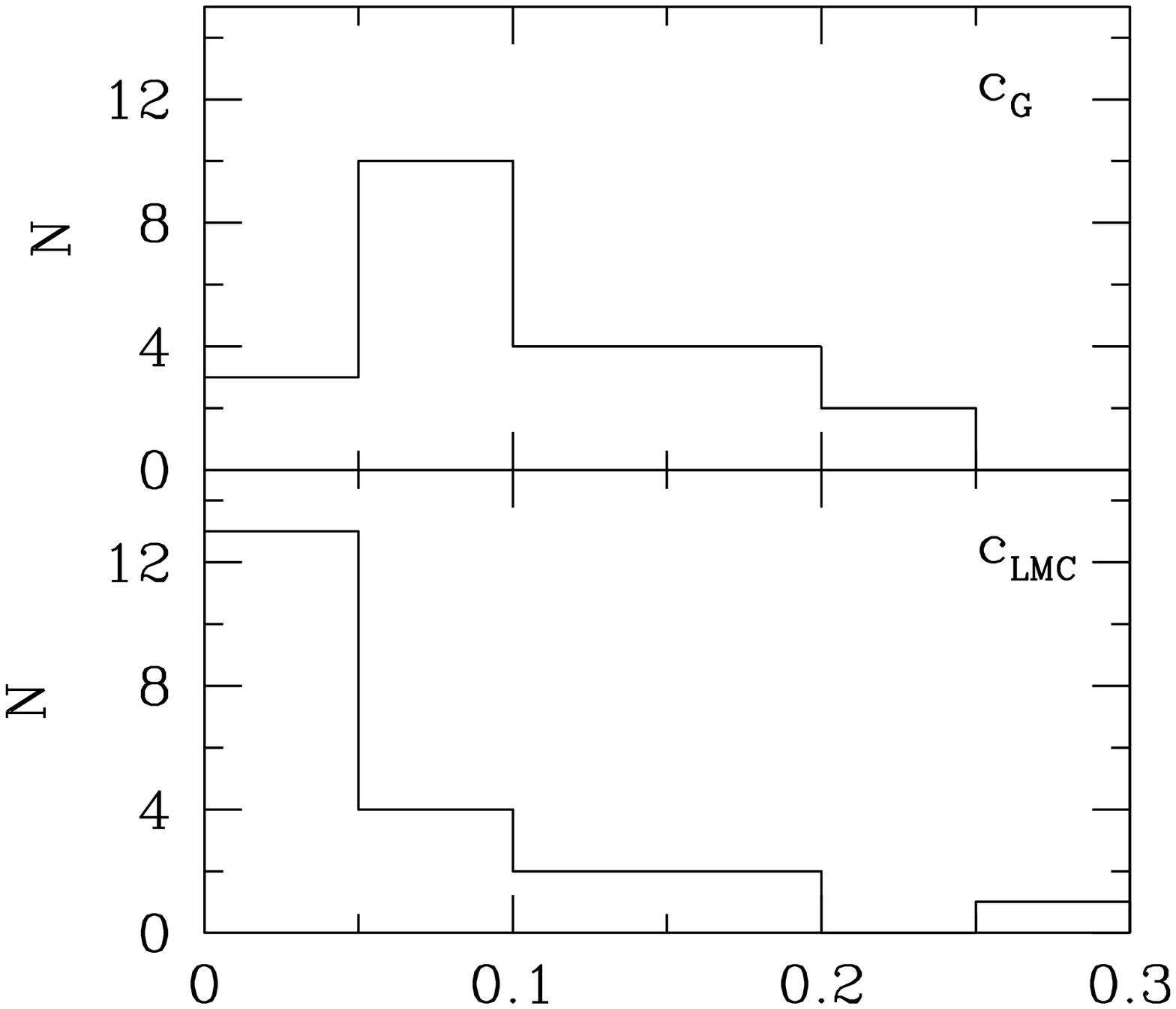}
\newpage
\plotone{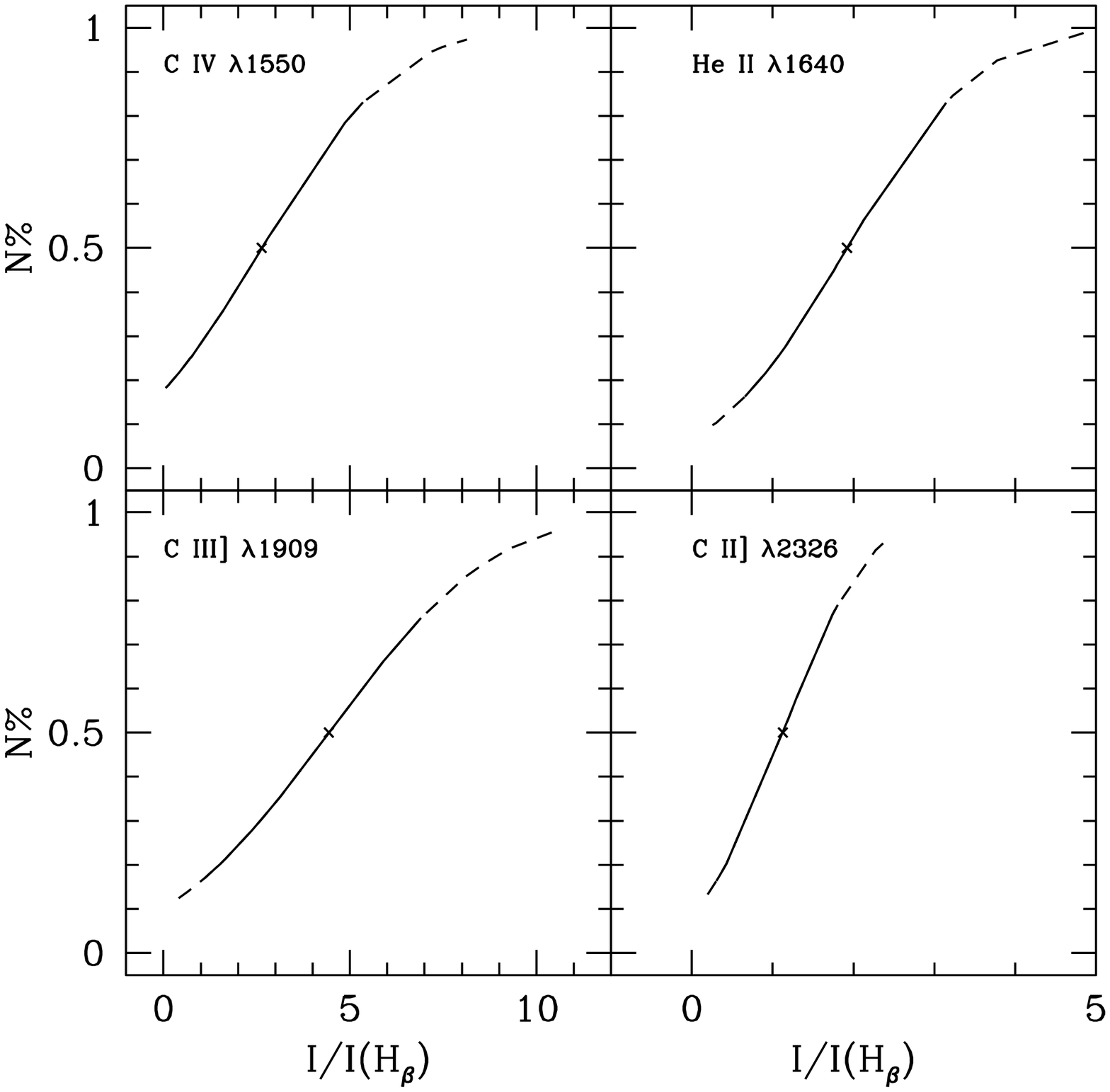}
\newpage
\plotone{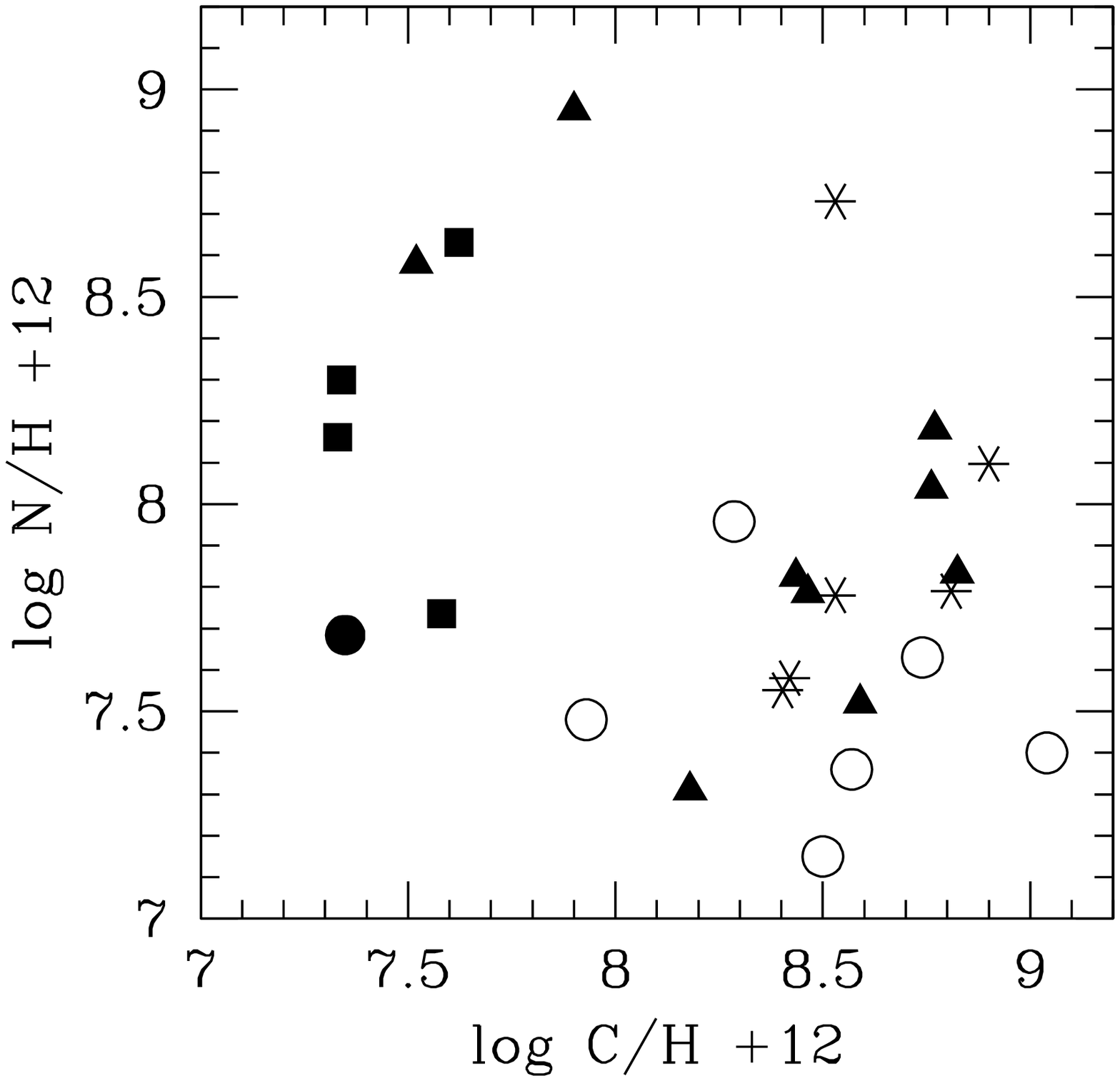}
\newpage
\plotone{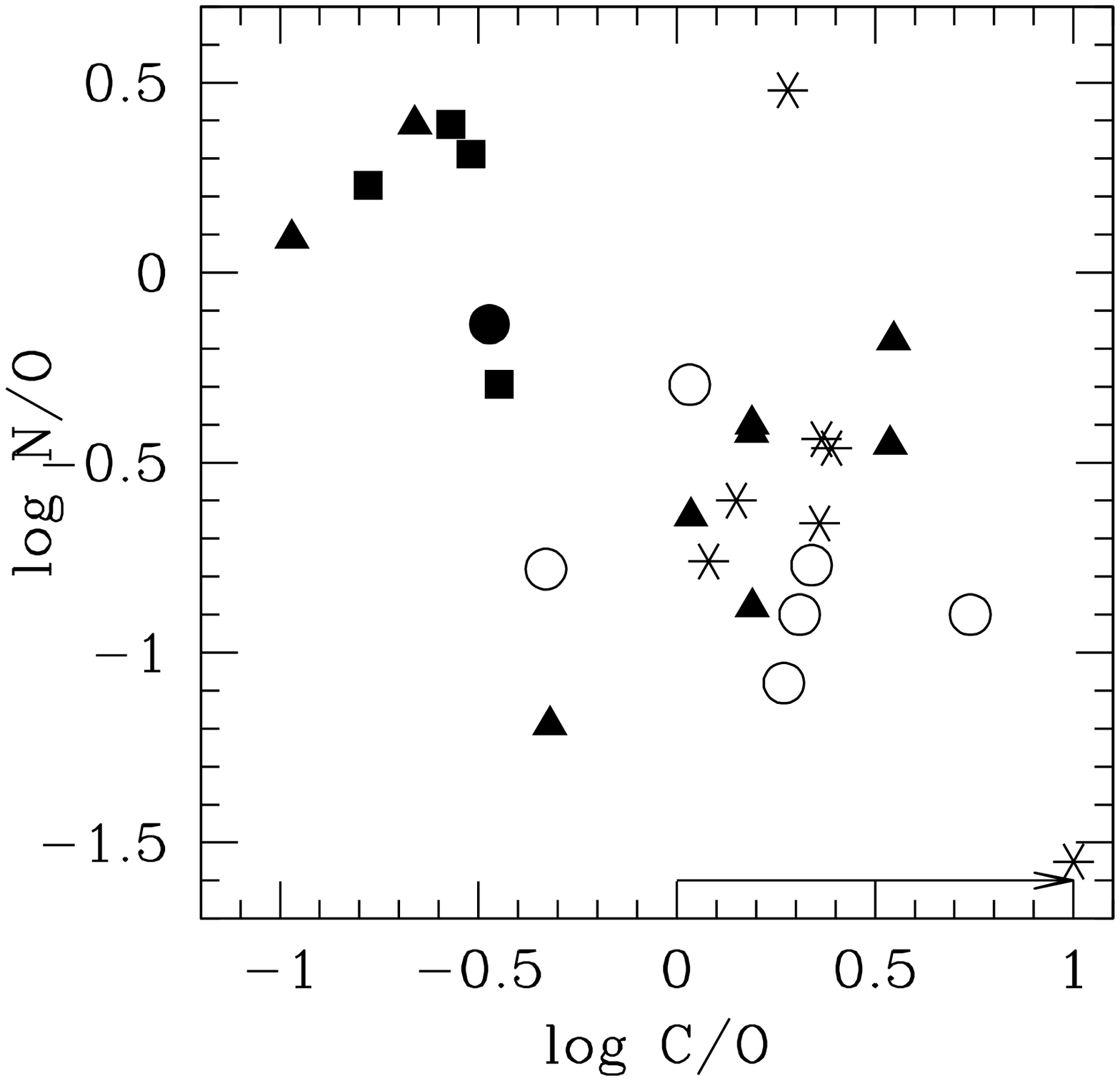}
\newpage
\plotone{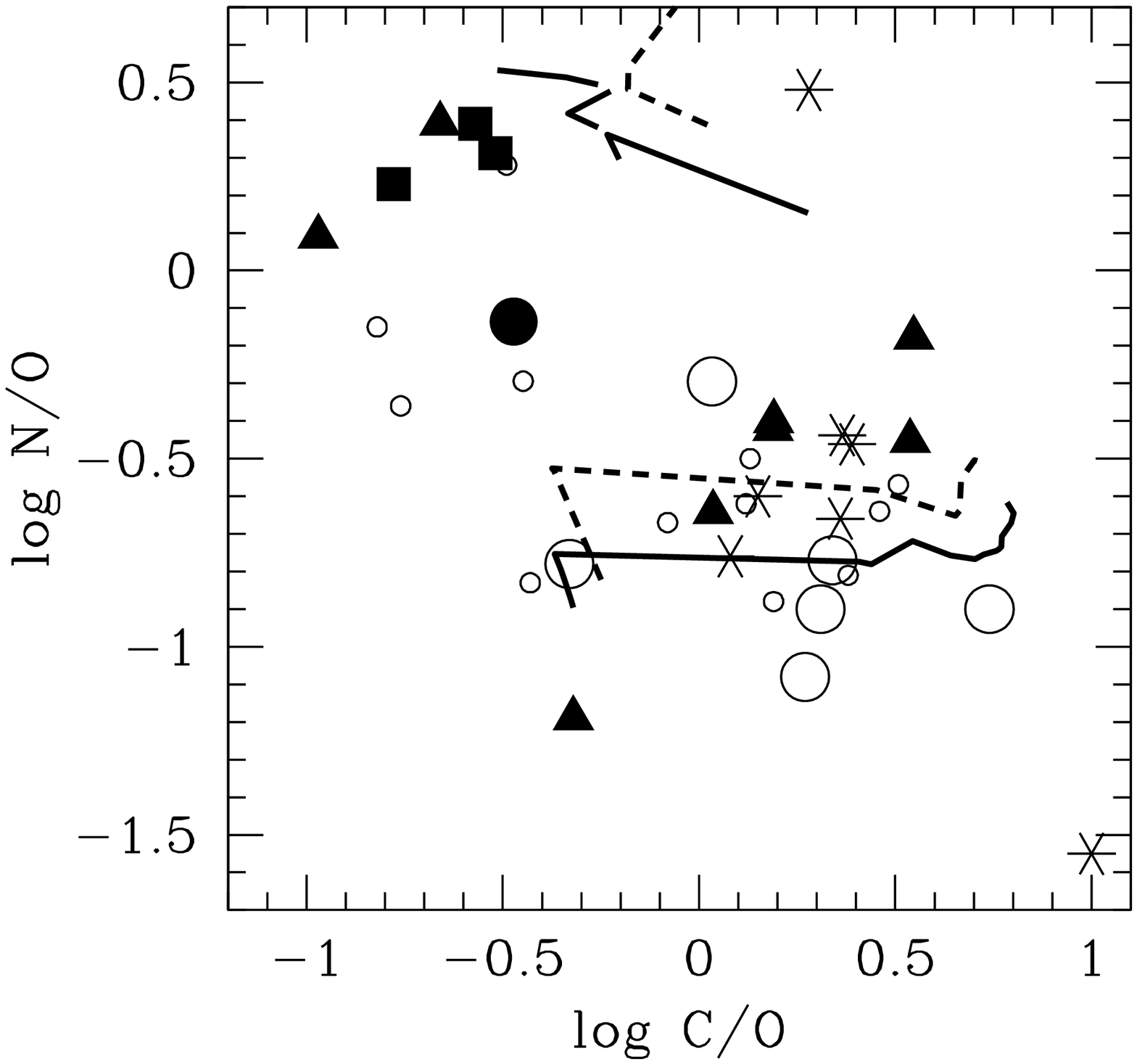}
\newpage
\plotone{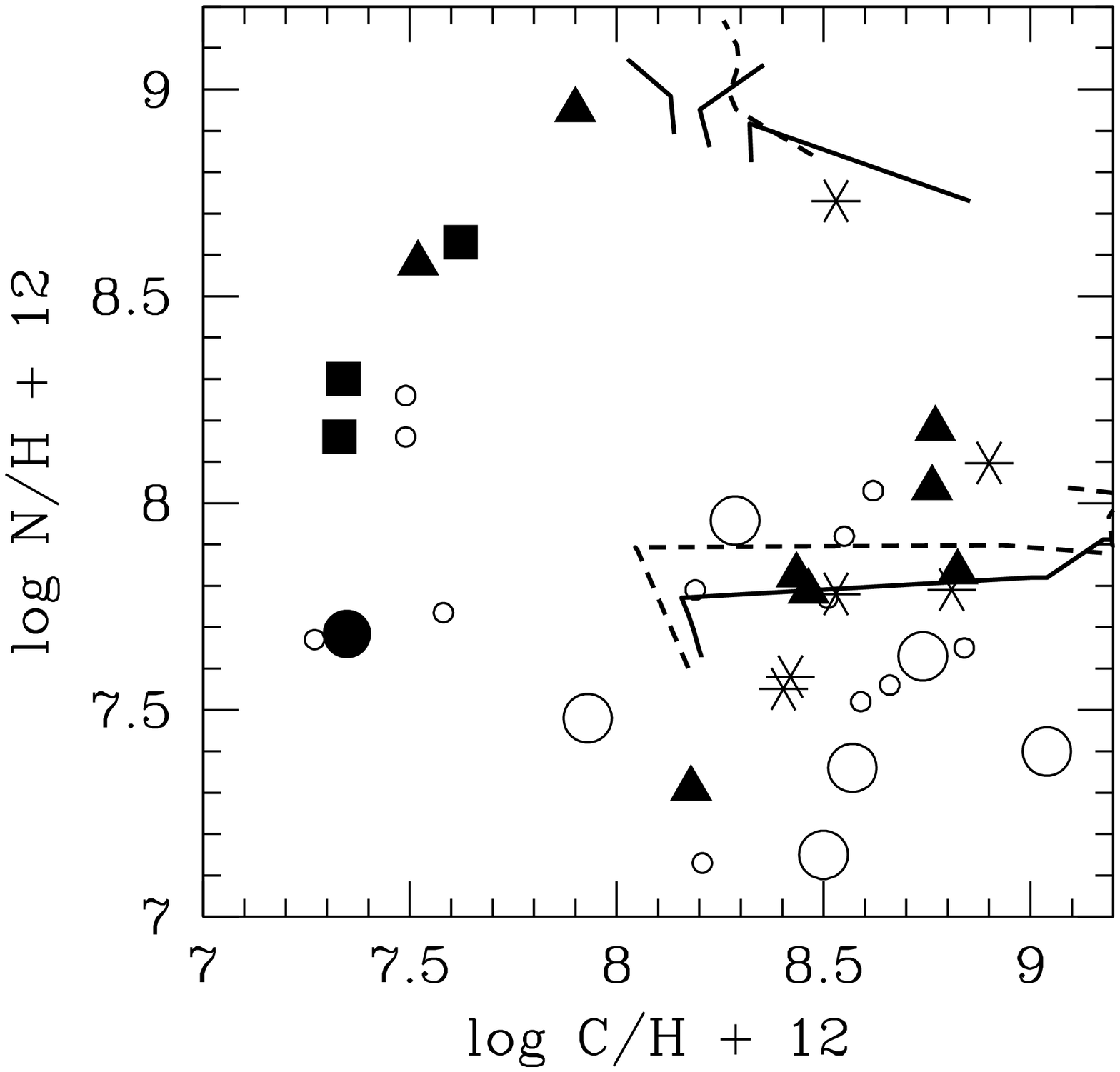}
\newpage
\plotone{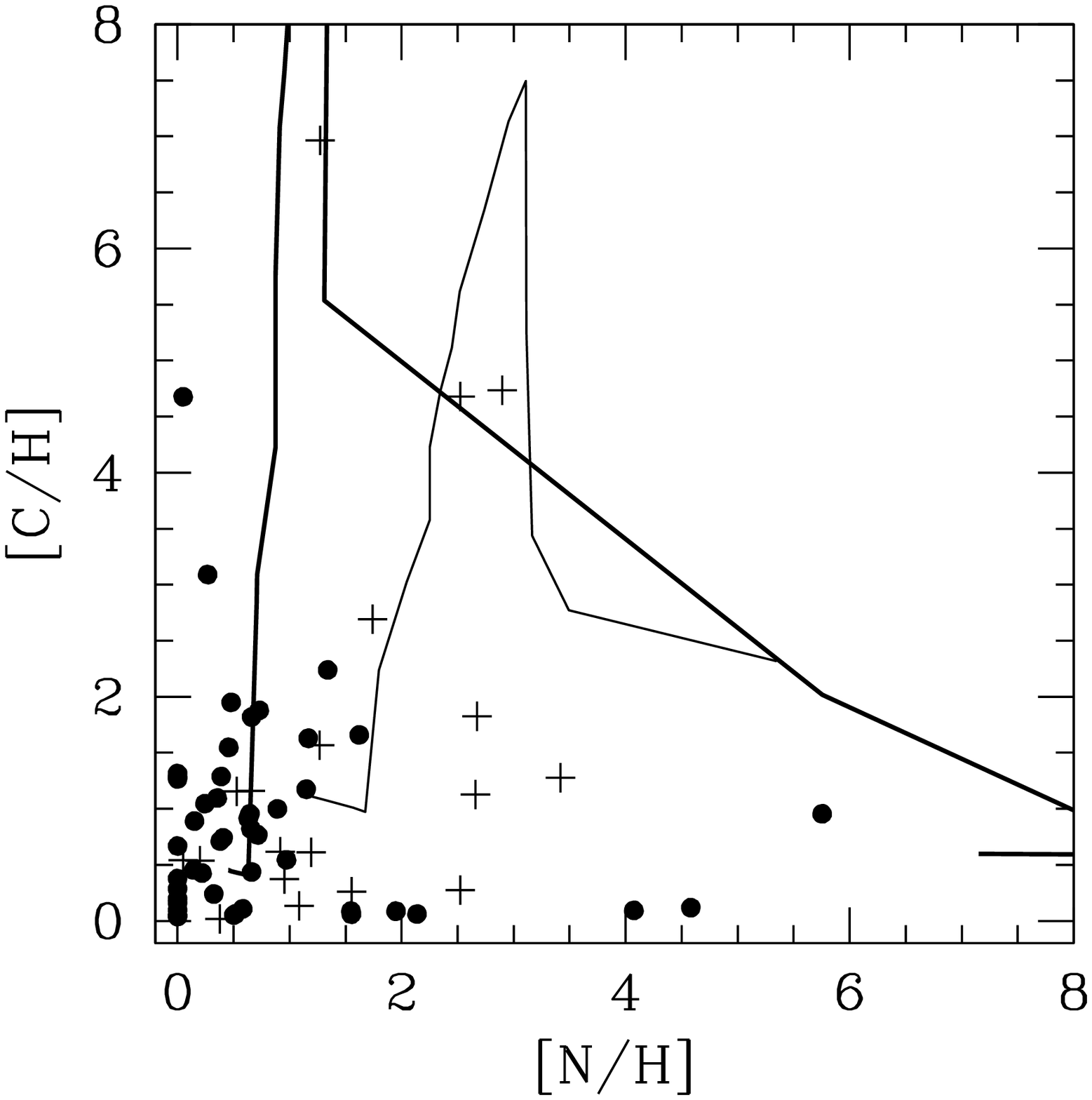}
\newpage
\plotone{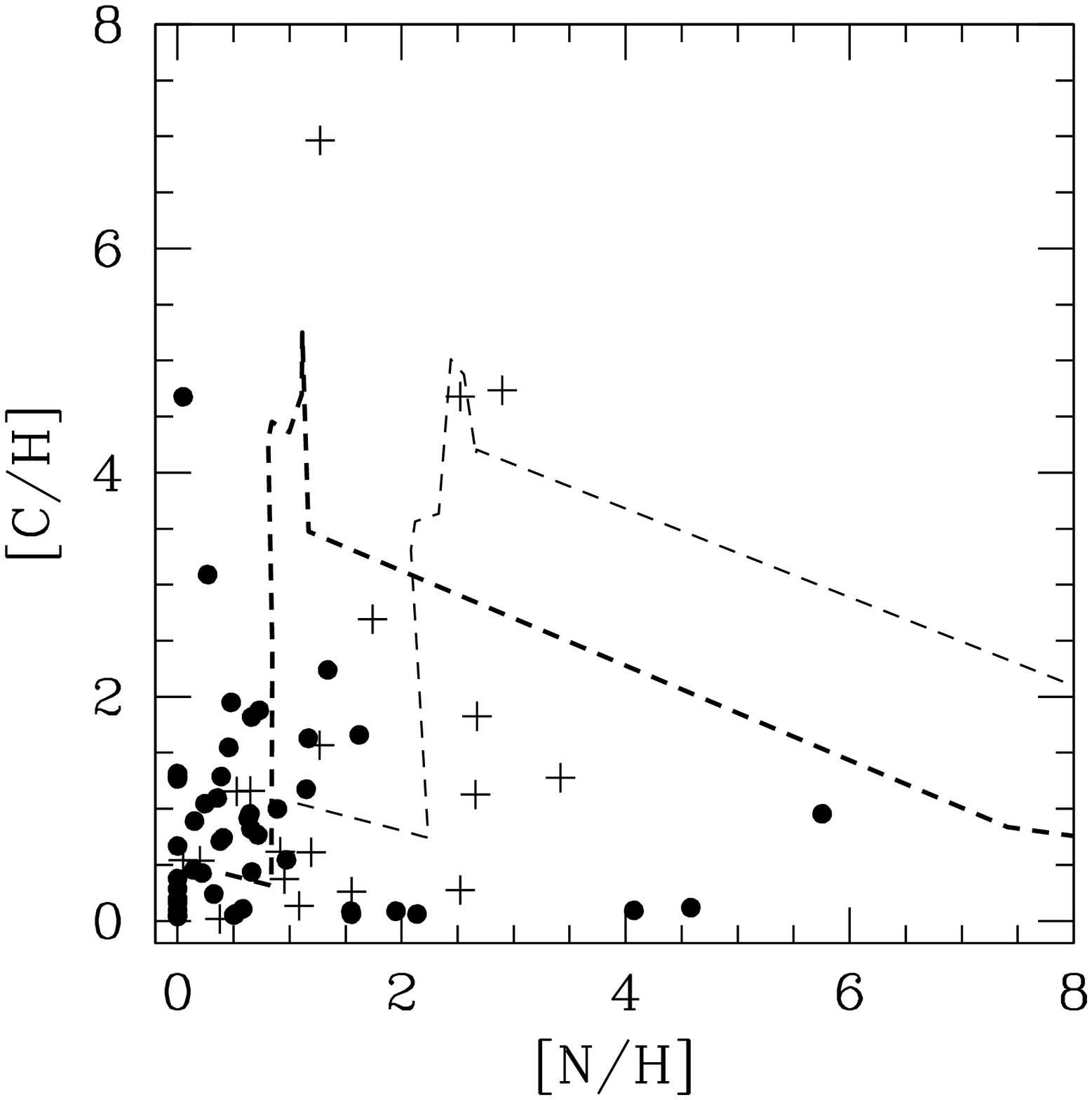}
\newpage
\plotone{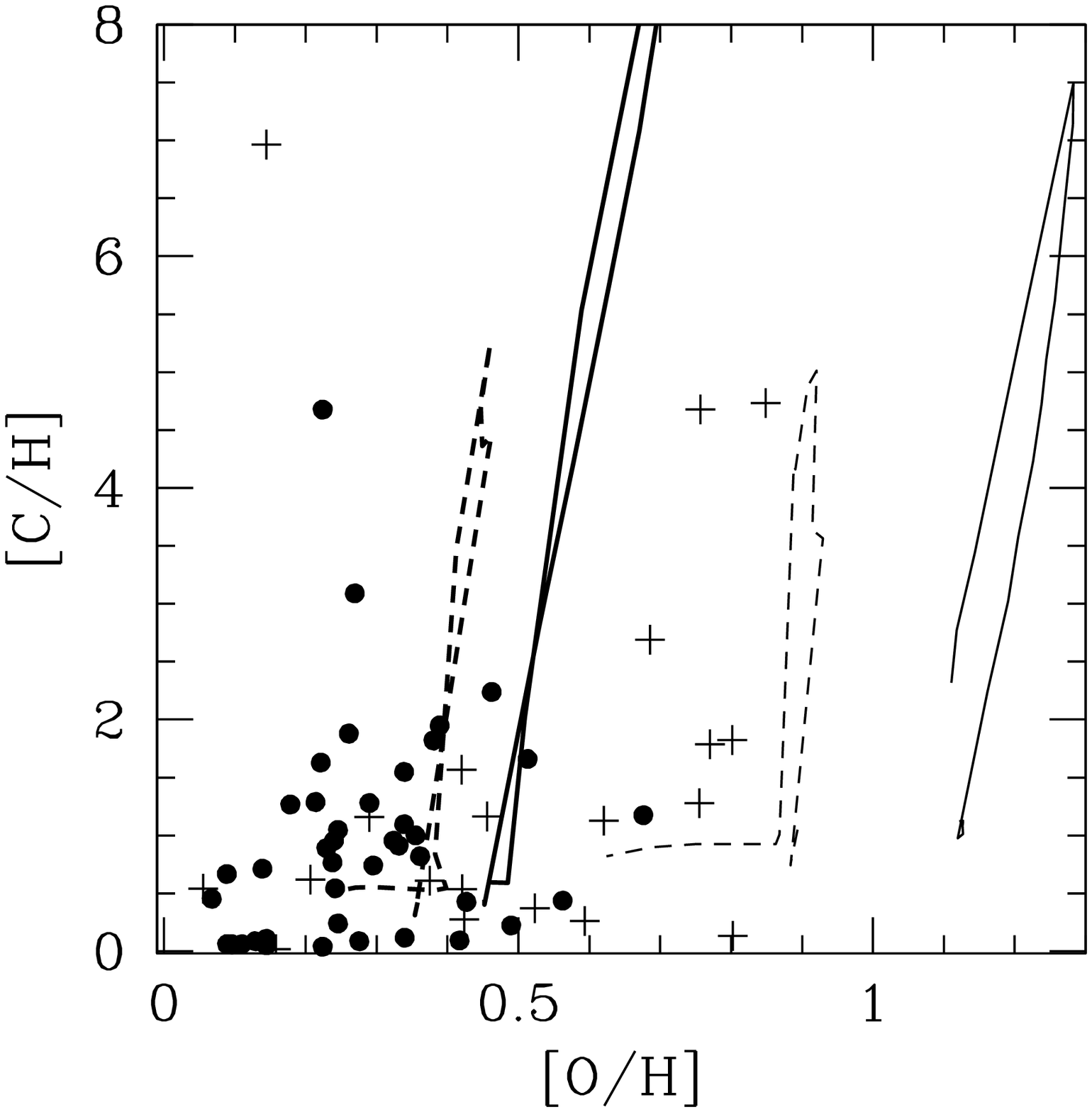}

\end{document}